\documentclass[a4paper,12pt]{article}
\usepackage[T2A]{fontenc}
\usepackage[utf8]{inputenc}
\usepackage{cmap}
\usepackage{slashed}
\usepackage{amsmath,amsthm}
\usepackage{amsfonts}
\usepackage{mathrsfs}
\usepackage{amssymb}
\usepackage{multirow}
\usepackage[normalem]{ulem}
\usepackage{physics}
\usepackage{bm} 

\newcommand{\beq}[1]{\begin{equation} \label{#1}}
\newcommand{\eeq}{\end{equation}}

\newcommand{\bea}[1]{\begin{eqnarray} \label{#1}}
\newcommand{\eea}{\end{eqnarray}}

\usepackage[svgnames,math]{xcolor}
\definecolor{refBlue}{rgb}{0.0, 0.0, 0.32}
\definecolor{urlBlue}{rgb}{0.0, 0.0, 0.46}
\usepackage{hyperref}
\hypersetup{
    colorlinks=true,
    linkcolor=refBlue,
    citecolor=refBlue,
    urlcolor=urlBlue,
    }

\numberwithin{equation}{section} 

\allowdisplaybreaks[1]

\usepackage{tcolorbox}

\newcommand{\bGray}{\begin{tcolorbox}[arc=0pt, colback=gray!10, boxrule=0pt] \it}
\newcommand{\eGray}{\end{tcolorbox}}

\newcommand{\bNew}{\vspace{-1mm}\begin{tcolorbox}[arc=0pt, colback=yellow!10, boxrule=0pt] \vspace{-2mm} \it}
\newcommand{\eNew}{\vspace{-2mm}\end{tcolorbox}\vspace{-1mm}}

\newcommand{\newpar}{\vspace{2mm}}

\makeatletter
\newcommand{\mathboldletters}[1]{\mathboldletters@process#1\@empty}
\def\mathboldletters@process#1{%
  \ifx\@empty#1
  \else
    \ifx#1s\relax
       {\text{\scalebox{1.15}[1.30]{$\scriptscriptstyle{\bm{s}}$}}}
    \else
      \ifx#1r\relax
         {\text{\scalebox{1.13}[1.27]{$\scriptscriptstyle{\bm{r}}$}}}
      \else
        \ifx#1p\relax
          \bm{\scriptscriptstyle{p}}%
        \else
          \ifcat\noexpand#1A
            \bm{\scriptscriptstyle{#1}}%
          \else
            \ifcat\noexpand#10\relax
              {\scriptscriptstyle #1}%
            \else
              \bm{#1}%
            \fi
          \fi
        \fi
      \fi
    \fi
    \expandafter\mathboldletters@process
  \fi
}
\newcommand{\mathboldall}[1]{\mathboldall@process#1\@empty}
\def\mathboldall@process#1{%
  \ifx\@empty#1
  \else
    \ifx#1s\relax
       {\text{\scalebox{1.25}[1.40]{$\scriptscriptstyle{\bm{s}}$}}}
    \else
      \ifx#1r\relax
         {\text{\scalebox{1.20}[1.35]{$\scriptscriptstyle{\bm{r}}$}}}
      \else
        \ifx#1p\relax
          \bm{\scriptscriptstyle{p}}%
        \else
          \ifcat\noexpand#1A
            \bm{\scriptscriptstyle{#1}}%
          \else
            \ifcat\noexpand#10\relax
              \bm{\scriptscriptstyle #1}%
            \else
              \bm{#1}
            \fi
          \fi
        \fi
      \fi
    \fi
    \expandafter\mathboldall@process
  \fi
}
\makeatother

\def\Bm{\mathboldletters} 
\def\BM{\mathboldall} 

\if{

}\fi

\newcommand{\HIDE}[1]{}

\newcommand{\gfX}[2][]{X^{{\scriptscriptstyle\gix{#1}}{\scriptscriptstyle\gix{#2}}}}

\newcommand{\MGen}[1]{{\color{genColor}\text{\small$ #1 $}}}
\newcommand{\MgfX}[1]{{\color{myGreen}\small $ #1 $}}
\newcommand{\MgfS}[1]{{\color{myGreen}\small$ #1 $}}
\newcommand{\MgfO}[1]{{\color{myGreen}\small$ #1 $}}

\newcommand{\vphII}{\vphantom{{\big|^I_g}^a}}

\begin{document}

\begin{titlepage}

\begin{center}
{\LARGE\bf Covariant quantization of totally}

\vspace{0.1cm}

{\LARGE\bf antisymmetric tensor-spinor field in $AdS_d$}

\vspace{0.5cm}
{\bf \large A.O.\hspace{2pt}Barvinsky$^{1,2}$, I.L.\hspace{2pt}Buchbinder$^{3,4,5}$, V.A.\hspace{2pt}Krykhtin$^{6}$, D.V.\hspace{2pt}Nesterov$^{1}$}

\vspace{0.5cm}
{\it $^1$I.E. Tamm Theory Department, P.N. Lebedev Physical Institute,\\
53 Leninsky Prospect, 119991, Moscow, Russia,\\
{\tt barvin@lpi.ru,\, nesterov@lpi.ru}\\
$^2$Institute for Theoretical and Mathematical Physics,
Moscow State University, Leninskie Gory, GSP-1, Moscow, 119991, Russia,\\
$^3$Bogoliubov Laboratory of Theoretical Physics,\\
 Joint Institute for Nuclear Research, 6, Joliot Curie, 141980 Dubna, Russia,\\
  {\tt buchbinder@theor.jinr.ru}\\
$^4$Center of Theoretical Physics, Tomsk State Pedagogical University,\\
$^5$National Research Tomsk State University,
Lenin Av. 36, 634050, Tomsk, Russia,\\
$^6$Tomsk Polytechnic University, Lenin Av., 30, 634050, Tomsk, Russia,\\
{\tt krykhtin@tpu.ru}}

\end{center}

\begin{abstract}
  We develop the quantization of a recently proposed model describing a totally antisymmetric rank-$p$ tensor-spinor field (a fermionic $p$-form theory) in $d$-dimensional anti–de Sitter (AdS) space. The model provides a new nontrivial example of a reducible gauge theory, in which gauge transformations are linearly dependent and the degree of reducibility increases with $p$. It is well known that in such cases the standard Faddeev-Popov-DeWitt prescription for the generating functional is not applicable. We quantize the fermionic $p$-form theory using the general Batalin-Vilkovisky (BV) formalism, employing two distinct gauge fermions associated with gauge-fixing functions of different admissible ranks,\HIDE{ thereby} confirming the independence\HIDE{ of the quantization} from the gauge choice. As a result, we obtain the quantum effective action\HIDE{ for the theory} in terms of a sequence of functional determinants corresponding to specific Dirac-like operators on AdS space.
\end{abstract}

{\small
  \setcounter{tocdepth}{2}
  \tableofcontents
}
\end{titlepage}


\definecolor{myComponent}{rgb}{0.2, 0.15, 0.1}
\definecolor{myGreen}{rgb}{0.0, 0.25, 0.0} 
\definecolor{myGray}{rgb}{0.2, 0.2, 0.0} 
\definecolor{myBrown}{rgb}{0.3, 0, 0.3}
\definecolor{genColor}{rgb}{0.5, 0.5, 1.0}  
\definecolor{mypaleBlue}{rgb}{0.2, 0.2, 0.4} 


\newcommand{\gix}[1]{{{\color{myBrown}{\BM{\scriptscriptstyle #1}} \hspace{.5pt}}}} 
\newcommand{\hgix}[1]{} 

\newcommand{\rix}[1]{{{\color{myGreen}\BM{\scriptscriptstyle #1} \hspace{.5pt}}}} 

\newcommand{\pix}[1]{{\!\!{\color{myGreen}\BM{\scriptscriptstyle #1} \hspace{.5pt}}}} 

\newcommand{\oix}[1]{{{\color{myGreen}\BM{\scriptscriptstyle #1}}}} 



\newcommand{\GG}{\textsl{g}\hspace{0.5pt}}  

\def\nfrac{\hspace{1pt}\tfrac}  

\newcommand{\sqrr}{{\hspace{-1pt}\sqrt{r_{\scriptscriptstyle{\hspace{-0.5pt}0}}\!}\hspace{1.5pt}}}

\newcommand{\mAdS}{m_{\scriptscriptstyle{\hspace{-0.5pt}0}}} 

\newcommand{\nnn}{n} 
\newcommand{\mmm}{m} 

\newcommand{\gvar}[1][]{{\color{Indigo}\operatorname\delta}_{#1}}
\newcommand{\glambda}{\lambda} 



\newcommand{\msp}{\mathchoice{{\vcenter{\hbox{$\scriptstyle{(-)}$}}}}{{\raisebox{2pt}{\ensuremath$\scriptstyle{(-)}$}}}{\scriptscriptstyle{(-)}}{(-)}}
\newcommand{\esm}{\hspace{0.5pt}{\epsilon}_{}}

\newcommand{\dcsc}[2][]{ {\color{myGray}{\dc{\alpha}}} ^{{\rix{#2}}} _{{\rix{#1}}} }


\newcommand{\anc}[2][]{ {\color{myGray}{a ^{{\color{myComponent}\rix{#2}}} _{{\color{myComponent}\rix{#1}}} }}}  
\newcommand{\bnc}[2][]{ {\color{myGray}{b ^{{\color{myComponent}\rix{#2}}} _{{\color{myComponent}\rix{#1}}} }}}  
\newcommand{\cnc}[2][]{ {\color{myGray}{c ^{{\color{myComponent}\rix{#2}}} _{{\color{myComponent}\rix{#1}}} }}}  


\newcommand{\defeq}{\mathrel{\color{Navy}=}} 

\newcommand{\tleq}{{\:\leq\:}}
\newcommand{\tgeq}{{\:\geq\:}}

\newcommand{\rker}{\,\mathop{\mathrm{rker}\hspace{1pt}}}
\newcommand{\im}{\,\mathop{\mathrm{im}\hspace{1pt}}}

\newcommand{\gh}[1]{\mathop{gh}{(#1)}}  
\newcommand{\af}{\star}


\newcommand{\hc}[1]{{#1}^+} 
\newcommand{\dc}[1]{\bar{#1}} 
\newcommand{\dcl}[1]{\overline{\vphantom{|^l}#1}} 


\newcommand{\Al}{{\scriptscriptstyle [\hspace{0.2pt} }} 
\newcommand{\Ar}{{ \hspace{0.2pt} \scriptscriptstyle ] }} 


\newcommand{\ro}{\mathchoice {r_{\!\circ\hspace{-2pt}}}
                             {r_{\!\circ\hspace{-2pt}}} {\BM{r}_{{\hspace{-1.5pt}\text{\scalebox{0.55}{$\circ$}}}\hspace{-0.5pt}}} {\BM{r}_{{\hspace{-1.5pt}\text{\scalebox{0.55}{$\circ$}}}\hspace{-0.5pt}}}}   

\newcommand{\lo}{\mathchoice {l_{\circ\hspace{-1pt}}}
                             {l_{\circ\hspace{-1pt}}} {l_{{\hspace{-0.2pt}\text{\scalebox{1.0}{$\circ$}}}\hspace{-0.5pt}}} {l_{{\hspace{-0.2pt}\text{\scalebox{0.8}{$\circ$}}}\hspace{-0.5pt}}}}   

\renewcommand{\lll}{l} 

\newcommand{\iBV}{\scriptscriptstyle{BV}}
\newcommand{\lead}{\!\!\rix{\blacktriangledown}} 
\newcommand{\sublead}{\!\!\rix{\vartriangle}} 


\newcommand{\FIELD}[1]{\mathcal{#1}}
\newcommand{\IRRC}[1]{{\hspace{0.5pt}\scalebox{0.85}[0.95]{\text{$\mathrm{#1}$}}}} 

\newcommand{\dcpsi}{\dc{\psi}}

\newcommand{\CPsi}[1][]{{\color{myComponent}{\text{\scalebox{0.9}{$\Psi$}}}\hspace{-0.5pt}} } \newcommand{\dcCPsi}[1][]{{\color{myComponent}\dc{\text{\scalebox{0.9}{$\Psi$}}}\hspace{-0pt}}}


\newcommand{\gC}{\FIELD{C}}   
\newcommand{\gB}{\FIELD{B}}   

\newcommand{\dcgC}{\dc{\FIELD{C}}}   
\newcommand{\dcgB}{\dc{\FIELD{B}}}   

\newcommand{\gCC}{\IRRC{C}}   
\newcommand{\gCB}{\IRRC{B}}   

\newcommand{\dcgCC}{\dc{\IRRC{C}}}   
\newcommand{\dcgCB}{\dc{\IRRC{B}}}   

\newcommand{\CPhi}{{\color{myComponent} \scalebox{1.25}{\text{$\mathtt{\Phi}$}} }}


\newcommand{\fA}{\FIELD{A}}  
\newcommand{\fCA}{\IRRC{A}}  

\newcommand{\dcfA}{\FIELD{A}}  
\newcommand{\dcfCA}{\IRRC{A}}  

\newcommand{\tB}{\FIELD{B}}  
\newcommand{\tCB}{\IRRC{B}}  



      \newcommand{\Lil}{{\scriptscriptstyle \color{myBlue} \hspace{-0.3pt}( }}
      \newcommand{\Lir}{{\scriptscriptstyle \color{myBlue} \hspace{-0.6pt}) }}

\newcommand{\zza}[1]{{{{\color{DarkBlue}\alpha}_{ \Bm{#1}}}}}  
\newcommand{\zzb}[1]{{{{\color{DarkBlue}\beta}_{ \Bm{#1}}}}}  
\newcommand{\zzc}[1]{{{\color{DarkBlue}\gamma}_{ \Bm{#1}}}}  


      \newcommand{\Mil}{{\scriptscriptstyle \color{mypaleBlue} [ }}
      \newcommand{\Mir}{{\scriptscriptstyle \color{mypaleBlue} ] }}

\newcommand{\Zk}{{\varkappa}} 
\newcommand{\Zl}{{\lambda}}   
\newcommand{\Zm}{{\mu}}       
\newcommand{\Zn}{{\nu}}       

\newcommand{\Zp}{\pi}     
\newcommand{\Zr}{{\hspace{-.2pt}\varrho\hspace{.1pt}}}  
\newcommand{\Zs}{\sigma}  
\newcommand{\Zt}{\tau}    

\newcommand{\Zo}{{\color{gray}{{\cdot}}}}   


\newcommand{\ZZm}[1]{{\mu \Mil {\color{mypaleBlue} \Bm{#1} } \Mir}} 
\newcommand{\ZZn}[1]{{\nu \Mil {\color{mypaleBlue} \Bm{#1} } \Mir}}
\newcommand{\ZZk}[1]{{\varkappa \Mil {\color{mypaleBlue} \Bm{#1} } \Mir}}
\newcommand{\ZZl}[1]{{\lambda \Mil {\color{mypaleBlue} \Bm{#1}} \Mir}}


\section{Introduction}
 \label{Sect:Introduction}
  \hspace{\parindent}

Totally antisymmetric tensor fields which are described geometrically by $p$-forms attract much attention in classical and quantum field theory due specific gauge structure and diverse dualities. Such fields naturally arise and are very broadly used in supergravity and string/brane theory (see \cite{J,O,FVP} and references therein). A remarkable feature of the theory of totally antisymmetric tensor fields is that their corresponding gauge transformations are defined up to additional gauge transformations. The latter are defined up to new gauge transformations, and so on up to some finite order which is called the order of reducibility. In general, the theories with reducible gauge transformations are called reducible and, as we now understand, for their covariant quantization simple Faddeev-Popov and DeWitt quantization rules are not applicable.

Dynamical model of free antisymmetric second rank tensor field in four dimensions (notoph theory) has been proposed for the first time by Ogievetsky and Polubarinov \cite{OP} and then rediscovered in \cite{Kalb-Ramond} in string context.\footnote{Various aspects of the pioneering paper \cite{OP} are discussed in \cite{EIvanov}.}  It was shown that in $4d$ this theory is off-shell dual to free scalar field model in flat space. Non-abelian generalization of the notoph theory has been given in \cite{FT} and it was proved that such a theory is dual to nonlinear sigma-model. The notoph theory can also be formulated in curved space where it is dual to $4d$ free non-conformal scalar field theory. The theory under consideration obviously admits the addition of a mass term, which leads to duality with the massive vector field theory (see \cite{BKP} and references therein). Natural generalization of the $4d$ notoph theory is a free $p$-form model in arbitrary $d$ dimensions, where it is dual to $(d{-}p{-}2)$-form model.\footnote{See \cite{J,O,FVP} and references therein. Proof of duality between $p$- and $(d{-}p{-}2)$-form models is discussed\HIDE{ e.g.} in \cite{FVP}.} By now, a large number of reducible gauge models have been constructed, including specific supersymmetric models containing $p$-forms, and their aspects of quantum duality have been studied as well. A practically complete list of relevant references is given in the work \cite{KR} to which we refer the reader. All such models are the reducible gauge theories as well as the $4d,\, p {\hspace{1pt}=\hspace{1pt}} 2$ model.

A distinctive feature of all reducible gauge theories is that the generators of gauge transformations are not independent. For this reason, the standard Faddeev-Popov-DeWitt rules of covariant quantization are not applicable to such theories and for their quantization one should use more general approaches. Quantization of $4d$, $p{\,=\,}2$ abelian gauge model in curved space has been carried out for the first time by A.S.~Schwarz \cite{Sch-1,Sch-2} using decomposition of the $p$-form into irreducible components in the functional integral. Note also the work by W.~Siegel \cite{Siegel} where correct ghost content for quantum $2$- and $3$-forms in curved space-time was firstly found. Later, various methods were developed for quantization of this theory and the other reducible gauge models.\footnote{Generalization of the Faddeev-Popov-DeWitt procedure for quantization of some superfield abelian reducible gauge theories was considered in \cite{BK}.} Aspects of quantum equivalence of classically dual theories were studied as well. The current state of art in this area is presented in detail in \cite{KR}. As a result, there are now consistent and well-developed procedures for quantizing the $p$-form models. However all of the above concerns mainly boson theories.

Recently a novel class of Lagrangian reducible gauge theories has been proposed in the papers \cite{Buchbinder:2009pa,Zinoviev:2009wh}. These theories are described by totally antisymmetric rank $p$ tensor-spinors (fermionic $p$-forms) with reducible gauge transformations depending on $p$ and space-time dimension $d$.\footnote{See also \cite{CFMS} where multi-symmetric reducible fermionic field models have been studied.} Remarkable feature of these theories is that they are consistently formulated only if $2p {\,<\,} d$ and a consistent coupling to curved space is possible only for $AdS_d$ space.\footnote{In \cite{Buchbinder:2009pa} a massive fermionic $p$-form theory was constructed, which becomes gauge invariant in the massless limit.} Note that the fermionic theories of a similar structure arise in some exotic models of extended supergravity in six dimensions (see \cite{CFMS,H-1,H-2,W,H-3,B,HLL,HLMP,MSZ,BHHS,C,G} and references therein), which recently attracted much attention.

The purpose of this paper is to carry out a covariant quantization of the totally antisymmetric tensor-spinor field theory introduced in \cite{Buchbinder:2009pa}  and construct its quantum effective actions. We start from the case $p{\,=\,}3$ and then generalize it to arbitrary $p$. Since these models are the reducible gauge theories, we use for quantization the Batalin-Vilkovisky (BV) method which is a powerful and most general covariant quantization method for arbitrary gauge systems with dependent generators \cite{Batalin:1981jr,Batalin:1983ggl} (see also reviews \cite{Hen,Henneaux:1992ig,Gomis:1994he}).\footnote{A simplified quantization method based on an adjustment of the Faddeev-Popov procedure was proposed in \cite{BBKN4:2025} and applied for deriving the effective action for the case $p=2$. The result coincides with that obtained on the basis of the general method. While this simplified approach can be generalized to the quantization of tensor-spinor theories with larger values of $p$, the procedure becomes technically very cumbersome.} Note that an application of BV method for quantization of the model \cite{Buchbinder:2009pa} in flat space was recently considered in \cite{Lekeu:2021oti}. It seems to us that the quantization of a free theory in a flat space, although interesting in its own way, is not very indicative, since in this case the effective action does not depend on any parameters and is simply a constant.\footnote{The paper \cite{Lekeu:2021oti} also discusses the gravitational anomaly of a fermionic $p$-form theory. However, the physical interpretation of this anomaly remains unclear from a general perspective, as the classical theory itself is ill-defined in an arbitrary curved spacetime and is only consistent in AdS space.}

The paper is organized as follows. In Section~\ref{Sect:Brief}, we provide a brief overview of the model of the totally antisymmetric tensor-spinor field of rank $p$ (the fermionic $p$-form) in $d$ dimensions, including the aspects of reducibility, and establish basic conventions for the compact index notation used throughout the paper. In Section~\ref{Sect:p=3}, we apply the BV quantization technique to the model of the fermionic $3$-form, which constitutes a gauge theory with two stages of reducibility, and derive its effective action in terms of functional determinants of relevant Dirac-type differential operators. In Section~\ref{Sect:p=p}, we extend the above analysis to arbitrary spinorial $p$ forms, carry out quantization,  derive the effective action within the BV formalism for reducible gauge theories and obtain a general expression for effective action. Special attention is paid to two types of covariant gauges. In Section~\ref{Sect:concl}, we summarize the results obtained, address related open questions, and discuss the suggestions for further research. In Appendix~\ref{Sect:BVst}, we review in detail the general framework of the Batalin–Vilkovisky (BV) quantization procedure for gauge theories with $p{-}1$ stages of reducibility.


\section{Fermionic $p$-form field model, conventions, and key result}
 \label{Sect:Brief}
%
  \subsubsection*{Conventions for products of gamma-matrices and covariant derivatives}
   \label{SSect:Conventions}
    \hspace{\parindent}
Throughout this paper, we adopt the following conventions for the gamma-matrix algebra:
 \begin{equation}
    \{\gamma^\Zm,\gamma^\Zn\}
     \,=\, - 2 \GG^{\Zm\Zn}
    \,,
    \qquad
    \dc{\psi}
    \,\defeq\, \psi^\dagger \gamma^0
    \,,
    \qquad
    \dc{\gamma}^\Zm
    \,\defeq\, (\gamma^0)^{-1}(\gamma^\Zm)^\dagger \gamma^0
    \,=\, \gamma^\Zm
    \,,
 \label{conventions_gamma}
 \end{equation}
where an overbar denotes Dirac conjugation, $\gamma^0$ is a constant hermitian matrix in a vielbein basis, and the metric $\GG^{\Zm\Zn}$ has the mostly-plus signature.
\if{
 \footnote{
  The standard hermiticity conditions: $\hc{(\gamma^0)}=\gamma^0$,\, $\hc{(\gamma^a)}=-\gamma^a$, together with the choice of $\gamma^0$ as the hermitizing matrix and the convention $\dc{\gamma}^\Zm=\gamma^\Zm$, imply the relation $\esm\eta_{00}=1$. thereby fixing the connection between $\esm$ and the metric signature.
 }
}\fi
 %
In what follows, we work exclusively with gamma matrices in a coordinate basis on a curved background manifold. Coordinate (tangent and cotangent) indices are raised and lowered using the metric $\GG_{\Zm\Zn}$ and its inverse $\GG^{\Zm\Zn}$.

We introduce a special notation for completely antisymmetrized products of gamma matrices:
 \begin{equation}
   \gamma^{\Zm\Zn}
   \,\defeq\,
   \nfrac12 [\gamma^\Zm,\gamma^\Zn]
   \,,  \quad \ldots \quad,\;\;
   \gamma^{\Zm_1\Zm_2\ldots\Zm_r}
   \,\defeq\,
    \gamma^{\Al\Zm_1} \gamma^{\Zm_2} ... \gamma^{\Zm_r\Ar}
   \,,
   \qquad
   (r \leq d)
   \,,
 \label{def_mmg}
 \end{equation}
which are nontrivial only for degrees $r$ not exceeding the spacetime dimension $d$.

The covariant derivative acting on a spinor field $\psi$ (with no tensor indices) is defined as
 \begin{eqnarray}
   \nabla_{\!\Zl} \psi
   \,\defeq\,
    \partial_\Zl \psi + \hat{\varGamma}_\Zl \psi
   \,,
  \qquad  
    \dc{\psi}\, \dc{\nabla}_{\!\Zl}
   \,\defeq\,
   ( \dcl{\nabla_{\!\Zl} \psi} )
   \:=\:
    (\partial_\Zl \dc{\psi}\, - \,\dc{\psi} \hat{\varGamma}_\Zl)
   \,,
 \label{nabla_psi}
 \end{eqnarray}
where the spinor connection $\hat{\varGamma}_\Zl$ is chosen such that the gamma matrices are covariantly constant:\footnote{
 The spinor connection is usually defined in terms of the vielbein spin-connection
 $\omega_\Zl^{\Zm\Zn}$:\;
 $\hat{\varGamma}_\Zl {\,=\,}
 {-} 
 \nfrac{1}{8} \omega_\Zl^{\Zm\Zn} [\gamma_\Zm,\gamma_\Zn] $.
}
 \begin{eqnarray}
   [\nabla_{\!\Zl}, \gamma^\Zm]
   \;\equiv\;
   \partial_\Zl \gamma^\Zm + \varGamma^\Zm_{\Zl\Zn} \gamma^\Zn + [\hat{\varGamma}_\Zl, \gamma^\Zm]
   \,=\, 0
   \,,
 \label{nabla_mg}
 \end{eqnarray}
with $\varGamma^\Zm_{\Zl\Zn}$ denoting the Christoffel symbols of the Levi-Civita connection. Throughout this article, the barred covariant derivative denotes differentiation acting to the left on Dirac-conjugated spinors.\footnote{For spinor-valued forms when using compact index notation, this convention yields more symmetric formulas with fewer sign factors arising from index permutations.} Finally, the spinorial curvature is related to the Riemann tensor via
 \begin{eqnarray}
  [\nabla_{\!\Zm} , \nabla_{\!\Zn}] \,\psi
  \:=\:
   {-}\nfrac{1}{4} 
   R_{\Zk\Zl\hspace{1pt}\Zm\Zn} \gamma^{\Zk\Zl} \psi
   \,.
 \label{nabla_nabla_gen}
 \end{eqnarray}

  \subsubsection*{Compact notation for antisymmetric multiindices}
   \hspace{\parindent}
    \label{SSect:Multiindex}
In this article we study the gauge theory of spinor-valued $p$-forms
 $ 
    \psi_{\Zm_1\ldots\Zm_p} = \psi_{\Al\Zm_1\ldots\Zm_p\Ar} 
    .
 $ 
In the Batalin–Vilkovisky extension of the configuration space, one encounters also spinor-valued forms of lower degrees.
To streamline formulas we use compact multiindex notation for antisymmetrized groups of indices.
A multiindex $\ZZm{r}$ denotes $r$ indices to be antisymmetrized:
 \begin{equation}
  f_{\ZZm{r}}
  \,\defeq\,
  f_{\Al\Zm_1\ldots\Zm_r\Ar}
  \,.
 \end{equation}
Antisymmetrization over the group of $r$ indices is defined as the normalized sum over all permutations, with the conventional factor ${1}/(r!)$\,. We extend antisymmetrization to cover indices belonging to the same group across different factors in a term, for instance
 \begin{equation}
  f_{\ZZm{r}} h_{\ZZm{k}}
   \,\defeq\,
  f_{\Al\Zm_1\ldots\Zm_r} h_{\Zm_{r+1}\ldots \Zm_{r+k}\Ar}
  \,,\qquad
  \nabla_{\!\Zm} h_{\ZZm{k}}
   \,\defeq\,
  \nabla_{\!\Al \Zm_1} h_{\Zm_{2}\ldots \Zm_{k+1}\Ar}
  \,.
 \end{equation}
When a group consists of a single index, we suppress the number:
 \begin{equation}
  f_{\ZZm{1}\ZZn{k}} \:\equiv\:  f_{\Zm\ZZn{k}}
  \,,
  \qquad
  \nabla_{\!\ZZk{1}} \:\equiv\: \nabla_{\!\Zk}
  \,.
 \end{equation}
These rules apply equally to lower and upper indices.
Index groups of the same kind in contravariant and covariant positions are implicitly contracted:
 \begin{equation}
  f^{\ZZm{r}} h_{\ZZm{r}}
  \,\defeq\,
  f^{\Al\Zm_1\ldots\Zm_r\Ar} h_{\Al\Zm_{1}\ldots \Zm_{r}\Ar}
  \,,
 \label{def_contraction_pp}
 \end{equation}
The lower and upper indices of the same groups are always assumed in the natural unidirectional index order.

For partial contractions $f^{\ZZm{r}} h_{\ZZm{k}}$ with $r\neq k$, one must specify how the smaller group is embedded into the larger one as a subgroup, which may affect an overall sign. By default we use the convention
 \begin{equation}
   f^{\ZZm{r}} h_{\ZZm{k}}
   \:\defeq\:
   f^{\Al\Zm_1\ldots\Zm_r\Ar} h_{\Al\Zm_{1}\ldots \Zm_{k}\Ar}
    \quad (r\geq k)
   \,,\qquad
   f^{\ZZm{r}} h_{\ZZm{k}}
   \:\defeq\:
   f^{\Al\Zm_{k-r+1}\ldots\Zm_k\Ar} h_{\Al\Zm_{1}\ldots \Zm_{k}\Ar}
    \quad (r\leq k)
   \,.
 \label{def_contraction_order_convention}
 \end{equation}
This ambiguity disappears when each group is fully contracted, for example:
 \begin{equation}
   f^{\ZZm{r}} g_{\ZZm{r-k}} h_{\ZZm{k}}
    \:=\:
   f^{\Al\Zm_1\ldots\Zm_r\Ar} g_{\Al\Zm_{1}\ldots \Zm_{r-k}} h_{\Al\Zm_{r-k+1}\ldots \Zm_{r}\Ar}
    \,,\qquad
   g_{\ZZm{r-k}} f^{\ZZm{r}} h_{\ZZm{k}}
    \:=\:
   g_{\Al\Zm_{1}\ldots \Zm_{r-k}} f^{\Al\Zm_1\ldots\Zm_r\Ar} h_{\Al\Zm_{r-k+1}\ldots \Zm_{r}\Ar}
   \,.
 \label{def_contraction_complete}
 \end{equation}

We also introduce notations for the unit operator and for covariant operators, which raise and lower antisymmetrized index groups in the space of degree-$r$ forms:
 \begin{eqnarray}
  \begin{array}{l@{\hspace{6pt}}c@{\hspace{9pt}}l@{\hspace{7pt}}c@{\hspace{9pt}}l}
    \delta_{\ZZm{r}}^{\:\ZZn{r}}
   &\!\defeq\!&
    (\delta_{\Zm}^{\:\Zn})^{r}
   &\!=\!&
    \delta_{\Zm_1}^{\,\Al\Zn_1} \delta_{\Zm_2}^{\:\Zn_2} \ldots \delta_{\Zm_r}^{\:\Zn_r\Ar}
   \,,
  \\[0.2em]
    \GG_{\ZZm{r}\ZZn{r}}
   &\!\defeq\!&
    (\GG_{\Zm\Zn})^{r}
   &\!=\!&
    \delta_{\ZZm{r}}^{\:\ZZk{r}}\,
    \GG_{\Zk_1\Zm_1}\GG_{\Zk_2\Zm_2}\ldots\GG_{\Zk_{r}\Zm_{r}}
   \,,
  \\[0.2em]
    \GG^{\ZZm{r}\ZZn{r}}
   &\!\defeq\!&
    (\GG^{\Zm\Zn})^{r}
   &\!=\!&
    \GG^{\Zm_1\Zk_1}\GG^{\Zm_2\Zk_2}\ldots\GG^{\Zm_{r}\Zk_{r}}
    \,\delta_{\ZZk{r}}^{\:\ZZn{r}}
   \,,
  \end{array}
 \label{def_index_operators} 
 \end{eqnarray}
where $\GG_{\Zm\Zn}$ is the spacetime metric and $\GG^{\Zm\Zn}$ its inverse.

  \subsubsection*{Totally antisymmetric tensor-spinor gauge field theory}
  \label{SSect:Setup_p=p}
    \hspace{\parindent}
The theory of a totally antisymmetric tensor–spinor field of rank $p$ (a fermionic $p$-form) was introduced in \cite{Buchbinder:2009pa} and is defined by the action

 \begin{equation}
  S_0[\psi, \dcpsi]
   \:=\:
    \!\int\!d^dx
     \,i\sqrt{|\GG|}
     \,\dcpsi_{\ZZm{p}}
     \gamma^{\ZZm{2p+1}}\,D_{\Zm}\,\psi_{\ZZm{p}}
   \:\equiv\:
    \!\int\!d^dx
     \,i\sqrt{|\GG|}
     \,\dcpsi_{\Zn_1\Zn_2\ldots\Zn_{p}}
     \gamma^{\Zn_1\Zn_2\ldots\Zn_{p} \hspace{0.5pt} \Zr \hspace{1pt} \Zm_1\Zm_2\ldots\Zm_{p}} \,D_\Zr\, \psi_{\Zm_1\Zm_2\ldots\Zm_{p}}
   \,,
 \label{def_S0}  
 \end{equation}
where $\psi_{\ZZm{p}} \defeq \psi_{\Zm_1\Zm_2...\Zm_p}$ and its Dirac conjugate are totally antisymmetric spinor-valued tensor fields and the multigamma $\gamma^{\ZZm{2p+1}}$ denotes the completely antisymmetrized product of $(2p{\,+\,}1)$ gamma matrices $\gamma^{\Al\Zm_1}\gamma^{\Zm_2}...\gamma^{\Zm_{2p}}\gamma^{\Zm_{2p+1}\Ar}$, (\ref{def_mmg}).\footnote{
 Throughout Section~\ref{Sect:p=p} and partly in Section~\ref{Sect:p=3} we use the multiindex notation reviewed in Section~\ref{SSect:Multiindex}. Here, for clarity, we also provide explicit index representations.
}
Operator $D_\Zm$ is a massive covariant operator of the form $D_\Zm = \nabla_{\!\Zm} + i \mAdS \gamma_\Zm $.

We restrict the theory to $AdS_d$ space, since its gauge properties crucially relies on this background \cite{Buchbinder:2009pa}. The $AdS_d$ curvature tensor is maximally symmetric with positive parameter $r_{\hspace{-1pt}\scriptscriptstyle{0}}$:
 \begin{equation}
   R^{\Zm\Zn}{}_{\!\Zk\lambda}
    \,=\, {-} 
    r_{\hspace{-1pt}\scriptscriptstyle{0}}\, (\delta^\Zm_\Zk\delta^\Zn_\lambda-\delta^\Zn_\Zk\delta^\Zm_\lambda),
    \quad\;\;
   r_{\hspace{-1pt}\scriptscriptstyle{0}} > 0
   \,.
 \label{AdS_d} 
 \end{equation}
For the special mass values $\mAdS = \pm \frac{1}{2}\, \sqrr$,\, the operator $D_\Zm$ becomes nilpotent
 \begin{equation}
  D_\Zm
   \:=\:
   \nabla_{\!\Zm} \pm \nfrac{i}{2}\, \sqrr\,\gamma_{\Zm}
   \::
   \qquad\;\;
   D_{\Zm} D_{\Zm} f_{\ZZm{k}}
   = 0
   \quad \Leftrightarrow \quad
   D_{\Al\Zm_{1}} D_{\Zm_{2}} f_{\Zm_{3}\ldots\Zm_{k+2}\Ar}
   = 0
   \,,
 \label{Dmu} 
 \end{equation}
implying that the action (\ref{def_S0}) possesses  a $(p{-}1)$-stage reducible gauge symmetry.
Introducing $p$ levels of gauge parameters --- spinor-valued forms $\glambda_{\ZZn{r}}$ --- the reducible gauge transformations take the form
 \begin{eqnarray}
  \begin{array}{llr@{\hspace{6pt}}c@{\hspace{6pt}}llr@{\hspace{6pt}}c@{\hspace{6pt}}l}
  &&
  \gvar \psi_{\ZZm{p}}
  &=&
  D_{\Zm}  \glambda_{\ZZm{p-1}}
  \,,
  &&
  \gvar \dcpsi_{\ZZm{p}}
  &=&
    \dc{\glambda}_{\ZZm{p-1}} \dc{D}_{\Zm}
  \,,
  \qquad
  \\[0.5em]
  &&
  \gvar \glambda_{\ZZm{p-1}}
  &=&
  D_{\Zm} \glambda_{\ZZm{p-2}}
  \,,
  &&
  \gvar \dc{\glambda}_{\ZZm{p-1}}
  &=&
   \dc{\glambda}_{\ZZm{p-2}} \dc{D}_{\Zm}
  \,,
  \qquad
  \\
  &&
  \ldots && \quad \ldots
  &&
  \ldots && \quad \ldots
  \\[0.2em]
  &&
  \gvar \glambda_{\Zm}
  &=&
   D_{\Zm} \glambda_{\Zo}
  \,,
  &&
  \gvar \dc{\glambda}_{\Zm}
  &=&
  \dc{\glambda}_{\Zo}   \dc{D}_{\Zm}
  \,,
  \\[0.5em]
  &&
   \gvar \glambda_{\Zo}
  &=&
  0
  \,,
  &&
   \gvar \dc{\glambda}_{\Zo}
  &=&
  0
  \,,
  \end{array}
 \label{gauge_transfs}
 \end{eqnarray}
where the conjugate derivative \emph{acting to the left} is defined by
 \begin{equation}
  \dc{D}_{\Zm}
  \:\defeq\:  \dc{\nabla}_{\!\Zm} \mp \nfrac{i}{2}\, \sqrr\,\gamma_{\Zm}
  \,.
 \label{dcDmu} 
 \end{equation}

\vspace{-2mm}

\subsubsection*{The key result}
 \hspace{\parindent}
The key result of this work is the computation of the effective action for the spin-form gauge theory of rank $p$. Expressed in terms of functional determinants over spaces of irreducible spinor-form fields\HIDE{ of degrees $s$}, the corresponding generation functional takes the compact form
 \begin{equation}
  Z_{}^{(p)}
  \;=\;
  {\displaystyle \prod_{n=0}^{[p/2]} (\Delta_{\oix{p-2n}})^{2n+1} }{\displaystyle \prod_{m=0}^{[(p-1)/2]} \!\!\!\! (\Delta_{\oix{p-1-2m}})^{-2m-2} }
   \;=\;
  \frac{\Delta_{\oix{p}}\;\Delta_{\oix{p-2}}^3\;\Delta_{\oix{p-4}}^5 \cdots \Delta_{\oix{p-2[p/2]}}^{2[p/2]+1}}
  {\Delta_{\oix{p-1}}^2 \;\Delta_{\oix{p-3}}^4 \cdots \Delta_{\oix{(p+1)-2[(p+1)/2]}}^{2[(p+1)/2]} \vphantom{\big|^b}}
  \,,
  \label{arb_p_result_EA_Intro} 
 \end{equation}
where $[N/2]$ denotes the integer part of a number $N/2$, and, with $\slashed{\nabla} {\:\equiv\:} \gamma^\Zm \nabla_{\!\Zm}$,
 \begin{eqnarray}
  \Delta_{\oix{s}}
   \defeq
  {\rm Det}\Bigl[\slashed{\nabla} \,
  \mp 
  \nfrac{i}{2}\,\sqrr (d{\,-\,}2s)\Bigr]
 \end{eqnarray}
is the functional determinant of the corresponding operator acting on \emph{irreducible} spinor-form fields $\fCA_{\ZZm{s}}$ of degrees $s$ constrained by conditions $\gamma^\Zm\fCA_{\ZZm{s}} {\,=\,} 0$\; $\Leftrightarrow$\; $\gamma^{\Zm_s}\fCA_{\Al\Zm_1...\Zm_{s}\Ar} {\,=\,} 0$.
As a byproduct, we derive intermediate representations of the gauge-fixed action in terms of irreducible components for two covariant gauges, valid for arbitrary rank $p$, with all normalization factors determined explicitly.

  \subsubsection*{Description of method}
   \hspace{\parindent}
We employ the powerful Batalin–Vilkovisky (BV) technique\HIDE{ and its generalizations} to derive (\ref{arb_p_result_EA_Intro}).
For a general gauge theory, the BV formalism provides a systematic prescription for constructing a nondegenerate gauge-fixed action and the corresponding quantum generating functional, which is guaranteed by construction to be gauge-independent \cite{Batalin:1981jr,Henneaux:1992ig,Gomis:1994he}. The method extends naturally to gauge theories whose symmetries are reducible, meaning that the gauge generators possess nontrivial zero modes. These modes are spanned by reducibility generators, which may themselves have further zero modes of lower range, and so on \cite{Batalin:1983ggl,Gomis:1994he}.

The BV procedure enlarges the original field content by introducing ghost fields and associating each field with a corresponding antifield. This extended field–antifield space is $\mathbb{Z}_2$- and $\mathbb{Z}$-graded space of even dimension, equipped with an odd symplectic structure known as the \emph{antibracket}. The dynamics on this extended space is encoded in the BV action\footnote{This functional is referred to as the \emph{master action} in the original papers \cite{Batalin:1981jr,Batalin:1983ggl}.} $S^{\iBV}[\varPhi,\varPhi^\af]$, which extends the classical action with antifield- and ghost-dependent terms \cite{Batalin:1981jr,Batalin:1983ggl}. The master action satisfies the nilpotency condition in terms of the antibracket:  $ (S^{\iBV},S^{\iBV}) = 0 $
with appropriate boundary conditions on the action's extension that encode the gauge structure of the original theory. Nilpotency of the antibracket ensures that the BV action possesses both extended local BV symmetry and global BRST symmetry. The latter, via cohomological methods, provides precise control over the gauge structure, including possible quantum anomalies.

When moving to quantum theory, the local symmetry of BV action should be broken. This is realized by gauge fixing, which leads to a nondegenerate action without residual local symmetry. In BV formalism, gauge fixing is carried out by means of a universal prescription involving the gauge fermion $\varPsi[\varPhi]$, which specifies a gauge. This fermion generates explicit conditions that eliminate antifields in favor of the fields,  $\varPhi^\af_I = \frac{\var \varPsi}{\var \varPhi^I}$. The BV construction guarantees that the resulting gauge-fixed action yields a quantum generating functional and observables independent of the specific choice of $\varPsi$.

A brief review of the BV formalism for $(p{-}1)$-stage reducible gauge theories is given in Appendix~\ref{Sect:BVst}. The spinorial $p$-form gauge models, considered in Sections~\ref{Sect:p=3} and \ref{Sect:p=p} present explicit examples which demonstrate the use of an alternative gauge prescriptions and compare them with the original ones advocated in the seminal papers \cite{Batalin:1981jr,Batalin:1983ggl}.


\section{Quantization and effective action of fermionic $3$-form gauge theory}
 \label{Sect:p=3}
  \hspace{\parindent}
In this section, we apply the BV quantization procedure to the gauge model of totally antisymmetric rank-$3$ tensor-spinor field in $AdS_d$ spacetime. The theory is defined by the action (\ref{def_S0}), which for $p=3$ takes the form
 \begin{equation}
    S^{\scriptscriptstyle(3)}_0[\psi, \dcpsi]
    \;=\;
     \!\int\!d^dx
      \,i\sqrt{|\GG|}
      \,\dcpsi_{\Zs\Zp\Zt}
      \gamma^{\Zs\Zp\Zt{}\Zr{}\Zm\Zn\Zk} \,D_\Zr\,\psi_{\Zm\Zn\Zk}
      \,,
 \label{S0_p=3} 
 \end{equation}
where $\psi_{\Zm\Zn\Zk}\equiv \psi_{\Al\Zm\Zn\Zk\Ar}$ is the spinor-valued 3-form field and\footnote{Such multigamma combination is nontrivial in dimensions $d \tgeq 7$, which we assume to hold.} $\gamma^{\Zs\Zp\Zt{}\Zr{}\Zm\Zn\Zk} \equiv \gamma^{\Al\Zs}\gamma^{\Zp}\gamma^{\Zt}\gamma^{\Zr}\gamma^{\Zm}\gamma^{\Zn}\gamma^{\Zk\Ar}$.
The covariant differential operator
 $
  D_{\Zm}
  {\:=\:}
  \nabla_{\!\Zm} \pm \nfrac{i}{2}\, \sqrr\,\gamma_{\Zm}
 $, (\ref{Dmu}),
in AdS spacetime with the curvature
 $
  R^{\Zm\Zn}{}_{\!\Zk\Zl}
   {\:=\:}
   {-} 
   r_{\scriptscriptstyle \hspace{-1pt} 0} (\delta^\Zm_\Zk\delta^\Zn_\Zl-\delta^\Zn_\Zk\delta^\Zm_\Zl)\,,
 $
acquires the nilpotence property
 $
  D_{\Al\Zm} D_{\Zn} f_{\Zk\ldots\Zl\Ar}
   {\:=\:} 0
 $,
which implies the $2$-stage reducible gauge structure of action (\ref{S0_p=3}).
Introducing three levels of gauge parameters --- spinor-valued forms $\glambda_{\Zm\Zn} {\,\equiv\,} \glambda_{\Al\Zm\Zn\Ar}$, $\glambda_{\Zm}$, and $\glambda_{\Zo}$ --- the reducible symmetry of the theory takes the form
 \begin{eqnarray}
  \begin{array}{llr@{\hspace{6pt}}c@{\hspace{6pt}}llr@{\hspace{6pt}}c@{\hspace{6pt}}l}
   &&
    \gvar \psi_{\Zm\Zn\Zk}
    &=&
      D_{\Al\Zm}  \glambda_{\Zn\Zk\Ar}
    \,,
   &&
    \gvar \dcpsi_{\Zm\Zn\Zk}
    &=&
      \dc{\glambda}_{\Al\Zm\Zn} \dc{D}_{\Zk\Ar}
    \,,
    \qquad
   \\[0.5em]
   &&
    \gvar \glambda_{\Zm\Zn}
    &=&
    D_{\Al\Zm} \glambda_{\Zn\Ar}
    \,,
   &&
    \gvar \dc{\glambda}_{\Zm\Zn}
    &=&
     \dc{\glambda}_{\Al\Zm} \dc{D}_{\Zn\Ar}
    \,,
    \qquad
   \\[0.5em]
   &&
    \gvar \glambda_{\Zm}
    &=&
    D_{\Zm} \glambda_\Zo
    \,,
   &&
    \gvar \dc{\glambda}_{\Zm}
    &=&
    \dc{\glambda}_{\Zo}  \dc{D}_{\Zm}
    \,,
  \end{array}
 \label{gauge_transfs_p=3} 
 \end{eqnarray}
where the Dirac-conjugated operator
 $
  \dc{D}_{\Zm}
  \defeq  \dc{\nabla}_{\!\Zm} \mp \nfrac{i}{2}\, \sqrr\,\gamma_{\Zm}
 $,
(\ref{dcDmu}), in which $\dc{\nabla}_{\!\Zm}$ acts \emph{to the left} on the Dirac-conjugated fields or gauge parameters,
is also nilpotent:
 $
  \dc{f}_{\Al\Zm\ldots\Zn} \dc{D}_{\Zk} \dc{D}_{\Zl\Ar}
   = 0
 $.

The transformations of fields in (\ref{gauge_transfs_p=3}) define gauge generators
 $R_{\Zm\Zn\Zk}^{\;\Zs\Zp}$,
 $\dc{R}^{\Zs\Zp}_{\,\Zm\Zn\Zk}$,
whereas the rest define reducibility generators
 $Z_{\Zm\Zn}^{\,\Zl}$,
 $\dc{Z}^{\Zl}_{\,\Zm\Zn}$
and
 $Z_{\Zm}^{\Zo}$,
 $\dc{Z}^{\Zo}_{\Zm}$.
Correspondence with the general BV scheme of Appendix \ref{Sect:BVst} implies $\psi^i  {\:\mapsto\:} \psi_{\Zm\Zn\Zk}\,,\dcpsi_{\Zm\Zn\Zk}$ and
 \begin{equation}
   \begin{array}{l@{\hspace{12pt}}c@{\hspace{12pt}}r@{\hspace{12pt}}l}
     R^i_{\,\zza{1}} &\mapsto&
     R_{\Zm\Zn\Zk}^{\;\Zs\Zp}
      {\defeq}\,  D{}_{\Al\Zm} \delta{}_{\Zn\Zk\Ar}^{\,\Zs\Zp}
      \,,&
     \dc{R}^{\Zs\Zp}_{\,\Zm\Zn\Zk}
      {\defeq}\,  \delta_{\,\Al\Zm\Zn}^{\Zs\Zp} \dc{D}_{\Zk\Ar}
      \,;
    \\[0.5em]
     Z^{\zza{1}}_{\,\zza{2}}
     &\mapsto&
     Z_{\Zm\Zn}^{\,\Zl}
      {\defeq}\,  D_{\Al\Zm} \delta_{\Zn\Ar}^{\,\Zl}
      \,,&
     \dc{Z}^{\Zl}_{\,\Zm\Zn}
      {\defeq}\,  \delta_{\,\Al\Zm}^{\Zl} \dc{D}_{\Zn\Ar}
      \,;
    \\[0.5em]
     Z^{\zza{2}}_{\,\zza{3}}
     &\mapsto&
     Z_{\Zm}^{\Zo}
      {\defeq}\,  D_{\Zm}
      \,,&
     \dc{Z}^{\Zo}_{\Zm}
      {\defeq}\,  \dc{D}_{\Zm}
      \,.
   \end{array}
 \label{gauge_generators_p=3} 
 \end{equation}
The gauge structure is \emph{abelian} since the gauge and reducibility generators are field-independent.

Spinor-valued 3-form fields carry
 $\frac16 d (d{-}1) (d{-}2)$
spinor components, so that original fields describe
 $\nnn_0 {\:=\:} \frac13 d (d{-}1) (d{-}2)$
spinors\footnote{We count degrees of freedom of spinor form fields as numbers of spinor components, so that numbers refer to ranges of values of tensor indices.}. Reducible gauge symmetry (\ref{gauge_transfs_p=3}) leaves
 $\frac16 d(d{-}1)(d{-}2) - \frac12 (d{-}1)(d{-}2)  {\:=\:} \frac16 (d{-}1)(d{-}2)(d{-}3) $
non-gauge spinor components for each field, so that
 $\mmm_0 = \frac13 (d{-}1)(d{-}2)(d{-}3)$
\cite{FVP}.
Analogously, each of the 2-form spinor gauge parameters $\glambda_{\Zm\Zn}$, $\dc{\glambda}_{\Zm\Zn}$ carry
 $\frac12 d(d{-}1)$
spinors, so that
 $\nnn_1 {\:=\:} d(d{-}1)$.
But since not all of them actually contribute to gauge transformation due to reducibility, the actual number of spinor-valued parameters transforming original fields is
 $\mmm_1 {\:=\:} (d{-}1)(d{-}2)$.
Each of spinorial gauge parameters $\glambda_{\Zm}$, $\dc{\glambda}_{\Zm}$ carry $d$ spinors, so that $\nnn_2 {\:=\:} 2 d$. But only $d{-}1$ of them lessen the number of actual components of $\glambda_{\Zm\Zn}$, $\dc{\glambda}_{\Zm\Zn}$, thus $\mmm_2 {\:=\:}2(d{-}1)$.



  \subsection{BV extension}
   \label{SSect:BV extension_p=3}
    \hspace{\parindent}
The BV extension\footnote{The overview of the Batalin-Vilkovisky method is given in Appendix~\ref{Sect:BVst}. In this section, we largely avoid discussion of the BV structures and their properties, referring the reader to Appendix~\ref{Sect:BVst} or standard reviews for details.} (\ref{BV_master_action}--\ref{BV_aux_action}) of the gauge model (\ref{S0_p=3})
introduces a set of \emph{minimal} ghosts $\gC^{\gix{0}}_{\ZZm{3-l}} ,\, \dcgC^{\gix{0}}_{\ZZm{3-l}}$ of three levels $1 \tleq l \tleq 3$\,:\,
   $\gC^{\gix{0}}_{\Zm\Zn},\,\dcgC^{\gix{0}}_{\Zm\Zn}$,\,
   $\gC^{\gix{0}}_{\Zm},\,\dcgC^{\gix{0}}_{\Zm}$,
   and $\gC^{\gix{0}}_{\Zo},\dcgC^{\gix{0}}_{\Zo}$,
\if{
 \begin{equation}
  \gC^{\gix{0}}_{\Zm\Zn},\,\dcgC^{\gix{0}}_{\Zm\Zn},\,\;\;
  \gC^{\gix{0}}_{\Zm},\,\dcgC^{\gix{0}}_{\Zm}\,,\;\;
  \gC^{\gix{0}}_{\Zo},\dcgC^{\gix{0}}_{\Zo}\,,\;\;
 \end{equation}
}\fi
corresponding to gauge parameters (\ref{gauge_transfs_p=3}).
For the gauge-fixing purposes we also introduce the set of \emph{auxiliary} fields --- pairs of the auxiliary ghost fields $\gC^{\gix{k}}_{\ZZm{3-l}} ,\, \dcgC^{\gix{k}}_{\ZZm{3-l}}$ and so-called Lagrange-multiplier fields $\gB^{\gix{k}}_{\ZZm{3-l}} ,\, \dcgB^{\gix{k}}_{\ZZm{3-l}}$. At each level $l$ there are $l$ \emph{generations} of these double pairs. Generations of the auxiliary fields are indexed by integers $k$: $1 \tleq k \tleq l$ and in the field's notation are denoted by the upper numerical index. In particular, the auxiliary ghosts of generation $k{\,=\,}1$ are called \emph{antighosts} and assigned with negative ghost numbers $-l$. Ghosts with $k \tgeq 2$ are referred as \emph{extraghosts}.
All ghost fields of level $l$ are spinor forms of degree $r = 3{\,-\,}l$.

The complete set of BV fields and antifields with assigned ghost numbers is summarized in Table~\ref{Table:p=3-Improved}.
\def \TblSpinThreeFormGhNumDNIB
{
\vspace{-0mm}
  \begin{center}
    \begin{tabular}{c@{\hspace{2pt}}c|c|c@{\hspace{2pt}} c@{\hspace{2pt}}||c|c@{\hspace{2pt}} c@{\hspace{2pt}}|}
    \cline{3-8}
    General BV
    && \quad$\varPhi^I$\quad  & $\gh{\varPhi^I}$
    && \quad$\varPhi^\af_I$\!\quad  & $\gh{\varPhi^\af_I}$ &
    \vphantom{\Big|}
    \\[0.1em]
    \cline{3-8}
    \multirow{10}{0cm}{}
    &
    & \multicolumn{6}{c|}{\textrm{\small minimal sector:}}
    \\
    \multirow{10}{0cm}{}
    &
    & \multicolumn{6}{c|}{\small\textrm{original fields and minimal ghosts of levels $1 \tleq l \tleq 3$}}
    \\
    \cline{3-8}
     $\;\;\psi^i \;|\!|\; \psi^\af_i \,\;\; $
    & $ \;\; \mapsto \;\; $
    \vphantom{\Big|}
    &  $\psi_{\Zm\Zn\Zk}
        ,\;
        \dc{\psi}_{\Zm\Zn\Zk}
        $
       & $0$  &
    &  $\psi_{\af}^{\Zm\Zn\Zk}
        ,\;
        \dc{\psi}_{\af}^{\Zm\Zn\Zk}
        $
       & $-1$ &
    \\
     $\gC_{\gix{0}}^{\zza{l}} \;|\!|\; \gC^\af_{\gix{0}\zza{l}} $
    & $\;\; \mapsto \;\; $
    \vphantom{\Big|} 
    &  $\gC^{\gix{0}}_{\ZZm{3-l}}
        ,\,
        \dcgC^{\gix{0}}_{\ZZm{3-l}}
        $
      & $l$  &
    & $\gC_{\af}^{\gix{0} \ZZm{3-l}}
       ,\;
       \dcgC_{\af}^{\gix{0} \ZZm{3-l}}
      $
      & $-l{-}1$  &
    \\[0.5ex]
    \cline{3-8}
    \multirow{10}{0cm}{}
    &
    & \multicolumn{6}{c|}{\textrm{\small auxiliary sector:}}
    \\
    \multirow{10}{0cm}{}
    &
    & \multicolumn{6}{c|}{\small\textrm{auxiliary ghosts and Lagrange multiplier fields}}
    \\
    &
    \multirow{10}{0cm}{}
    & \multicolumn{6}{c|}{\small\textrm{of levels $1 \tleq l \tleq 3$, \,generations $k=1,...,\,l$}}
    \\
    \cline{3-8}
     $\gC_{\gix{k}}^{\zza{l}} \;|\!|\; \gC^\af_{\gix{k}\zza{l}} $
    & $ \;\; \mapsto \;\; $
    \vphantom{\Big|} 
    & \;\;$\gC^{\gix{k}}_{\ZZm{3-l}},\, \dcgC^{\gix{k}}_{\ZZm{3-l}}$\;\;
      & $ \msp^k(l{-}k{+}\tfrac12) - \tfrac12 $  &
    &  $\gC_{\af}^{\gix{k} \ZZm{3-l}} ,\, \dcgC_{\af}^{\gix{k} \ZZm{3-l}}$
      & $ -\,\msp^k(l{-}k{+}\tfrac12) - \tfrac12 $   &
    \\[0.5em]
     $\gB_{\gix{k}}^{\zza{l}} \;|\!|\; \gB^\af_{\gix{k}\zza{l}} $
    & $ \;\; \mapsto \;\; $
    \vphantom{\Big|} 
    & $\gB^{\gix{k}}_{\ZZm{3-l}},\,
       \dcgB^{\gix{k}}_{\ZZm{3-l}}$
      & $ \msp^k(l{-}k{+}\tfrac12) + \tfrac12 $   &
    &    &   &
    \\[0.5ex]
    \cline{3-8}
  \end{tabular}
 \end{center}
 \vspace{-5mm}
}
\begin{table}[h!]
 \centering
  \TblSpinThreeFormGhNumDNIB
 \caption{BV fields of the $2$-stage reducible gauge theory of a $3$-form spinor field}
 \label{Table:p=3-Improved}
\end{table}
To the left of the table we put the corresponding fields and antifields of the general BV method of Appendix~\ref{Sect:BVst}. Grassmann parities of the fields are opposite to the parity of their level and for this particular model coincide with the parity of their degrees as forms. We do not specify antifields to Lagrange-multiplier fields since they do not appear in the BV master action. Each field of level $l=3{\,-\,}r$ is a spinor-valued $r$-form and describes $\nnn_l=\binom{d}{r}$ independent spinor components.

\newpar

The gauge algebra of generators (\ref{gauge_generators_p=3}) is \emph{abelian} (and of course closed), thus the ghost part of minimal master sector extends has a simple form bilinear in antighosts and differentiated ghosts (cf. \ref{BV_master_action}). Thus for the $3$-form spinor field the nonminimal BV master action is
 \begin{equation}
  S^{\scriptscriptstyle(3)}_{\iBV}
   [\varPhi,\varPhi^\af]
  =
  S^{\scriptscriptstyle(3)}_{0}[\psi,\dcpsi]
   + S^{\scriptscriptstyle(3)}_{\text{gh}}[\varPhi_{min},\varPhi_{min}^\af]
   + S^{\scriptscriptstyle(3)}_{\text{aux}}[\varPhi_{aux},\varPhi_{aux}^\af]
  \,,
 \label{master_action_p=3} 
 \end{equation}
where the original action $S^{\scriptscriptstyle(3)}_{0}$ is defined in (\ref{S0_p=3}), and ghost part of the minimal action and action of auxiliary sector fields take the form (cf. \ref{BV_master_action}):
 \begin{align}
    S^{\scriptscriptstyle(3)}_{\text{gh}} 
    &\,=\,   
     \!\int\! d^dx
     \Bigl\{\,
       i \big( \psi_{\af}^{\Zr\Zm\Zn} D_{\Zr} \gC^{\gix{0}}_{\Zm\Zn}
             \!-
             \dcgC^{\gix{0}}_{\Zm\Zn} \dc{D}_{\Zr} \, \dcpsi_{\af}^{\Zm\Zn\Zr} \big)
      %
      + i \big( \gC_{\af}^{\gix{0} \Zr\Zm} D_{\Zr} \gC^{\gix{0}}_{\Zm}
            \!-
            \dcgC^{\gix{0}}_{\Zm} \dc{D}_{\Zr} \, \dcgC_{\af}^{\gix{0} \Zm\Zr} \big)
      +
      i \big( \gC_{\af}^{\gix{0} \Zr} D_{\Zr{}} \gC^{\gix{0}}_{\Zo}
            \!-
            \dcgC^{\gix{0}}_{\Zo} \dc{D}_{\Zr} \, \dcgC_{\af}^{\gix{0} \Zr} \big)
      \Bigr\}
      \,,
    \nonumber
    \\[.5em]
    S^{\scriptscriptstyle(3)}_{\text{aux}} 
    &\,=\,  
     \!\int\!d^dx
     \Bigl\{
      \big(
      \gC_{\af}^{\gix{1} \Zm\Zn} \gB^{\gix{1}}_{\Zm\Zn}
      \!+ \dcgB^{\gix{1}}_{\Zm\Zn} \dcgC_{\af}^{\gix{1} \Zm\Zn}
      \big)
    +
      \sum_{k=1}^{2}
      \big(
      \gC_{\af}^{\gix{k} \Zm} \gB^{\gix{k}}_{\Zm}
      \!+ \dcgB^{\gix{k}}_{\Zm} \dcgC_{\af}^{\gix{k} \Zm}
      \big)
    +
     \sum_{k=1}^{3}
      \big(
      \gC_{\af}^{\gix{k} \Zo} \gB^{\gix{k}}_{\Zo}
      \!+ \dcgB^{\gix{k}}_{\Zo} \dcgC_{\af}^{\gix{k} \Zo}
      \big)
     \Bigr\}
     \,,
 \label{gh_aux_action_p=3}
 \end{align}
in which $D_{\Zm}$ act to the right and $\dc{D}_{\Zm}$ act to the left. The sum of\HIDE{ the original action} $S^{\scriptscriptstyle(3)}_{0}$ and $S^{\scriptscriptstyle(3)}_{\text{gh}} [\varPhi_{min},\varPhi_{min}^\af]$ form the \emph{minimal} master action, which is the solution of the master equation on the minimal set of BV fields, and $S^{\scriptscriptstyle(3)}_{\text{aux}} [\varPhi_{aux},\varPhi_{aux}^\af]$ is the pure gauge contribution from the auxiliary-sector fields.


  \subsection{Gauge fixing and quantization}
 \label{SSSect:Gauge-fermion_p=3}
    \hspace{\parindent}
The BV action (\ref{master_action_p=3}) possesses the extended local symmetry (\ref{BV_NoetherIdmaS}) which has to be gauge-fixed prior to quantization.
Standard BV gauge fixing (\ref{BV_gauge_fixing}) implies excluding antifields via
 $ 
    \varPhi^\af_I {\:=\,} \frac{\var^{} \varPsi}{\,\var \,\varPhi^I}
    \,,\,
    \dc{\varPhi}^\af_I {\:=\,} \frac{\var^{} \varPsi}{\,\var \,\dc{\varPhi}^I \vphantom{|^{l}}}
    \,,
 $ 
with an appropriate choice of the gauge fermion $\varPsi[\varPhi]$.
We choose linear gauge-fixing functions for original fields: $\chi^{\Zs\Zp}(\psi) = X^{\Al\Zs\Zp\Ar \hspace{.5pt} \Al\Zm\Zn\Zk\Ar} \psi_{\Zm\Zn\Zk}$  and $\dc{\chi}^{\Zs\Zp}(\dcpsi) = \dcpsi_{\Zm\Zn\Zk} \dc{X}^{\Al\Zm\Zn\Zk\Ar \hspace{.5pt} \Al\Zs\Zp\Ar}$\HIDE{ appropriate for free theories}, so that these
and other gauge-fixing operators are field-independent. The gauge fermion for delta-function gauge fixing (\ref{BV_gauge_fermion}) takes the form:
 \begin{align}
   \varPsi_{\delta}[\varPhi]
   \; & =  
   \displaystyle
   \int \!d^dx\,
    \Bigl\{\,
    \dcgC^{\gix{1}}_{\Zs\Zp} X^{\Al\Zs\Zp\Ar \hspace{.5pt}  \Al\Zm\Zn\Zk\Ar} \psi_{\Zm\Zn\Zk}
    +
    \dcpsi_{\Zm\Zn\Zk} \dc{X}^{\Al\Zm\Zn\Zk\Ar \hspace{.5pt} \Al\Zs\Zp\Ar}  \gC^{\gix{1}}_{\Zs\Zp}
   \nonumber
   \\[0.2em]
   & 
   \displaystyle
    \qquad\qquad\:
    + 
    \sum_{k=1}^{2}
    \big(
    \dcgC^{\gix{k}}_{\Zl} X^{\Zl \hspace{.5pt} \Al\Zm\Zn\Ar} \gC^{\gix{k-1}}_{\Zm\Zn}
    \!+
    \dcgC^{\gix{k-1}}_{\Zm\Zn} \dc{X}^{\Al\Zm\Zn\Ar \hspace{.5pt} \Zl} \gC^{\gix{k}}_{\Zl}
    \big)
    + 
    \sum_{k=1}^{3}
    \big(
    \dcgC^{\gix{k}}_{\Zo} X^{\Zm{}} \gC^{\gix{k-1}}_{\Zm{}}
    \!+
    \dcgC^{\gix{k-1}}_{\Zm{}} \dc{X}^{\Zm{}} \gC^{\gix{k}}_{\Zo}
    \big)
    \Bigr\}
     \,,
 \label{gauge_fermion_delta_p=3} 
 \end{align}
where we chose gauge-fixing operators $X$ and $\dc{X}$ to be the same (at each particular level) for all generations of ghosts fields.

The structure of such gauge fermion, as well as various structures of the BV action,\footnote{
 More comments on triangular diagram in Fig.~\ref{fig:BVst_Triangular_diagram} and its correspondence with action structures in general BV case can be found in Appendix \ref{Sect:BVst}.
}
may be illustrated by the triangular diagram in Fig.~\ref{fig:BVp=3_Triangular_diagram}, on which we assume implicit presence of the Dirac-conjugated partner for each field:

\def \TriangleDNp
{
\hspace{-175pt}
\begin{picture}(300,250)
\put(171,222){\text{\small $\psi_{\Zm\Zn\Zk}$}}
\put(151,191.3){\vector(3,4){15}}
\put(171,218){\line(-3,-4){15}}
\put(200,190){\line(-3,4){21}}
%
\put(122,181)
  {\text{\small $\gC^{\gix{1}}_{\Zm\Zn}$,\,$\gB^{\gix{1}}_{\Zm\Zn}$}}
\put(199,181)
  {\text{\small $\gC^{\gix{0}}_{\Zm\Zn}$}}
\put(123,154){\vector(3,4){15}}
\put(141,178){\vector(-3,-4){15}}
\put(183,154){\vector(3,4){15}}
\put(201,178){\line(-3,-4){15}}
\put(229,151){\line(-3,4){19}}
%
\put(98,142)
  {\text{\small $\gC^{\gix{2}}_{\Zm{}}$,\,$\gB^{\gix{2}}_{\Zm{}}$}}
\put(159,142)
  {\text{\small $\gC^{\gix{1}}_{\Zm{}}$,\,$\gB^{\gix{1}}_{\Zm{}}$}}
\put(228,142)
  {\text{\small $\gC^{\gix{0}}_{\Zm{}}$}}
\put(93,114){\vector(3,4){15}}
\put(111,138){\vector(-3,-4){15}}
\put(153,114){\vector(3,4){15}}
\put(171,138){\vector(-3,-4){15}}
\put(213,114){\vector(3,4){15}}
\put(231,138){\line(-3,-4){15}}
\put(259,111){\line(-3,4){20}}
%
\put(71,102) {\text{\small $\gC^{\gix{3}}_{\Zo}$\!,\,$\gB^{\gix{3}}_{\Zo}$}}
\put(131,102) {\text{\small $\gC^{\gix{2}}_{\Zo}$\!,\,$\gB^{\gix{2}}_{\Zo}$}}
\put(191,102) {\text{\small $\gC^{\gix{1}}_{\Zo}$\!,\,$\gB^{\gix{1}}_{\Zo}$}}
\put(258,102)  {\text{\small $\gC^{\gix{0}}_{\Zo}$}}
    \put(350,242) {\text{level \quad \;degree}}
    \multiput(350,225)(-5,0){22}{\line(1,0){3}}
    \put(360,225) {\text{\small $0$ \qquad\quad $\;3$}}
    \multiput(350,186)(-5,0){17}{\line(1,0){3}}
    \put(360,186) {\text{\small $1$ \qquad\quad\; $\!\!2$}}
    \multiput(350,147)(-5,0){12}{\line(1,0){3}}
    \put(360,147) {\text{\small $2$ \qquad\quad\; $\!\!1$}}
    \multiput(350,107)(-5,0){7}{\line(1,0){3}}
    \put(360,107) {\text{\small $3$ \qquad\quad\; $\!\!0$}}
%
\end{picture}
 \vspace{-100pt}
}
\begin{figure}[htbp]
\centering
  \TriangleDNp  
  \caption{\hspace{1ex} Triangular diagram for fermionic $3$-form model \hspace{7ex} }
 \label{fig:BVp=3_Triangular_diagram}
\end{figure}
%
%
\noindent
With each down-left edge (line) between vertices with ghosts of degrees $r{\,-\,}1$ and $r$ one associates gauge-fixing operators\footnote{Here we use compact index notation for antisymmetrized groups of indices.} $X^{\ZZm{r}\ZZn{r+1}}$, $\dc{X}^{\ZZm{r}\ZZn{r+1}}$.
The gauge fermion (\ref{gf_action_p=3}) is the sum over terms, in which these gauge-fixing matrices couple
ghost fields $ \gC_{\ZZm{r}},\, \dcgC_{\ZZm{r}} $ and $ \gC_{\ZZm{r+1}},\, \dcgC_{\ZZm{r+1}} $  from the incident vertices into combinations $\dcgC\, X\, \gC $ and $\dcgC\, \dc{X}\, \gC$ by contracting corresponding groups of indices.\footnote{The $X^{\ZZm{r}\ZZn{r-1}}$ operators couple the lower-degree Dirac conjugated ghost with a higher-degree ghost, whereas $\dc{X}^{\ZZm{r}\ZZn{r-1}}$ operators couple higher-degree conjugated ghosts with lower-degree ghosts. In what follows we will use particular gauges for which $\dc{X}^{\ZZm{r}\ZZn{r-1}}$ is just an operator $X^{\ZZn{r-1}\ZZm{r}}$ with transposed groups of indices.}
This rule includes the top vertex with original fields $  \psi_{\ZZn{3}},\,\dcpsi_{\ZZn{3}}   \,\leftrightarrow\, \gC^{\gix{0}}_{ \ZZn{3}},\,\dcgC^{\gix{0}}_{ \ZZn{3}} $.

\newpar

For gauge fermion (\ref{gauge_fermion_delta_p=3}), the gauge fixing excludes antifields via
 \begin{eqnarray}
   \varPhi^\af_I {\:=\,} \frac{\var^{} \varPsi_{\delta}}{\,\var \,\varPhi^I}
    \,,
    \qquad
   \dc{\varPhi}^\af_I {\:=\,} \frac{\var^{} \varPsi_{\delta}}{\,\var \,\dc{\varPhi}^I \vphantom{|^{l}}}
    \,,
 \label{gauge_fixing_p=3} 
 \end{eqnarray}
so that in the minimal sector one gets substitutions
 \begin{equation}
  \hspace{-5mm}
   \begin{array}{l@{\hspace{1pt}}c@{\hspace{5pt}}l@{\hspace{10pt}}c@{\hspace{10pt}}}
      \psi_\af^{\Zm\Zn\Zk}
      &\mapsto&
       \dcgC^{\gix{1}}_{\Zs\Zp} X^{\Al\Zs\Zp\Ar \hspace{.5pt} \Al\Zm\Zn\Zk\Ar}
       \vphII\,,
     \\[0.2em]
      \gC_\af^{\gix{0} \Zm\Zn}
      &\mapsto&
       \dcgC^{\gix{1}}_{\Zl} X^{\Zl \hspace{.5pt} \Al\Zm\Zn\Ar} \,,
     \\[0.2em]
      \gC_\af^{\gix{0} \Zm}
      &\mapsto&
       \dcgC^{\gix{1}}_{\Zo} X^{\Zo \hspace{.5pt} \Zm} \,,
     \\[0.2em]
      \gC_\af^{\gix{0} \Zo}
      &\mapsto&
       0  \,,
     \\[0.2em]
    \end{array}
    \begin{array}{l@{\hspace{1pt}}c@{\hspace{5pt}}l@{\hspace{10pt}}l@{\hspace{1pt}}}
      \dcpsi_\af^{\Zm\Zn\Zk}
      &\mapsto&
       \dc{X}^{\Al\Zm\Zn\Zk\Ar \hspace{.5pt} \Al\Zs\Zp\Ar} \gC^{\gix{1}}_{\Zs\Zp} \,,
      &  {\small{\text{ $(\,l{\,=\,}0,\,k{\,=\,}0\,)$}}}
      ;
     \\[0.2em]
       \dcgC_\af^{\gix{0} \Zm\Zn}
      &\mapsto&
         \dc{X}^{\Al\Zm\Zn\Ar \hspace{.5pt} \Zl} \gC^{\gix{1}}_{\Zl} \,,
       &  {\small{\text{ $(\,l{\,=\,}1,\, k{\,=\,}0\,)$}}}
      ;
     \\[0.2em]
       \dcgC_\af^{\gix{0} \Zm}
      &\mapsto&
         \dc{X}^{\Zm \hspace{.5pt} \Zo} \gC^{\gix{1}}_{\Zo} \,,
       &  {\small{\text{ $(\,l{\,=\,}2,\, k{\,=\,}0\,)$}}}
      ;
     \\[0.2em]
      \dcgC_\af^{\gix{0} \Zo}
      &\mapsto&
       0 \,,
      &  {\small{\text{ $(\,l{\,=\,}3,\,k{\,=\,}0\,)$}}}
      ;
     \\[0.2em]
   \end{array}
 \label{gf_antifields_min_p=3}
 \end{equation}
whereas for the antifields in the auxiliary sector one gets:\footnote{
 Gauge-fixing rules (\ref{gf_antifields_min_p=3}) and (\ref{gf_antifields_aux_p=3}) prescribed by the gauge fermion (\ref{gauge_fermion_delta_p=3}) can be compactly encoded as:
 \begin{equation}
   \begin{array}{|lcl@{\hspace{1pt}}c@{\hspace{1pt}}}
      \gC_{\af}^{\gix{k} \ZZm{3-l}}
      &\mapsto&
       \dcgC^{\gix{k-1}}_{\ZZn{4-l}} \dc{X}^{\ZZn{4-l}\ZZm{3-l}}
       +
       \dcgC^{\gix{k+1}}_{\ZZn{2-l}} X^{\ZZn{2-l}\ZZm{3-l}}
       \,,
      &
     \\[0.3em]
      %
      \dcgC_{\af}^{\gix{k}\ZZm{3-l}}
      &\mapsto&
       X^{\ZZm{3-l}\ZZn{4-l}} \gC^{\gix{k-1}}_{\ZZn{4-l}}
       +
       \dc{X}^{\ZZm{3-l}\ZZn{2-l}} \gC^{\gix{k+1}}_{\ZZn{2-l}}
       \,,
      &
   \end{array}
   \quad {\footnotesize{ \text{( $0\tleq k\tleq l$,\,\; $0 \tleq l\tleq 3$ )}}}
  \nonumber
 \end{equation}
 implying ``boundary rules'' that the ghosts of level $0$: $\gC^{\gix{0}}_{\ZZn{3}},\,\dcgC^{\gix{0}}_{\ZZn{3}}$, are the original fields $\psi_{\ZZn{3}},\,\dcpsi_{\ZZn{3}}$, and the ghosts $\gC^{\gix{k}}_{\ZZn{r}},\,\dcgC^{\gix{k}}_{\ZZn{r}}$ of levels $r<0$,\, $r>3$ and generations $k<0$,\, $k>3{\,-\,}r$ are zeros.
}
 \begin{equation}
  \hspace{0mm}
   \begin{array}{l@{\hspace{1pt}}c@{\hspace{5pt}}l@{\hspace{10pt}}c@{\hspace{10pt}}}
      \gC_\af^{\gix{1} \Zm\Zn}
      &\mapsto&
       \dcpsi_{\Zs\Zp\Zt} \dc{X}^{\Al\Zs\Zp\Zt\Ar \hspace{.5pt} \Al\Zm\Zn\Ar}
       {\,+\,}
       \dcgC^{\gix{2}}_{\Zl} X^{\Zl \hspace{.5pt} \Al\Zm\Zn\Ar}
       ,
     \\[0.3em]
      \gC_\af^{\gix{k} \Zl}
      &\mapsto&
       \dcgC^{\gix{k-1}}_{\Zm\Zn} \dc{X}^{\Al\Zm\Zn\Ar \hspace{.5pt} \Zl}
       {\,+\,}
       \dcgC^{\gix{k+1}}_{\Zo} X^{\Zo \hspace{.5pt} \Zl}
       ,
     \\[0.3em]
      \gC_\af^{\gix{k} \Zo}
      &\mapsto&
       \dcgC^{\gix{k-1}}_{\Zm} \dc{X}^{\Zm \hspace{.5pt} \Zo}
       \,,
    \end{array}
    \begin{array}{l@{\hspace{1pt}}c@{\hspace{5pt}}l@{\hspace{10pt}}r@{\hspace{1pt}}}
      \dcgC_\af^{\gix{1} \Zm\Zn}
      &\mapsto&
       X^{\Al\Zm\Zn\Ar \hspace{.5pt} \Al\Zs\Zp\Zt\Ar} \psi_{\Zs\Zp\Zt}
       {\,+\,}
       \dc{X}^{\Al\Zm\Zn\Ar \hspace{.5pt} \Zn} \gC^{\gix{2}}_{\Zn } 
       ,
       &  {\small{\text{  $(\,l{\,=\,}1,\, k{\,=\,}1\,)$}}}
       ;
     \\[0.3em]
      \dcgC_\af^{\gix{k} \Zl}
      &\mapsto&
       X^{\Zl \hspace{.5pt} \Al\Zm\Zn\Ar} \gC^{\gix{k-1}}_{\Al\Zm\Zn\Ar}
       {\,+\,}
       \dc{X}^{\Zl \hspace{.5pt} \Zo} \gC^{\gix{k+1}}_{\Zo} 
       ,
       &  {\small{\text{  $(\,l{\,=\,}2,\, 1 \tleq k \tleq 2\,)$}}}
       ;
     \\[0.3em]
      \dcgC_\af^{\gix{k} \Zo}
      &\mapsto&
       X^{\Zo \hspace{.5pt} \Zm} \gC^{\gix{k-1}}_{\Zm}
      \,,
      &  {\small{\text{ $(\,l{\,=\,}3,\; 1\tleq k\tleq 3 \,)$}}}
      .
   \end{array}
 \label{gf_antifields_aux_p=3}
 \end{equation}

\newpar

Gauge fixing (\ref{gf_antifields_min_p=3}) and (\ref{gf_antifields_aux_p=3}) leads to the new representation of the BV theory (\ref{master_action_p=3}), defined on the space of BV fields $\varPhi^I$ only. It is described by the gauge-fixed action
 \begin{eqnarray}
  \hspace{-5mm}
   &&S^{\scriptscriptstyle(3)}_{\delta} [\varPhi]
    \;=\;
     S^{\scriptscriptstyle(3)}_{0} [\psi,\dcpsi]
     +
     S^{\scriptscriptstyle(3)}_{\delta\,\text{gh}} [\varPhi] 
    +
    S^{\scriptscriptstyle(3)}_{\delta\,\text{aux}} [\varPhi]
    \,,
 \label{gf_action_p=3}  
 \end{eqnarray}
in which the original action term (\ref{S0_p=3}) does not change,  the gauge-fixed ghost part of the minimal action takes the form
 \begin{eqnarray}
  &&
  \hspace{-8mm}
   \begin{array}{lllllll}
    S^{\scriptscriptstyle(3)}_{\delta\,\text{gh}} [\varPhi]
    &\!\!=\!\!&
    \displaystyle
    \!\!\int \!d^dx\,
       \, \Bigl\{ i 
      \big(
          \dcgC^{\gix{1}}_{\Zk\Zl} X^{\Al\Zk\Zl\Ar \hspace{.5pt} \Al\Zr\Zm\Zn\Ar}
           D_{\Zr} \gC^{\gix{0}}_{\Zm\Zn}
         -
          \dcgC^{\gix{0}}_{\Zk\Zl} \dc{D}_{\Zr}
           \dc{X}^{\Al\Zk\Zl\Zr\Ar \hspace{.5pt} \Al\Zm\Zn\Ar} \gC^{\gix{1}}_{\Zm\Zn} \big)
    \\[.5em]
     && \hspace{10mm} \displaystyle
       +\,
      i 
      \big(
           \dcgC^{\gix{1}}_{\Zl} X^{\Zl \hspace{.5pt} \Al\Zr\Zm\Ar}
           D_{\Zr} \gC^{\gix{0}}_{\Zm}
         -
           \dcgC^{\gix{0}}_{\Zl} \dc{D}_{\Zr}
           \dc{X}^{\Al\Zl\Zr\Ar \hspace{.5pt} \Zm} \gC^{\gix{1}}_{\Zm}
     \big)
       +\,
      i 
      \big( \dcgC^{\gix{1}}_{\Zo} X^{\Zo \hspace{.5pt}  \Zr}
           D_{\Zr} \gC^{\gix{0}}_{\Zo}
         -
          \dcgC^{\gix{0}}_{\Zo} \dc{D}_{\Zr}
           \dc{X}^{\Zr \hspace{.5pt} \Zo} \gC^{\gix{1}}_{\Zo} \big)
     \Bigr\}
     \,,
   \end{array}
 \label{gf_gh action_p=3}
 \end{eqnarray}
becoming the functional of the minimal ghosts and antighosts only,
and the auxiliary part reads
 \begin{eqnarray}
  &&
  \hspace{-8mm}
   \begin{array}{lllllll}
    S^{\scriptscriptstyle(3)}_{\delta\,\text{aux}} [\varPhi]
    &\!\!=\!\!&
    \displaystyle
   \!\!\int \!d^dx\,
     \Bigl\{\,
      \big(
      \dcpsi_{\Zs\Zp\Zt} \dc{X}^{\Al\Zs\Zp\Zt\Ar \hspace{.5pt} \Al\Zm\Zn\Ar}
       {+\,}
       \dcgC^{\gix{2}}_{\Zl} X^{\Zl \hspace{.5pt} \Al\Zm\Zn\Ar}
       \big)
      \gB^{\gix{1}}_{\Zm\Zn}
      +
      \dcgB^{\gix{1}}_{\Zm\Zn}
      \big( X^{\Al\Zm\Zn\Ar \hspace{.5pt} \Al\Zs\Zp\Zt\Ar} \psi_{\Zs\Zp\Zt}
       {+}
       \dc{X}^{\Al\Zm\Zn\Ar \hspace{.5pt} \Zl} \gC^{\gix{2}}_{\Zl}
      \big)
    \\[-.1em]
    &&\hspace{14mm} \displaystyle
      +
      \sum_{k=1}^{2}
      \big(
      (
      \dcgC^{\gix{k-1}}_{\Zk\Zl} \dc{X}^{\Al\Zk\Zl\Ar \hspace{.5pt} \Zm}
       {+\,}
       \dcgC^{\gix{k+1}}_{\Zo} X^{\Zo \hspace{.5pt} \Zm}
       )
      \gB^{\gix{k}}_{\Zm}
      +
      \dcgB^{\gix{k}}_{\Zm}
      ( X^{\Zm \hspace{.5pt} \Al\Zk\Zl\Ar} \gC^{\gix{k-1}}_{\Zk\Zl}
       {+}
       \dc{X}^{\Zm \hspace{.5pt} \Zo} \gC^{\gix{k+1}}_{\Zo}
      )
      \big)
    \\[-.1em]
    &&\hspace{14mm} \displaystyle
      +
      \sum_{k=1}^{3}
      \big(
       \dcgC^{\gix{k-1}}_{\Zm} \dc{X}^{\Zm \hspace{.5pt} \Zo}
       \gB^{\gix{k}}_{\Zo}
      +\,
       \dcgB^{\gix{k}}_{\Zo}
       X^{\Zo \hspace{.5pt} \Zm} \gC^{\gix{k-1}}_{\Zm}
      \big) \,
     \Bigr\}
      \,.
   \end{array}
 \label{gf_aux_action_p=3} 
 \end{eqnarray}
 %
The fields of the minimal and auxiliary sectors after gauge fixing become entangled. In particular, original fields $\psi_{\Zm\Zn\Zk},\,\dcpsi_{\Zm\Zn\Zk}$ enter the gauge-fixed auxiliary action.

\newpar

The gauge fixing (\ref{gauge_fermion_delta_p=3}),(\ref{gauge_fixing_p=3}) leads to the \emph{nondegenerate} gauge-fixed action when certain rank conditions are satisfied. First, the conditions of transversality of the gauge and reducibility generators (\ref{gauge_generators_p=3}) with respect to the gauge-fixing surface, are equivalent to maximal-rank property of Faddeev-Popov operators $X^{\ZZn{r} \hspace{.5pt} \ZZm{r+1}} D_{\Zm}$,\;  $\dc{D}_{\Zm} \dc{X}^{\ZZm{r+1} \hspace{.5pt} \ZZn{r}}$,\, $2 \tgeq r \tgeq 0$\,:
 \begin{equation}
    X^{\Al\Zk\Zl\Ar \hspace{.5pt} \Al\Zr\Zm\Zn\Ar} D_{\Zr}
    \,,\;\;
    \dc{D}_{\Zr} \dc{X}^{\Al\Zm\Zn\Zr\Ar \hspace{.5pt} \Al\Zk\Zl\Ar}
    \,,\;\;
    X^{\Zk \hspace{.5pt} \Al\Zr\Zm\Ar} D_{\Zr}
    \,,\;\;
    \dc{D}_{\Zr} \dc{X}^{\Al\Zm\Zr\Ar \hspace{.5pt} \Zk}
    \,,\;\;
    X^{\Zo \hspace{.5pt} \Zr} D_{\Zr}
    \,,\;\;
    \dc{D}_{\Zr} \dc{X}^{\Zr \hspace{.5pt} \Zo}
 \label{}
 \end{equation}
for residual dynamical fields. Second, gauge-fixing conditions should fix appropriate number of degrees of freedom, which guarantee that all local gauge freedom of BV theory (\ref{master_action_p=3}) are fixed.
For the gauge fermion (\ref{gauge_fermion_delta_p=3}) this condition constraints ranks of $X^{\ZZm{r} \hspace{.5pt} \ZZn{r+1}}$ and $\dc{X}^{\ZZn{r+1} \hspace{.5pt} \ZZm{r}}$, which may be read off the maximal rank property of the linear system
 \begin{equation}
  \hspace{0mm}
   \begin{array}{|l@{\hspace{3pt}}c@{\hspace{5pt}}l@{\hspace{10pt}}c@{\hspace{10pt}}}
      \frac{ \var S^{\scriptscriptstyle(3)}_{\delta}}{\var \gB^{\gix{1}}_{ \Zm\Zn} }
      &\sim&
       \dcpsi_{\Al\Zs\Zp\Zt\Ar} \dc{X}^{\Al\Zs\Zp\Zt\Ar \hspace{.5pt} \Al\Zm\Zn\Ar}
       {\,+\,}
       \dcgC^{\gix{2}}_{ \Zl} X^{\Zl \hspace{.5pt} \Al\Zm\Zn\Ar}
       = 0
       \,,
     \\[0.4em]
      \frac{ \var S^{\scriptscriptstyle(3)}_{\delta}}{\var \gB^{\gix{k}}_{ \Zl} }
      &\sim&
       \dcgC^{\gix{k-1}}_{ \Zm\Zn} \dc{X}^{\Al\Zm\Zn\Ar \hspace{.5pt} \Zl}
       {\,+\,}
       \dcgC^{\gix{k+1}}_{ \Zo} X^{\Zo \hspace{.5pt} \Zl}
       = 0
       \,,
     \\[0.4em]
      \frac{ \var S^{\scriptscriptstyle(3)}_{\delta}}{\var \gB^{\gix{k}}_{ \Zo} }
      &\sim&
       \dcgC^{\gix{k-1}}_{ \Zm} \dc{X}^{\Zm \hspace{.5pt} \Zo}
       = 0
       \,,
    \end{array}
    \begin{array}{|l@{\hspace{3pt}}c@{\hspace{5pt}}l@{\hspace{10pt}}r@{\hspace{1pt}}}
      \frac{ \var S^{\scriptscriptstyle(3)}_{\delta}}{\var \dcgB^{\gix{1}}_{ \Zm\Zn} \vphantom{{|^I}^b} }
      &\sim&
       X^{\Al\Zm\Zn\Ar \hspace{.5pt} \Al\Zs\Zp\Zt\Ar} \psi_{\Al\Zs\Zp\Zt\Ar}
       {\,+\,}
       \dc{X}^{\Al\Zm\Zn\Ar \hspace{.5pt} \Zl} \gC^{\gix{2}}_{ \Zl} 
       = 0
       \,, \hspace{-1em}
       &  {\small{\text{  $(\, 
       k{\,=\,}1\,)$}}}
       ;
     \\[0.4em]
      \frac{ \var S^{\scriptscriptstyle(3)}_{\delta}}{\var \dcgB^{\gix{k}}_{ \Zl} \vphantom{{|^I}^b}}
      &\sim&
       X^{\Zl \hspace{.5pt} \Al\Zm\Zn\Ar} \gC^{\gix{k-1}}_{ \Zm\Zn}
       {\,+\,}
       \dc{X}^{\Zl \hspace{.5pt} \Zo} \gC^{\gix{k+1}}_{ \Zo} 
       = 0
       \,,
       &  {\small{\text{
           $(\, 
                 k = 2,\, 1\,  
           \,)$}}}
       ;
     \\[0.4em]
      \frac{ \var S^{\scriptscriptstyle(3)}_{\delta}}{\var \dcgB^{\gix{k}}_{ \Zo} \vphantom{{|^I}^b}}
      &\sim&
       X^{\Zo \hspace{.5pt} \Zm} \gC^{\gix{k-1}}_{ \Zm}
       = 0
       \,,
      & \hspace{-2em}
      {\small{\text{
       $(\, 
            k = 3,\,2,\,1
       \,)$}}}
      .
   \end{array}
 \label{gf_conditions_aux_p=3} 
 \end{equation}
implying
the rank conditions
 \begin{eqnarray}
  \begin{array}{|lr}
   \rank
    \begin{pmatrix}
       X^{\Al\Zs\Zp\Ar \hspace{.5pt}  \Al\Zm\Zn\Zk\Ar}
       \;\big|\;
       \dc{X}^{\Al\Zs\Zp\Ar  \hspace{.5pt} \Zl}
    \end{pmatrix}
   \,=\,
   \rank
    \begin{pmatrix}
       \dc{X}^{\Al\Zm\Zn\Zk\Ar \hspace{.5pt} \Al\Zs\Zp\Ar}
       \\
       X^{\Zl \hspace{.5pt} \Al\Zs\Zp\Ar}
    \end{pmatrix}
   \,=\,
    \nfrac{d(d{-}1)}{2}\,;
  \\[0.8em]
   \rank
    \begin{pmatrix}
       X^{\Zl \hspace{.5pt} \Al\Zm\Zn\Ar}
       \; \big| \;
       \dc{X}^{\Zl \hspace{.5pt} \Zo}
    \end{pmatrix}
   \,=\,
   \rank
    \begin{pmatrix}
       \dc{X}^{\Al\Zm\Zn\Ar \hspace{.5pt} \Zl}
       \\
       X^{\Zo \hspace{.5pt} \Zl}
    \end{pmatrix}
   \,=\,
     d \,,
   \\[1em]
   \rank X^{\Zo \hspace{.5pt} \Zm} \;=\; \rank \dc{X}^{\Zm \hspace{.5pt} \Zo} \,=\, 1
   \,,
  \end{array}
 \label{GF_rank_conds_p=3} 
 \end{eqnarray}
where rank (as before) counts the rank of an object with two groups of tensor indices, each component of which is an endomorphism in the space of spinors.

\newpar

Provided these rank conditions are satisfied, the gauge-fixed action (\ref{gf_action_p=3}) is nondegenerate and suitable for quantization: quantum properties of the theory are described by the generating functional
 \begin{eqnarray}
   Z^{\scriptscriptstyle(3)}
   \:=\: \!\int\! D\varPhi\, \exp \Bigl\{ i S^{\scriptscriptstyle(3)}_{\delta}[\varPhi] \Bigr\}
   \,.
 \label{Z_with_Sdelta_p=3} 
 \end{eqnarray}
The BV quantization scheme guarantees the independence of the physical results of the particular choice of the gauge fermion.

For the delta-function gauge fixing, fields $\gB^{\gix{k}}_{\ZZm{r}}$ and $\dcgB^{\gix{k}}_{\ZZm{r}}$ play a role of Lagrange multipliers in the gauge-fixed action (\ref{gf_action_p=3}), entering only its auxiliary part. They can be integrated out in (\ref{Z_with_Sdelta_p=3}) generating delta functions, encoding conditions (\ref{gf_conditions_aux_p=3}), which are often referred as gauge fixing conditions for ghosts and original fields. Integrating out Lagrange multiplier fields and resolving delta functions effectively provide reduction
 \begin{equation}
  S^{\scriptscriptstyle(3)}_{\delta}[\varPhi]
  \;\;\to\;\; \breve{S}^{\scriptscriptstyle(3)}_{\delta}[\breve{\varPhi}]
  \,,
 \label{reduction_p=3_in_action} 
 \end{equation}
generating a contribution to the measure, which is ultralocal for algebraic, nondifferential gauge-fixing.\footnote{For abelian gauge theory and linear gauge-fixing this ultralocal contribution to the measure is field-independent.}
Alternatively, for such gauge fixing when operators $X^{\ZZm{r} \hspace{.5pt} \ZZn{r+1}}$ and $\dc{X}^{\ZZn{r+1} \hspace{.5pt} \ZZm{r}}$ are purely algebraic matrices, the reduction (\ref{reduction_p=3_in_action}) could be performed at the classical level by exclusion of this set of fields on the reduction surface $\varPhi=\varPhi(\breve{\varPhi})$ in the action leading to equivalent reduced representation $\breve{S}^{\scriptscriptstyle(3)}_{\delta}[\breve{\varPhi}] = S^{\scriptscriptstyle(3)}_{\delta}[\varPhi(\breve{\varPhi})]$. The constraint surface appears as the solution of the subset of equations of motion with respect to these auxiliary fields, allowing to algebraically express them in terms of residual fields \cite{Henneaux:1992ig}. In this case, the quantization prescription (\ref{Z_with_Sdelta_p=3}) can be applied after the reduction in the classical action directly for $\breve{S}^{\scriptscriptstyle(3)}_{\delta}[\breve{\varPhi}]$.

\newpar

In any case, to proceed with the reduction one needs to resolve conditions (\ref{gf_conditions_aux_p=3}), which implies specifying the gauge-fixing matrices.
Below we consider two gauges: one defined by full-rank matrices\;
 $ \rank X^{\ZZm{r} \hspace{.5pt} \ZZn{r+1}}
  = \rank \dc{X}^{\ZZn{r+1} \hspace{.5pt} \ZZm{r}}
  = \binom{d}{r}
 $,
and another defined by matrices of minimal possible ranks:
 $ \rank X^{\ZZm{r} \hspace{.5pt} \ZZn{r+1}}
  = \rank \dc{X}^{\ZZn{r+1} \hspace{.5pt} \ZZm{r}}
  = \binom{d-1}{r}
 $,
allowable by (\ref{GF_rank_conds_p=3}).
To specify gauge-fixing matrices in the latter case we need to decompose fields into their algebraically irreducible components. Also, transition to the basis of irreducible fields allows to solve (\ref{gf_conditions_aux_p=3}) (for both gauges) and to perform subsequent explicit reduction in the gauge-fixed action.


  \subsection{Basis of irreducible fields, two gauges, and reduction}
 \label{SSect:Gauge_fixing_p=3}
    \subsubsection{Component decomposition of fields}
 \label{SSSect:Gauge-fixing_in_components_p=3}
    \hspace{\parindent}
The original $3$-form spinor field can be algebraically decomposed into $\gamma$-irreducible components
 \begin{equation}
   \begin{array}{|clllllll}  
    &   \displaystyle
    \psi_{\Zm\Zn\Zk}
    &\!=\!&
     \CPsi_{\Zm\Zn\Zk}
    + \gamma_{\Al\Zm} \CPsi_{\Zn\Zk\Ar}
    + \gamma_{\Al\Zm\Zn} \CPsi_{\Zk\Ar}
    + \gamma_{\Zm\Zn\Zk} \CPsi_{\Zo}
    \,,
    &\;&
    \gamma^{\Zk} \CPsi_{\Zm\Zn\Zk}
    = \gamma^{\Zn} \CPsi_{\Zm\Zn}
    = \gamma^{\Zm} \CPsi_{\Zm}
    = 0
    \,,
    \\[0.5em]
    &   \displaystyle
    \dcpsi_{\Zm\Zn\Zk}
    &\!=\!&
      \dcCPsi_{\Zm\Zn\Zk}
     +  \dcCPsi_{\Al\Zm\Zn} \gamma_{\Zk\Ar}  
     -  \dcCPsi_{\Al\Zm} \gamma_{\Zn\Zk\Ar}  
     -  \dcCPsi_{\Zo} \gamma_{\Zm\Zn\Zk}   
     \,,
    &\;&
    \dcCPsi_{\Zm\Zn\Zk}\gamma^{\Zm}
    = \dcCPsi_{\Zm\Zn}\gamma^{\Zm}
    = \dcCPsi_{\Zm}\gamma^{\Zm}
    = 0
    \,,
   \end{array}
 \label{psi_decomp_p=3} 
 \end{equation}
We choose sign factors
at the Dirac-conjugated irreducible fields to provide $\dcCPsi_{\ZZm{s}}= \dcl{\CPsi_{\ZZm{s}}}$ for $\dcpsi_{\Zm\Zn\Zk}= \dcl{\psi_{\Zm\Zn\Zk}}$.
In the basis of irreducible components, the action (\ref{S0_p=3}) of the original model gets the form
 \begin{equation}
  \hspace{-8mm}
   \begin{array}{lllllll}
    S^{\scriptscriptstyle(3)}_{0} [\psi(\CPsi),\dcpsi(\dcCPsi)]
    &\!=\!&
     S^{\scriptscriptstyle(3) \text{D}}_{0} [\CPsi,\dcCPsi]
    +S^{\scriptscriptstyle(3) \text{OD}}_{0} [\CPsi,\dcCPsi]
    \,,
   \end{array}
  \label{So_in_components_p=3} 
 \end{equation}
where the diagonal and off-diagonal contributions are
 \begin{align}
   S^{\scriptscriptstyle(3) \text{D}}_{0} [\CPsi,\dcCPsi]
   & \:=\: 
     \HIDE{+} 
     \int \!d^dx\,
     i\sqrt{|\GG|}
    \,\Bigl\{\,
     \HIDE{+} 
     6
     \,\dcCPsi^{\Zm\Zn\Zk}
     \,\slashed{\nabla}_{\!\oix{3}}
     \CPsi_{\Zm\Zn\Zk}
     - 
    \, 2 \nfrac{(d{-}5)!}{(d{-}7)!}
     \,\dcCPsi^{\Zm\Zn}
     \,\slashed{\nabla}_{\!\oix{2}}^*
     \CPsi_{\Zm\Zn}
   \nonumber
   \\
   &
    \displaystyle
     \qquad\qquad\qquad\quad\;
    + 
    \, \nfrac{(d{-}3)!}{(d{-}7)!}
     \,\dcCPsi^{\Zm}
     \,\slashed{\nabla}_{\!\oix{1}}
     \CPsi_{\Zm}
    - 
    \, \nfrac{(d{-}1)!}{(d{-}7)!}
     \,\dcCPsi^{\Zo}
     \,\slashed{\nabla}_{\!\oix{0}}^*
     \CPsi_{\Zo}
    \,\Bigr\}
    \,,
   \nonumber
   \\
   S^{\scriptscriptstyle(3) \text{OD}}_{0} [\CPsi,\dcCPsi]
   & \:=\: 
    \displaystyle
     \int \!d^dx\,
     i\sqrt{|\GG|}
     \Bigl\{\,
    \HIDE{+} 
     6 \nfrac{(d{-}6)!}{(d{-}7)!}
      \,\big(
       \,\dcCPsi^{\Zr\Zm\Zn}
       \,{\nabla}_{\!\Zr}
       \CPsi_{\Zm\Zn}
       + 
       \,\dcCPsi^{\Zm\Zn}
       \,{\nabla}^{\Zr}
       \CPsi_{\Zm\Zn\Zr}
      \big)
   \nonumber
   \\
   &
    \displaystyle
     \qquad\qquad\qquad\quad\;
     - 
    \, 2 \nfrac{(d{-}4)!}{(d{-}7)!}
      \,\big(
       \,\dcCPsi^{\Zr\Zm}
       \,{\nabla}_{\!\Zr}
       \CPsi_{\Zm}
      - 
       \,\dcCPsi^{\Zm}
       \,{\nabla}^{\Zr}
       \CPsi_{\Zm\Zr}
      \big)
   \nonumber
   \\
   &
    \displaystyle
     \qquad\qquad\qquad\quad\;
     + 
      \, \nfrac{(d{-}2)!}{(d{-}7)!}
      \,\big(
       \,\dcCPsi^{\Zr}
       \,{\nabla}_{\!\Zr}
       \CPsi_{\Zo}
      + 
       \,\dcCPsi^{\Zo}
       \,{\nabla}^{\Zr}
       \CPsi_{\Zr}
      \big)
   \Bigr\}
   \,,
 \label{So_O_OD_in_components_p=3} 
 \end{align}
where in the diagonal part, which aggregates bilinear terms with irreducible fields of the same degree, we introduced shorthand notation for massive Dirac-like operators
 \begin{equation}
  \slashed{\nabla}_{\!\oix{s}}
   = \slashed{\nabla}
   \mp 
   \nfrac{i}{2}\, \sqrr\,(d{-}2s)
  \,,
  \qquad
  \slashed{\nabla}_{\!\oix{s}}^*
   = \slashed{\nabla}
     \pm 
     \nfrac{i}{2}\, \sqrr\,(d{-}2s)
  \,,
 \label{nabla-s_nabla*-s}
 \end{equation}
where $\slashed{\nabla} \equiv \gamma^\Zm \nabla_{\!\Zm}$.

\newpar

Analogously we decompose ghosts $\gC^{\gix{k}}_{\ZZm{r}}$,\,$\dcgC^{\gix{k}}_{\ZZm{r}}$ and Lagrange multiplier fields $\gB^{\gix{k}}_{\ZZm{r}}$,\,$\dcgB^{\gix{k}}_{\ZZm{r}}$ of levels $l=3{\,-\,}r:\; 0\tleq r \tleq 2 \,$ and generations $k:\; 0\tleq k\tleq l$. For ghosts $\gC^{\gix{k}}_{\ZZm{r}}$,\,$\dcgC^{\gix{k}}_{\ZZm{r}}$ decompositions read
 \begin{equation}
  \begin{array}{|c@{\hspace{6pt}}l@{\hspace{18pt}}l@{\hspace{18pt}}l@{\hspace{0pt}}llll}
   &
    \gC^{\gix{k}}_{ \Zm\Zn}  \;=\;
      \gCC^{\gix{k}}_{\rix{2} \Zm\Zn}
     \!+ \gamma_{\Al\Zm} \gCC^{\gix{k}}_{\rix{2} \Zn\Ar}
     \!+ \gamma_{\Zm\Zn} \gCC^{\gix{k}}_{\rix{2} \Zo}
    \,, \;
   &
    \gC^{\gix{k}}_{ \Zm}  \;=\;
      \gCC^{\gix{k}}_{\rix{1} \Zm}
     \!+ \gamma_{\Zm} \gCC^{\gix{k}}_{\rix{1} \Zo}
    \,, \;
   &
    \gC^{\gix{k}}_{ \Zo}  \;\equiv\;
      \gCC^{\gix{k}}_{\rix{0} \Zo}
    \,, \;
    \\[0.5em]
   &
    \dcgC^{\gix{k}}_{ \Zm\Zn}  \;=\;
      \dcgCC^{\gix{k}}_{\rix{2} \Zm\Zn}
     \!- \dcgCC^{\gix{k}}_{\rix{2} \Al\Zm} \gamma_{\Zm\Ar}
     \!- \dcgCC^{\gix{k}}_{\rix{2} \Zo} \gamma_{\Zm\Zn}
    \,, \;
   &
    \dcgC^{\gix{k}}_{ \Zm}  \;=\;
      \dcgCC^{\gix{k}}_{\rix{1} \Zm}
     \!+ \dcgCC^{\gix{k}}_{\rix{1} \Zo} \gamma_{\Zm}
    \,, \;
   &
    \dcgC^{\gix{k}}_{ \Zo}  \;\equiv\;
      \dcgCC^{\gix{k}}_{\rix{0} \Zo}
    \,, \;
    \\
  \end{array}
 \label{C_decomp_p=3} 
 \end{equation}
where irreducible components $\gCC^{\gix{k}}_{\rix{r} \ZZm{s}}$,\, $\dcgCC^{\gix{k}}_{\rix{r} \ZZm{s}}$ satisfy
 \begin{equation}
  \begin{array}{lll}
   &
    \gamma^{\Zn} \gCC^{\gix{k}}_{\rix{2} \Zm\Zn}
    =
    \gamma^{\Zn} \gCC^{\gix{k}}_{\rix{2} \Zn}
    =
    \gamma^{\Zn} \gCC^{\gix{k}}_{\rix{1} \Zn}
    = 0
    \,, \;
   &\;\;
    \dcgCC^{\gix{k}}_{\rix{2} \Zm\Zn} \gamma^{\Zn}
    =
    \dcgCC^{\gix{k}}_{\rix{2} \Zn} \gamma^{\Zn}
    =
    \dcgCC^{\gix{k}}_{\rix{1} \Zn} \gamma^{\Zn}
    = 0
    \,. \;
  \\
  \end{array}
 \label{CC_irreducibility} 
 \end{equation}
The new lower numerical subscript denotes the rank of the initial field, from which the component originates. Sign factors
in front of the components of degree $s$ in decomposition of the Dirac-conjugated field of degree $r$ in (\ref{C_decomp_p=3}) are chosen so that $\dcgCC^{\gix{k}}_{\rix{r} \ZZm{s}} = \dcl{\gCC^{\gix{k}}_{\rix{r} \ZZm{s}}}$ for $\dcgC^{\gix{k}}_{\ZZm{r}} = \dcl{\gC^{\gix{k}}_{\ZZm{r}}}$. Decomposition of the Lagrange multiplier fields $\gB^{\gix{k}}_{\ZZm{r}}$,\,$\dcgB^{\gix{k}}_{\ZZm{r}}$ of levels $l=3{\,-\,}r:\; 0\tleq r \tleq 2 $\,, and generations $k:\; 1\tleq k\tleq l$\,, into irreducible components $\gCB^{\gix{k}}_{\rix{r} \ZZm{s}}$,\,$\dcgCB^{\gix{k}}_{\rix{r} \ZZm{s}}$ is identical to (\ref{C_decomp_p=3}), (\ref{CC_irreducibility}).

\newpar

Field $\psi_{\Zm\Zn\Zk}$ of degree $r{\,=\,}3$ (corresponding to level $0$) describes $\binom{d}{3}=\frac16 d(d{-}1)(d{-}2)$ spinor components whereas each irreducible component $\CPsi_{\ZZn{s}}$ of degree $s$ describes $\binom{d}{s-1}$ independent spinor components. Analogously a ghost $\gC^{\gix{k}}_{\ZZn{r}}$ of degree $r{\,=\,}0,\,1,\,2,$ contains $\binom{d}{r}$ spinor components, and its irreducible components $\gCC^{\gix{k}}_{\rix{r} \ZZn{s}}$ of degree $s$  are the spinor-valued $s$-forms subject to $\binom{d}{s-1}$ spinor-valued conditions $\gamma^{\Zm}\gCC^{\gix{k}}_{\rix{r} \ZZm{s}}{\,=\,}0$,\, ($s \tgeq 1$), and thus contain
 $
  \binom{d}{s}-\binom{d}{s-1}
  = \frac{d-2s+1}{d-s+1}\binom{d}{s}
 $
independent spinor components (assuming $\binom{d}{-1}=0$). The same counting applies to the Dirac-conjugated fields and their components as well as to the Lagrange multiplier fields.

\newpar  

To draw a parallel with the general $p$ case, which we consider in the next Section~\ref{Sect:p=p}, we introduce projectors onto basis elements of decompositions (\ref{psi_decomp_p=3}), (\ref{C_decomp_p=3})
 \begin{equation}
  \begin{array}{|@{\hspace{1pt}}clllll}  
    &
      P_\pix{s} {}_{\ZZm{3}}^{\:\ZZn{3}} \psi_{\ZZn{3}}
       =
       \gamma_{\ZZm{3-s}}\CPsi_{\ZZm{s}}
    \,,
    \\[0.7em]
    &
    \dcpsi_{\ZZn{3}} \dc{P}_\pix{s} {}^{\ZZn{3}}_{\,\ZZm{3}}
    =
    \dcsc[3]{s}  \dcCPsi_{\ZZm{s}}\gamma_{\ZZm{3-s}}
    \,,
  \end{array}
   \, \text{\small{( $0 \tleq s\tleq 3$ )}}
   \,,
  \quad
  \begin{array}{|@{\hspace{1pt}}clllll}  
    &
      P_\pix{s} {}_{\ZZm{r}}^{\:\ZZn{r}} \gC^{\gix{k}}_{ \ZZn{r}}
     =
     \gamma_{\ZZm{r-s}} \gCC^{\gix{k}}_{\rix{r} \ZZm{s}} \,,
    \\[0.7em]
    &
    \dcgC^{\gix{k}}_{\ZZn{r}} \dc{P}_\pix{s} {}_{\,\ZZm{r}}^{\ZZn{r}}
    =  
    \dcsc[r]{s} \, \dcgCC^{\gix{k}}_{\rix{r} \ZZm{s}} \gamma_{\ZZm{r-s}}
    \,,
  \end{array}
  \, \text{\small{( $r=1,\,2$;\;  $0 \tleq s\tleq r$ )}}
  \,,
 \label{def_Projectors_p=3,r=2,1} 
 \end{equation}
where the sign factors $\dcsc[r]{s} = \msp^{\frac{r(r-1)}{2}-\frac{s(s-1)}{2}}$ at the Dirac-conjugated components are those from (\ref{psi_decomp_p=3}) and (\ref{C_decomp_p=3}).\footnote{Sign factors $\dcsc[r]{s}$ appear in decompositions of Dirac-conjugated fields (\ref{psi_decomp_p=3}) and (\ref{C_decomp_p=3}) written in the form:
 \begin{equation}
   \begin{array}{clll}
    &   \displaystyle
    \dcpsi_{\Zm\Zn\Zk}
     \;=\;
      \dcsc[3]{3} \dcCPsi_{\Zm\Zn\Zk}
     +  \dcsc[3]{2} \dcCPsi_{\Al\Zm\Zn} \gamma_{\Zk\Ar}  
     +  \dcsc[3]{1} \dcCPsi_{\Al\Zm} \gamma_{\Zn\Zk\Ar}  
     +  \dcsc[3]{0} \dcCPsi_{\Zo} \gamma_{\Zm\Zn\Zk}   
     \,,
       \\[0.5em]
   &
    \dcgC^{\gix{k}}_{ \Zm\Zn}  \;=\;
      \dcsc[2]{2} \dcgCC^{\gix{k}}_{\rix{2} \Zm\Zn}
     \!+ \dcsc[2]{1} \dcgCC^{\gix{k}}_{\rix{2} \Al\Zm} \gamma_{\Zm\Ar}
     \!+ \dcsc[2]{0} \dcgCC^{\gix{k}}_{\rix{2} \Zo} \gamma_{\Zm\Zn}
    \,,
   \quad
    \dcgC^{\gix{k}}_{ \Zm}  \;=\;
      \dcsc[1]{1} \dcgCC^{\gix{k}}_{\rix{1} \Zm}
     \!+ \dcsc[1]{0} \dcgCC^{\gix{k}}_{\rix{1} \Zo} \gamma_{\Zm}
    \,,
   \quad
    \dcgC^{\gix{k}}_{ \Zo}  \;=\;
      \dcsc[0]{0} \dcgCC^{\gix{k}}_{\rix{0} \Zo}
    \,. \;
  \end{array}
 \nonumber 
 \end{equation}

}
 Explicit forms of these projectors is in fact not needed, the defining properties would be enough.
The Dirac bar over projectors just reflects the appropriate adjustment of indices:
 \begin{equation}
  \dc{P}_\pix{s} {}^{\ZZn{r}}_{\,\ZZm{r}}
    =
  \dcl{ P_\pix{s} {}_{\ZZn{r}}^{\:\ZZm{r}} }
    =
  P_\pix{s} {}^{\ZZn{r}}_{\,\ZZm{r}}
   \equiv
  \GG^{\ZZn{r}\ZZk{r}} P_\pix{s} {}_{\ZZk{r}}^{\:\ZZl{3}}\, \GG_{\ZZl{r}\ZZm{r}}
    \,,
    \qquad
  \dc{P}_\pix{s}^{\ZZm{r} \hspace{.5pt} \ZZn{r}}
  = P_\pix{s}^{\ZZm{r} \hspace{.5pt} \ZZn{r}}
  \,,
 \label{dcProj_p=3}
 \end{equation}
where $\GG^{\ZZn{3}\ZZk{3}}$ and $\GG_{\ZZl{3}\ZZn{3}}$ are defined in (\ref{def_index_operators}) being covariant index raising and lowering operators on a space of antisymmetric tensors.
Though expressions for projectors are known,
their explicit application can be avoided by relying on the defining properties (\ref{def_Projectors_p=3,r=2,1}).
Within covariant structures we deal with, it will be possible to express terms with projectors so, that a projector directly acts on a field by picking up a corresponding irreducible basis component.

    \subsubsection{Minimal-rank gauge fixing and reduction}
     \label{}
      \hspace{\parindent}
Projectors (\ref{def_Projectors_p=3,r=2,1}) allow to define the gauge-fixing operators  $X^{\ZZm{r} \hspace{.5pt} \ZZn{r+1}}$ and $\dc{X}^{\ZZn{r+1} \hspace{.5pt} \ZZm{r}}$
   \begin{align}
    &
    \begin{array}{|@{\hspace{1pt}}lllll}
      &X^{\Al\Zs\Zp\Ar \hspace{.5pt} \Al\Zm\Zn\Zk\Ar}
      &\!\!=\!\!&
     \displaystyle
       \gamma_{\Zl} P_\pix{2}^{\Al\Zs\Zp\Zl\Ar \hspace{.5pt} \Al\Zm\Zn\Zk\Ar}
       +
       \gamma_{\Zl} P_\pix{0}^{\Al\Zs\Zp\Zl\Ar  \hspace{.5pt} \Al\Zm\Zn\Zk\Ar}
      =
       P_\pix{2}^{\Al\Zs\Zp\Ar \hspace{.5pt} \Al\Zm\Zn}  \gamma^{\Zk\Ar}
       +
       P_\pix{0}^{\Al\Zs\Zp\Ar \hspace{.5pt} \Al\Zm\Zn}  \gamma^{\Zk\Ar}
      \,,
     \\[0.5ex]
     &\dc{X}^{\Al\Zm\Zk\Zr\Ar \hspace{.5pt} \Al\Zt\Zs\Ar}
      &\!\!=\!\!&
     \displaystyle
      P_\pix{2}^{\Al\Zm\Zk\Zr\Ar \hspace{.5pt} \Al\Zn\Zt\Zs\Ar} \gamma_{\Zn}
      +
      P_\pix{0}^{\Al\Zm\Zk\Zr\Ar \hspace{.5pt} \Al\Zn\Zt\Zs\Ar} \gamma_{\Zn}
      =
      \gamma^{\Al\Zm} P_\pix{2}^{\Zk\Zr\Ar \hspace{.5pt} \Al\Zt\Zs\Ar}
      +
      \gamma^{\Al\Zm} P_\pix{0}^{\Zk\Zr\Ar \hspace{.5pt} \Al\Zt\Zs\Ar}
      \,,
     \\[0.3ex]
    \end{array}
    \;\, \text{\small{$\rank : \; {\textstyle \binom{d-1}{2} = \tfrac{(d{-}1)(d{-}2)}{2}} $}}
    \,,
    \nonumber
    \\[0.5ex]
   &
    \begin{array}{|@{\hspace{1pt}}llll}
      &X^{\Zs \hspace{.5pt} \Al\Zm\Zk\Ar}
      &\!\!=\!\!&
     \displaystyle
      \gamma_{\Zl} P_\pix{1}^{\Al\Zs\Zl\Ar \hspace{.5pt} \Al\Zm\Zk\Ar}
      \,=\,
      P_\pix{1}^{\Zs \hspace{.5pt} \Al\Zm}  \gamma^{\Zk\Ar}
      \,,
      \\[0.5ex]
      &\dc{X}^{\Al\Zm\Zk\Ar \hspace{.5pt} \Zt}
      &\!\!=\!\!&
     \displaystyle
       P_\pix{1}^{\Al\Zm\Zk\Ar \hspace{.5pt} \Al\Zn\Zt\Ar} \gamma_{\Zn}
      \,=\,
       \gamma^{\Al\Zm} P_\pix{1}^{\Zk\Ar \hspace{.5pt} \Zt}
      \,,  
     \\[0.3ex]
    \end{array}
     \;\; \text{\small{$\rank : \; d{\,-\,}1  $}}
    \,,
    \quad
    \begin{array}{|@{\hspace{1pt}}llll}
      &X^{ \Zo \hspace{.5pt} \Zk}
      &\!\!=\!\!&
     \displaystyle
      \gamma_{\Zl} P_\pix{0}^{\Zl \hspace{.5pt} \Zk}
      \,=\,
        \gamma^{\Zk}
      \,,
     \\[0.5ex] %
      &\dc{X}^{\Zm \hspace{.5pt} \Zo }
      &\!\!=\!\!&
     \displaystyle
       P_\pix{0}^{\Zm \hspace{.5pt} \Zn} \gamma_{\Zn}
      \,=\,
       \gamma^{\Zm}
      \,,
    \end{array}
     \;\; \text{\small{$\rank : \; 1 $}}
    \,.
   \label{min-rank_gf_p=3} 
   \end{align}
satisfying (\ref{GF_rank_conds_p=3}) with minimal possible ranks: $\rank X^{\Al\Zs\Zp\Ar \hspace{.5pt} \Al\Zm\Zn\Zk\Ar} {\,=\,} \nnn_1-\nnn_2+\nnn_3 {\,=\,} \mmm_1 $,\, $\rank X^{\Zs \hspace{.5pt} \Al\Zm\Zk\Ar} {\,=\,} \nnn_2-\nnn_3 {\,=\,} \mmm_2 $,\, $\rank X^{\Zo \hspace{.5pt} \Zk} {\,=\,} \nnn_3 {\,=\,} \mmm_3$.\footnote{The counting of independent spinor components in the $p=3$ case was first addressed just before Section~\ref{SSect:BV extension_p=3}.}
Rank properties of gauge-fixing operators follow from the independence and rank properties of projectors
 $
  \rank P_\pix{s} {}_{\ZZm{r}}^{\:\ZZn{r}}
  =
  \rank \dc{P}_\pix{s} {}_{\ZZm{r}}^{\:\ZZn{r}}
  = \binom{d}{s}-\binom{d}{s-1}
 $,\,
The rank condition (\ref{GF_rank_conds_p=3}) for minimal-rank type of gauges imply complementarity of the neighbouring gauge-fixing operators, and projector representation makes it straightforward to check. For example, block matrix
 $ \begin{pmatrix}
       X^{\Al\Zs\Zp\Ar \hspace{.5pt}  \Al\Zm\Zn\Zk\Ar}
       \;\big|\;
       \dc{X}^{\Al\Zs\Zp\Ar  \hspace{.5pt} \Zl}
    \end{pmatrix}
 $
obviously disentangles into three complementary blocks projecting onto $s=2$, $s=0$, and $s=1$ irreducible components, which span the whole space of spinor-forms of degree $2$. This follows from right representation for $X^{\Al\Zs\Zp\Ar \hspace{.5pt}  \Al\Zm\Zn\Zk\Ar}$ and the left one for $\dc{X}^{\Al\Zs\Zp\Ar  \hspace{.5pt} \Zl}$, (\ref{min-rank_gf_p=3}).

For such choice of $X^{\ZZm{r} \hspace{.5pt} \ZZn{r+1}}$ and $\dc{X}^{\ZZn{r+1} \hspace{.5pt} \ZZm{r}}$ the gauge fermion (\ref{gauge_fermion_delta_p=3}) in terms of irreducible fields takes the form
 \begin{align}
   \varPsi[\varPhi(\CPhi)]
  & \,=\, 
    \displaystyle
   - 
   \!\int \!d^dx\,
     \Bigl\{\,
       \HIDE{+} 
         \nfrac{d{-}4}{3}
         \big( \dcgCC^{\gix{1} \Zm\Zn}_{\rix{2}} \CPsi_{\Zm\Zn}
                                      + \dcCPsi^{\Zm\Zn} \gCC^{\gix{1}}_{\rix{2} \Zm\Zn} \big)
       \,+\, 
         d(d{-}1)(d{-}2) 
         \big( \dcgCC^{\gix{1} \Zo}_{\rix{2}} \CPsi_{\Zo}
                                      + \dcCPsi^{\Zo} \gCC^{\gix{1}}_{\rix{2} \Zo} \big)
   \nonumber
   \\[-0.5ex]
    & 
    \qquad\qquad\;  \displaystyle
     +\, 
       \nfrac{(d{-}2)}{2}  \sum_{k=1}^{2} 
        \big( {-} \HIDE{\msp^{r}}  \dcgCC^{\gix{k} \Zm}_{\rix{1}} \gCC^{\gix{k-1}}_{\rix{2} \Zm}
         + \dcgCC^{\gix{k} \Zm}_{\rix{2}} \gCC^{\gix{k-1}}_{\rix{1} \Zm}  \big)
    +\, 
     d\, \sum_{k=1}^{3} 
          \big( \dcgCC^{\gix{k} \Zo}_{\rix{0}} \gCC^{\gix{k-1}}_{\rix{1} \Zo}
            + \dcgCC^{\gix{k} \Zo}_{\rix{1}} \gCC^{\gix{k-1}}_{\rix{0} \Zo} \big)
    \Bigr\}
    \,,
 \label{gaugeFermion_min_rank_p=3} 
 \end{align}
and the ghost part of the gauge-fixed action (\ref{gf_action_p=3}) reads
 \begin{equation}
  \hspace{-8mm}
   \begin{array}{lllllll}
    S^{\scriptscriptstyle(3)}_{\delta\,\text{gh}} [\Phi(\CPhi)]
    &\!\!=\!\!&
     S^{\scriptscriptstyle(3) \text{D}}_{\delta\,\text{gh}} [\CPhi]
    +S^{\scriptscriptstyle(3) \text{OD}}_{\delta\,\text{gh}} [\CPhi]
    \,,
   \end{array}
 \label{gf_gh action_p=3_min-rank_in_components}
 \end{equation}
in which the diagonal part is
 \begin{align}
    S^{\scriptscriptstyle(3) \text{D}}_{\delta\,\text{gh}} [\CPhi]
    &\,=\, 
    \displaystyle
    \!\!\int \!d^dx
       \, \Bigl\{   
       i 
       \dcsc[2]{2} \nfrac13
      \big(
        \dcgCC^{\gix{1} \Zm\Zn}_\rix{2} \slashed\nabla_{\!\oix{2}}  \gCC^{\gix{0}}_{\rix{2} \Zm\Zn}
       -
        \dcgCC^{\gix{0} \Zm\Zn}_\rix{2} \dc{\slashed{\nabla}}_{\!\oix{2}}  \gCC^{\gix{1}}_{\rix{2} \Zm\Zn}
      \big)
      - i 
      \dcsc[2]{0} \, (d{-}1)(d{-}2) 
      \big( \dcgCC^{\gix{1} \Zo}_\rix{2} \slashed\nabla_{\!\oix{0}} \gCC^{\gix{0}}_{\rix{2} \Zo}
         -
          \dcgCC^{\gix{0} \Zo}_\rix{2} \dc{\slashed{\nabla}}_{\!\oix{0}} \gCC^{\gix{1}}_{\rix{2} \Zo}
      \big)
   \nonumber
   \\[0.0ex]
    & 
    \hspace{10ex} \displaystyle 
     -\, i 
      \dcsc[1]{1} \nfrac12 \big(
          \dcgCC^{\gix{1} \Zm}_\rix{1} \slashed\nabla_{\!\oix{1}} \gCC^{\gix{0}}_{\rix{1}_\Zm}
         -
          \dcgCC^{\gix{0} \Zm}_\rix{1} \dc{\slashed{\nabla}}_{\!\oix{1}}  \gCC^{\gix{1}}_{\rix{1}_\Zm}
        \big)
      +\, i 
      \dcsc[0]{0} \big(
          \dcgCC^{\gix{1} \Zo}_\rix{0} \slashed\nabla_{\!\oix{0}}  \gCC^{\gix{0}}_{\rix{0}_\Zo}
         -
          \dcgCC^{\gix{0} \Zo}_\rix{0} \dc{\slashed{\nabla}}_{\!\oix{0}} \gCC^{\gix{1}}_{\rix{0}_\Zo}
        \big)
     \Bigr\}
     \,,
 \label{gf_gh action_p=3_min-rank_diag}
 \end{align}
with operators $\slashed{\nabla}_{\!\oix{s}}$ defined in (\ref{nabla-s_nabla*-s}), and the off-diagonal part takes the form
 \begin{align}
  \hspace{-3ex}
  S^{\scriptscriptstyle(3) \text{OD}}_{\delta\,\text{gh}} [\CPhi]
  & \,=\, 
    \displaystyle
    \!\!\int \!d^dx
        \, \Bigl\{  
      \HIDE{+} i 
      \nfrac{d{-}4}{3}
      \big(
        \dcsc[2]{2} \dcgCC^{\gix{1} \Zr\Zm}_\rix{2} \nabla_{\!\Zr}  \gCC^{\gix{0}}_{\rix{2} \Zm}
       \!-
        \dcsc[2]{1} \dcgCC^{\gix{0} \Zm}_\rix{2} \dc{\nabla}^{\Zr}  \gCC^{\gix{1}}_{\rix{2} \Zm\Zr}
      \big)
     + i 
     (d{-}2)
      \big(
        \dcsc[2]{0} \dcgCC^{\gix{1} \Zo}_\rix{2} \nabla^{\Zr}  \gCC^{\gix{0}}_{\rix{2} \Zr}
       \!-
        \dcsc[2]{1} \dcgCC^{\gix{0} \Zr}_\rix{2} \dc{\nabla}_{\!\Zr}  \gCC^{\gix{1}}_{\rix{2} \Zo}
      \big)
     \!\!\!\! \!\!\!\!
   \nonumber
   \\[0.0ex]
    & 
    \hspace{10ex} \displaystyle  
      - i 
      \nfrac{d-2}{2}
      \big(
          \dcsc[1]{1}
          \dcgCC^{\gix{1} \Zr}_\rix{1} \nabla_{\!\Zr} \gCC^{\gix{0}}_{\rix{1} \Zo}
         \!-
          \dcsc[1]{0}
          \dcgCC^{\gix{0} \Zo}_\rix{1} \dc{\nabla}^{\Zr} \gCC^{\gix{1}}_{\rix{1} \Zr} \big)
     \Bigr\}
     \,.
 \label{gf_gh action_p=3_min-rank_off-diag}
 \end{align}
Off-diagonal part contains only sub- and super-diagonal terms, in which the degree of irreducible fields differ by one. It is the only nontrivial nondiagonal structure, which is constructed of antisymmetrized irreducible fields, multigammas and one one-index extra factor (here it is $\nabla$). If one of irreducible fields contains more then one extra indices compared to another field, then one of its indices inevitably contracts with index of gamma matrix and the term vanishes, whereas one extra index may nontrivially contract with the index of $\nabla$ and generate an off-diagonal contribution.

\newpar

The auxiliary part (\ref{gf_aux_action_p=3}) of the gauge-fixed action (\ref{gf_action_p=3}) is
 \begin{equation}
   \begin{array}{@{\hspace{0pt}}l@{\hspace{0pt}}lll}
    &S^{\scriptscriptstyle(3)}_{\delta\,\text{aux}} [\CPhi]
    &\!\!=\!\!&
    \displaystyle
    - 
    \!\!\int \!d^dx\,
      \Bigl\{
        \nfrac{d{-}4}{3} 
       (
        \dcCPsi^{\Zm\Zn} 
        \gCB^{\gix{1}}_{\rix{2} \Zm\Zn}
        \!+
        \dcsc[2]{3}
        \dcgCB^{\gix{1} \Zm\Zn}_{\rix{2}}
        \CPsi_{\Zm\Zn}
       )
      + 
        \nfrac{d{-}2}{2} 
       (
        \dcsc[1]{2} 
        \dcgCC^{\gix{2} \Zm}_{\rix{1}}
        \gCB^{\gix{1}}_{\rix{2} \Zm}
        +
        \dcgCB^{\gix{k} \Zm}_{\rix{2}}
        \gCC^{\gix{2}}_{\rix{1} \Zm}
       )
    \\
     &&& \displaystyle
      \hspace{24mm}
       + 
        d(d{-}1)(d{-}2) 
       (
        \dcCPsi^{\Zo} 
        \gCB^{\gix{k}}_{\rix{2} \Zo}
        +
        \dcsc[2]{3}  
        \dcgCB^{\gix{k} \Zo}_{\rix{2}}
        \CPsi_{\Zo} 
       )
     \Bigr\}
   \\[0.5em]
    &
    &&
   - 
    \displaystyle
   \!\! \int \!d^dx\,
    \sum_{k=1}^{2}
      \Bigl\{ 
       \HIDE{+} 
        \nfrac{d{-}2}{2} 
       (
        \dcgCC^{\gix{k-1} \Zm}_{\rix{2}}
        \gCB^{\gix{k}}_{\rix{1} \Zm}
        +
        \dcsc[1]{2}  
        \dcgCB^{\gix{k} \Zm}_{\rix{1}}
        \gCC^{\gix{k-1}}_{\rix{2} \Zm}
       )
      + 
        d 
       (
        \dcsc[0]{1} 
        \dcgCC^{\gix{k+1} \Zo}_{\rix{0}}
        \gCB^{\gix{k}}_{\rix{1} \Zo}
        +
        \dcgCB^{\gix{k} \Zo}_{\rix{1}}
        \gCC^{\gix{k+1}}_{\rix{0} \Zo}
       )
     \Bigr\} 
   \\[0.5em]
    &
    &&
   - 
    \displaystyle
   \!\!\int \!d^dx\,
    \sum_{k=1}^{3}
        d 
       \,(
        \dcgCC^{\gix{k-1} \Zo}_{\rix{1}}
        \gCB^{\gix{k}}_{\rix{0} \Zo}
        +
        \dcsc[0]{1}  
        \dcgCB^{\gix{k} \Zo}_{\rix{0}}
        \gCC^{\gix{k-1}}_{\rix{1} \Zo}
       )
    \,,
  \end{array}
 \label{gfAction_aux_min-rank_p=3}
 \end{equation}
where terms in the integral groups contributions from components of ghost and antighost from $l=1,2,3$ levels, sign factors
$\dcsc[r]{r+1} \defeq {\dcsc[r]{s}}/{\dcsc[r+1]{s}} = \msp^{r}$
originate from nontrivial factors in the decomposition of Dirac-conjugated fields.

Lagrange multiplier fields  $\gB^{\gix{k}}_{\ZZm{r}}$ and $\dcgB^{\gix{k}}_{\ZZm{r}}$  are present only in the auxiliary part (\ref{gf_aux_action_p=3}) of the gauge-fixed action, and being integrated out
generate delta-functional constraints (\ref{gf_conditions_aux_p=3}) on components of ghosts and original fields. For minimal-rank gauge-fixing (\ref{min-rank_gf_p=3}), in terms of irreducible fields (\ref{psi_decomp_p=3}), (\ref{C_decomp_p=3}) these constraints take the simple form
 \begin{eqnarray}
  \hspace{0mm}
   \begin{array}{|@{\hspace{1pt}}clll}
       &
       \CPsi_{\Zm\Zn}
       = 0
       ,\;
       &
       \gCC^{\gix{2}}_{\rix{1} \Zm}
       = 0 
       ,\;
       &
        \CPsi_{\Zo}
       = 0 
       \,,
     \\[0.4em]
       &
       \gCC^{\gix{k-1}}_{\rix{2} \Zm}
       = 0
       ,\;
       &
        \gCC^{\gix{k+1}}_{\rix{0} \Zo}
        = 0 
       \,,
     \\[0.4em]
       &
       \gCC^{\gix{k-1}}_{\rix{1} \Zo}
       =\: 0
       \,,
   \end{array}
   \quad
   \begin{array}{|@{\hspace{1pt}}clll}
       &
       \dcCPsi_{\Zm\Zn}
       = 0
       ,\;
       &
       \dcgCC^{\gix{2}}_{\rix{1} \Zm}
       = 0 
       ,\;
       &
        \dcCPsi_{\Zo}
       = 0 
       \,,
     \\[0.4em]
       &
       \dcgCC^{\gix{k-1}}_{\rix{2} \Zm}
       = 0
       ,\;
       &
        \dcgCC^{\gix{k+1}}_{\rix{0} \Zo}
        = 0 
       \,,
     \\[0.4em]
       &
       \dcgCC^{\gix{k-1}}_{\rix{1} \Zo}
       =\: 0
       \,,
   \end{array}
   \begin{array}{l@{\hspace{3pt}}c@{\hspace{5pt}}l@{\hspace{10pt}}|r@{\hspace{1pt}}}
       &
     \\[0.4em]
      &  {\small{\text{ $(\, 
                         2 \tgeq k \tgeq 1 \,)$}}} \,,
     \\[0.4em]
      &  {\small{\text{ $(\, 
                         3 \tgeq k \tgeq 1 \,)$}}} \,,
   \end{array}
 \label{gf_solutions_min_in_components_p=3} 
 \end{eqnarray}

Only \emph{original fields}, \emph{ghosts}, and \emph{antighosts} ($k{\,=\,}0,1$) though partially constrained on the reduction surface still have unfixed components:
 \begin{equation}
  \begin{array}{c}
   \text{Fixed}
    \\[0.2em]
   \hline \vphantom{\Big|}
     \CPsi_{\Zm\Zn} {\,=\,} \CPsi_{\Zo}
     {\,=\,} 0
     \,,\quad
     \dcCPsi_{\Zm\Zn} {\,=\,} \dcCPsi_{\Zo}
     {\,=\,} 0
     \,,
    \\[0.4em]
     \gCC^{\gix{k}}_{\rix{2} \Zm}
     {\,=\, 0}
     \,,\quad
     \dcgCC^{\gix{k}}_{\rix{2} \Zm}
     {\,=\, 0}
     \,,
    \\[0.4em]
     \gCC^{\gix{k}}_{\rix{1} \Zo}
     {\,=\, 0}
     \,,\quad
     \dcgCC^{\gix{k}}_{\rix{1} \Zo}
     {\,=\, 0}
     \,,
    \\[0.4em]
     {}
    \\
   \end{array}
   \!\!\!
  \begin{array}{cc}
   & \vphantom{\text{F}}
    \\[0.2em]
   \vphantom{\Big|}
     & \text{\small $\begin{array}{l}  (\,k=0\,;\; l=0\,;\; r=3\,)\,,\end{array}$}
    \\[0.4em]
     &  \text{\small $\begin{array}{l} (\,k = 1,\,0\, ;\;\; l=1\,;\;  r=2\,)\,,\end{array}$}
    \\[0.4em]
     &  \text{\small $\begin{array}{l} (\,k = 1,\,0\, ;\;\; l=2\,;\;  r=1\,)\,,\end{array}$}
    \\[0.4em]
     &  \text{\small $\begin{array}{l} (\,k = 1,\,0\, ;\;\; l=3\,;\;  r=0\,)\,,\end{array}$}
     \\
   \end{array}
   \begin{array}{c}
   \text{Free} \; (\varPhi_{\text{red}})
    \\[0.2em]
   \hline \vphantom{\Big|}
     \CPsi_{\Zm\Zn\Zl},\: \CPsi_{\Zm}
     ;\quad
     \dcCPsi_{\Zm\Zn\Zl},\: \dcCPsi_{\Zm}
     \,,
    \\[0.4em]
     \gCC^{\gix{k}}_{\rix{2} \Zm\Zn},\: \gCC^{\gix{k}}_{\rix{2} \Zo}
     \,,\,\;
     \dcgCC^{\gix{k}}_{\rix{2} \Zm\Zn},\: \dcgCC^{\gix{k}}_{\rix{2} \Zo}
     \,,
    \\[0.4em]
     \gCC^{\gix{k}}_{\rix{1} \Zm}
     \,,\,\;
     \dcgCC^{\gix{k}}_{\rix{1} \Zm}
     \,,
    \\[0.4em]
     \gCC^{\gix{k}}_{\rix{0} \Zo}
     \,,\,\;
     \dcgCC^{\gix{k}}_{\rix{0} \Zo}
     \,,
     \\
    \end{array}
 \label{gf_min-rank_reduction_surface_p=3} 
 \end{equation}
where in the table we also restored the level number $l$ for more transparent matching with triangular diagram in Fig.~\ref{fig:BVp=3_Triangular_diagram}, on which this subset of fields correspond to the two upper-right rows. The leading-degree irreducible components of these fields and components with degrees of the same parity remain unfixed by the constraints (\ref{gf_conditions_aux_p=3}). For the field of degree $r$ these are the components of degrees $r{\,-\,}2n$,\, $n{\,\in\,}\mathbb{N}_0$. At the same time, complementary components with degrees $r{\,-\,}1{\,-\,}2n$ are fixed to zero.
All \emph{extraghosts} ($k\tgeq2$) are completely fixed to zero:
 \begin{equation}
  \begin{array}{c}
   \text{Fixed}
    \\[0.2em]
   \hline \vphantom{\Big|}
     \gCC^{\gix{2}}_{\rix{1} \Zm}
       {\,=\,}
     \gCC^{\gix{2}}_{\rix{1} \Zo}
     {\,=\, 0}
     \,;
     \quad
     \dcgCC^{\gix{2}}_{\rix{1} \Zm}
       {\,=\,}
     \dcgCC^{\gix{2}}_{\rix{1} \Zo}
     {\,=\, 0}
     \,,\;
     \\[0.4em]
     \gCC^{\gix{k}}_{\rix{0} \Zo}
        =  0
     \,;
     \quad
     \dcgCC^{\gix{k}}_{\rix{0} \Zo}
        = 0
        \,,
     \\
  \end{array}
   \!\!\!\!\!\!\!
  \begin{array}{cc}
   & \vphantom{\text{F}}
    \\[0.2em]
   \vphantom{\Big|}
     & \text{\small $\begin{array}{l} (\,k=2\,;\;\; l=2\,;\;\;  r = 1\,)\,,\end{array}$}
    \\[0.4em]
     & \text{\small $\begin{array}{l} (\,k=3,\,2\,;\;\; l=3\,;\;\;  r = 0\,)\,.\end{array}$}
     \\
   \end{array}
 \label{gf_full-rank_reduction_surface_p=3_extraghosts}
 \end{equation}

    \subsubsection{Full-rank gauge and reduction}
     \label{}
      \hspace{\parindent}
Gauge-fixing operators of the gauge fermion (\ref{gauge_fermion_delta_p=3}) can alternatively be chosen in another covariant form
 \begin{equation} 
  \begin{array}{lllllll}
    & 
    X^{\ZZm{r} \hspace{.5pt} \ZZn{r+1}}
    \;=\;
    \gamma_{\Zl}\, \GG^{[\ZZm{r} \Zl]  \hspace{.5pt}  \ZZn{r+1}}
    \,,
    \quad
    & \dc{X}^{\ZZn{r+1} \hspace{.5pt} \ZZm{r}}
    \;=\; \GG^{\ZZn{r+1}  \hspace{.5pt} [\Zl \ZZm{r}]} \gamma_{\Zl}
    \,,
  \end{array}
 \label{full-rank_gf_p=3} 
 \end{equation}
for which they appear of maximal possible rank\HIDE{ (full rank)} $\binom{d}{r}$.\footnote{
The rank of gauge-fixing operators $X^{\ZZm{r} \hspace{.5pt} \ZZn{r+1}}$, $\dc{X}^{\ZZn{r+1} \hspace{.5pt} \ZZm{r}}$ equals $\binom{d}{r}$ when $r\tleq [(d{\,-\,}1)/2]$. This is true due to initial assumption $p = 3 \tleq[(d-1)/2]$ for which the theory exists.
}
In (\ref{full-rank_gf_p=3}), matrix operator $  \GG^{\ZZm{r+1}\ZZn{r+1}} $, (\ref{def_index_operators}), antisymmetrically raises indices by inverse metric tensors.
Gauge fermion (\ref{gauge_fermion_delta_p=3}) then acquires the simple\HIDE{ covariant} form
 \begin{align}
   \varPsi_{\delta}[\varPhi]
   &= \; 
   \!\!\int \!d^dx\,
    \Bigl\{ 
    \dcgC^{\gix{1} \Zm\Zn} \gamma^{\Zl} \psi_{\Zm\Zn\Zl}
    +
    \dcpsi^{\Zl\Zm\Zn} \gamma_{\Zl}  \gC^{\gix{1}}_{\Zm\Zn}
    \nonumber
   \\[-0.2em]
   & \qquad\qquad 
    + 
    \sum_{k=1}^{2} \!
    \big(
    \dcgC^{\gix{k} \Zm} \gamma^{\Zl} \gC^{\gix{k-1}}_{\Zm\Zl}
    \!+
    \dcgC^{\gix{k-1} \Zl\Zm} \gamma_{\Zl} \gC^{\gix{k}}_{\Zm}
    \big)
    + 
    \sum_{k=1}^{3} \!
    \big(
    \dcgC^{\gix{k} \Zo} \gamma^{\Zl} \gC^{\gix{k-1}}_{\Zl}
    \!+
    \dcgC^{\gix{k-1} \Zl} \gamma_{\Zl} \gC^{\gix{k}}_{\Zo}
    \big)
    \Bigr\}
     \,.
 \label{gauge_fermion_delta_p=3_full_rank} 
 \end{align}

For the full-rank gauge fixing (\ref{gf_action_p=3}) in the gauge-fixed action (\ref{gf_action_p=p}) besides the original action term (\ref{S0_p=3}) does not change being independent of gauge-fixing operators. The ghost part (\ref{gf_gh action_p=3}) now takes the form
 \begin{eqnarray}
  &&
  \hspace{-8mm}
   \begin{array}{lllllll}
    S^{\scriptscriptstyle(3)}_{\delta\,\text{gh}} [\varPhi]
    &\!\!=\!\!&
    \displaystyle
    \!\!\int \!d^dx\,
       \, \Bigl\{ i 
      \big( \dcgC^{\gix{1} \Zm\Zn} \gamma^{\Zk}
           D_{\Al\Zm} \gC^{\gix{0}}_{\Zn\Zk\Ar}
         -
          \dcgC^{\gix{0} \Al\Zm\Zn} \dc{D}^{\Zk\Ar}
           \gamma_{\Zm} \gC^{\gix{1}}_{\Zn\Zk} \big)
    \\[.6ex]
     && \hspace{10mm} \displaystyle
       +\,
      i 
      \big( \dcgC^{\gix{1} \Zm} \gamma^{\Zn}
           D_{\Al\Zm} \gC^{\gix{0}}_{\Zn\Ar}
         -
          \dcgC^{\gix{0} \Al\Zm} \dc{D}^{\Zn\Ar}
           \gamma_{\Zm} \gC^{\gix{1}}_{\Zn} \big)
       +\,
      i 
      \big( \dcgC^{\gix{1} \Zo} \gamma^{ \Zm{}}
           D_{\Zm} \gC^{\gix{0}}_{\Zo}
         -
          \dcgC^{\gix{0} \Zo} \dc{D}_{\Zm}
           \gamma^{\Zm} \gC^{\gix{1}}_{\Zo} \big)
     \Bigr\}
     \,,
   \end{array}
 \label{gf_gh action_p=3_full-rank}
 \end{eqnarray}
whose representation in components may be separated into diagonal and off-diagonal contributions
 \begin{equation}
  \hspace{-8mm}
   \begin{array}{lllllll}
    S^{\scriptscriptstyle(3)}_{\delta\,\text{gh}} [\Phi(\CPhi)]
    &\!\!=\!\!&
     S^{\scriptscriptstyle(3) \text{D}}_{\delta\,\text{gh}} [\CPhi]
    +S^{\scriptscriptstyle(3) \text{OD}}_{\delta\,\text{gh}} [\CPhi]
    \,,
   \end{array}
 \label{gf_gh action_p=3_full-rank_in_components}
 \end{equation}
 \begin{equation}
  \hspace{-8mm}
   \begin{array}{lllllll}
    S^{\scriptscriptstyle(3) \text{D}}_{\delta\,\text{gh}} [\CPhi]
    &\!\!=\!\!&
    \displaystyle
    \!\!\int \!d^dx
       \, \Bigl\{   
       i \dcsc[2]{2} \nfrac13
      \big(
        \dcgCC^{\gix{1} \Zm\Zn}_\rix{2} \slashed\nabla_{\!\oix{2}}  \gCC^{\gix{0}}_{\rix{2} \Zm\Zn}
       -
        \dcgCC^{\gix{0} \Zm\Zn}_\rix{2} \dc{\slashed{\nabla}}_{\!\oix{2}}  \gCC^{\gix{1}}_{\rix{2} \Zm\Zn}
      \big)
     - 
      i \dcsc[2]{1} \, \nfrac{d-3}{3}
      \big(
        \dcgCC^{\gix{1} \Zm}_\rix{2} \slashed\nabla_{\!\oix{1}}^*  \gCC^{\gix{0}}_{\rix{2} \Zm}
       -
        \dcgCC^{\gix{0} \Zm}_\rix{2} \dc{\slashed{\nabla}}_{\!\oix{1}}^*  \gCC^{\gix{1}}_{\rix{2} \Zm}
      \big)
   \\[0.7ex]
    &&\hspace{12mm} 
      -\, 
      i \dcsc[2]{0} \, (d{-}1)(d{-}2)  
      \big( \dcgCC^{\gix{1} \Zo}_\rix{2} \slashed\nabla_{\!\oix{0}} \gCC^{\gix{0}}_{\rix{2} \Zo}
         -
          \dcgCC^{\gix{0} \Zo}_\rix{2} \dc{\slashed{\nabla}}_{\!\oix{0}} \gCC^{\gix{1}}_{\rix{2} \Zo}
      \big)
   \\[0.7ex]
    && \hspace{12mm} \displaystyle 
     -\,
      i \dcsc[1]{1} \nfrac12 \big(
          \dcgCC^{\gix{1} \Zm}_\rix{1} \slashed\nabla_{\!\oix{1}} \gCC^{\gix{0}}_{\rix{1} \Zm}
         -
          \dcgCC^{\gix{0} \Zm}_\rix{1} \dc{\slashed{\nabla}}_{\!\oix{1}}  \gCC^{\gix{1}}_{\rix{1} \Zm} \big)
     - 
      i \dcsc[1]{0} (d{-}1) \big(
          \dcgCC^{\gix{1} \Zo}_\rix{1} \slashed\nabla_{\!\oix{0}}^* \gCC^{\gix{0}}_{\rix{1} \Zo}
         -
          \dcgCC^{\gix{0} \Zo}_\rix{1} \dc{\slashed{\nabla}}_{\!\oix{0}}^* \gCC^{\gix{1}}_{\rix{1} \Zo} \big)
   \\[0.7ex]
    && \hspace{12mm} \displaystyle 
      +\,
      i \dcsc[0]{0} \big(
          \dcgCC^{\gix{1} \Zo}_\rix{0} \slashed\nabla_{\!\oix{0}}  \gCC^{\gix{0}}_{\rix{0} \Zo}
         -
          \dcgCC^{\gix{0} \Zo}_\rix{0} \dc{\slashed{\nabla}}_{\!\oix{0}} \gCC^{\gix{1}}_{\rix{0} \Zo} \big)
     \Bigr\}
     \,,
   \\[.3em]
   \end{array}
 \label{gf_gh action_p=3_full-rank_diag}
 \end{equation}
 \begin{equation}
  \hspace{-8mm}
   \begin{array}{lllllll}
    S^{\scriptscriptstyle(3) \text{OD}}_{\delta\,\text{gh}} [\CPhi]
    &\!\!=\!\!&
    \displaystyle
    \!\!\int \!d^dx
    \textstyle
       \, \Bigl\{  
    - 
     i \nfrac23
      \big(
        \dcsc[2]{1} \dcgCC^{\gix{1} \Zm}_\rix{2} \nabla^{\Zr}  \gCC^{\gix{0}}_{\rix{2} \Zm\Zr}
       \!-
        \dcsc[2]{2} \dcgCC^{\gix{0} \Zr\Zm}_\rix{2} \dc{\nabla}_{\!\Zr}  \gCC^{\gix{1}}_{\rix{2} \Zm}
      \big)
     + 
      i \nfrac{d{-}4}{3}
      \big(
        \dcsc[2]{2} \dcgCC^{\gix{1} \Zr\Zm}_\rix{2} \nabla_{\!\Zr}  \gCC^{\gix{0}}_{\rix{2} \Zm}
       \!-
        \dcsc[2]{1} \dcgCC^{\gix{0} \Zm}_\rix{2} \dc{\nabla}^{\Zr}  \gCC^{\gix{1}}_{\rix{2} \Zm\Zr}
      \big)
   \\[0.7ex]
    && \hspace{2mm} 
     +\, 
     i \HIDE{1} (d{-}2)
      \big(
        \dcsc[2]{0} \dcgCC^{\gix{1} \Zo}_\rix{2} \nabla^{\Zr}  \gCC^{\gix{0}}_{\rix{2} \Zr}
       \!-
        \dcsc[2]{1} \dcgCC^{\gix{0} \Zr}_\rix{2} \dc{\nabla}_{\!\Zr}  \gCC^{\gix{1}}_{\rix{2} \Zo}
      \big)
     - 
      i \nfrac{(d{-}2)(d{-}3)}{3}
      \big(
        \dcsc[2]{1} \dcgCC^{\gix{1} \Zr}_\rix{2} \nabla_{\!\Zr}  \gCC^{\gix{0}}_{\rix{2} \Zo}
       \!-
        \dcsc[2]{0} \dcgCC^{\gix{0} \Zo}_\rix{2} \dc{\nabla}^{\Zr}  \gCC^{\gix{1}}_{\rix{2} \Zr}
      \big) \!\!\!\!
   \\[0.7ex]
    && \hspace{2mm} \displaystyle  
     -\, 
      i  \nfrac{d-2}{2} \big(
          \dcsc[1]{1}
          \dcgCC^{\gix{1} \Zr}_\rix{1} \nabla_{\!\Zr} \gCC^{\gix{0}}_{\rix{1} \Zo}
         \!-
          \dcsc[1]{0}
          \dcgCC^{\gix{0} \Zo}_\rix{1} \dc{\nabla}^\Zr \gCC^{\gix{1}}_{\rix{1} \Zr} \big)
     + 
      i  \big(
          \dcsc[1]{0}
          \dcgCC^{\gix{1} \Zo}_\rix{1} \nabla^\Zr \gCC^{\gix{0}}_{\rix{1} \Zr}
         \!-
          \dcsc[1]{1}
          \dcgCC^{\gix{0} \Zr}_\rix{1} \dc{\nabla}_{\!\Zr} \gCC^{\gix{1}}_{\rix{1} \Zo} \big)
     \Bigr\}
     \,.
   \\[.3em]
   \end{array}
 \label{gf_gh action_p=3_full-rank_off-diag}
 \end{equation}
As in (\ref{gf_gh action_p=3_min-rank_off-diag}), the off-diagonal part contains only sub- and super-diagonal terms. However, for the full-rank gauge (\ref{full-rank_gf_p=3}) all types of nontrivial nondiagonal terms from ghosts of degrees $2 \tgeq r \tgeq 0$ are present in the action.

\newpar

We will not explicate the auxiliary part (\ref{gf_aux_action_p=3}) of the gauge-fixed action (\ref{gf_action_p=3}) for the full-rank gauge here\footnote{
 We summarized the gauge-fixed action for $p=3$ case in Appendix~\ref{SSect:actions_p=3}, see (\ref{gfAction_aux_full-rank_p=3}).}
and rather directly solve constraints (\ref{gf_conditions_aux_p=3}) in terms of irreducible components.
Since (\ref{gf_aux_action_p=3}) structurally is the sum of constraints at these Lagrange multipliers, it in any case vanishes on the reduction surface.

Constraints (\ref{gf_conditions_aux_p=3}), imposed by the Lagrange multipliers on ghosts and original fields in for gauge fixing (\ref{full-rank_gf_p=3}) take the form
 \begin{equation}
  \hspace{0mm}
   \begin{array}{|@{\hspace{10pt}}lc@{\hspace{5pt}}l@{\hspace{10pt}}c@{\hspace{10pt}}}
       \dcpsi_{\Zk\Zm\Zn} \gamma^{\Zk}
       {\,+\,}
       \dcgC^{\gix{2}}_{\Al\Zm} \gamma_{\Zn\Ar}
       = 0
       \,,
     \\[0.4em]
       \dcgC^{\gix{k-1}}_{\Zk\Zm} \gamma^{\Zk}
       {\,+\,}
       \dcgC^{\gix{k+1}}_{\Zo} \gamma_{\Zm}
       = 0
       \,,
     \\[0.4em]
       \dcgC^{\gix{k-1}}_{\Zk} \gamma^{\Zk}
       = 0
       \,,
   \end{array}
   \qquad
   \begin{array}{|@{\hspace{8pt}}lc@{\hspace{5pt}}l@{\hspace{10pt}}|r@{\hspace{1pt}}}
       \gamma^{\Zk} \psi_{\Zm\Zn\Zk}
       {\,+\,}
       \gamma_{\Al\Zm} \gC^{\gix{2}}_{\Zn\Ar}
       \,=\: 0
       \,,
     \\[0.4em]
       \gamma^{\Zk} \gC^{\gix{k-1}}_{\Zm\Zk}
       {\,+\,}
       \gamma_{\Zm} \gC^{\gix{k+1}}_{\Zo}
       \,=\: 0
       \,,
     \\[0.4em]
       \gamma^{\Zk} \gC^{\gix{k-1}}_{\Zk}
       \,=\: 0
       \,,
   \end{array}
   \qquad
   \begin{array}{l@{\hspace{1pt}}c@{\hspace{5pt}}l@{\hspace{10pt}}|r@{\hspace{1pt}}}
       &
     \\[0.4em]
       &  {\small{\text{  $(\, 
                           2 \tgeq k \tgeq 1\,)$}}} \,,
     \\[0.4em]
      &  {\small{\text{ $(\, 
                         3 \tgeq k\tgeq 1 \,)$}}} \,,
   \end{array}
 \label{gf_equations_full_p=3} 
 \end{equation}
which being represented in irreducible components (\ref{psi_decomp_p=3}), (\ref{C_decomp_p=3}) gives
 \begin{eqnarray}
  \hspace{0mm}
   \begin{array}{|@{\hspace{10pt}}l}
      - 
       \nfrac13 (d{-}4) \CPsi_{\Zm\Zn}
      {\,-\,} 
       \nfrac23 (d{-}3) \gamma_{\Al\Zm} \CPsi_{\Zn\Ar}
      {\,-\,} 
       (d{-}2) \gamma_{\Zm\Zn} \CPsi_{\Zo}
      {\:+\:}
       \gamma_{\Al\Zm} \gCC^{\gix{2}}_{\rix{1} \Zn\Ar} 
      {\,+\,}
       \gamma_{\Zm\Zn} \gCC^{\gix{2}}_{\rix{1} \Zo}
       \,=\: 0
       \,,
     \\[0.4em]
     \HIDE{+} 
       \nfrac{1}{2} (d{-}2) \gCC^{\gix{k-1}}_{\rix{2} \Zm}
      {\,+\,} 
       (d{-}1) \gamma_\Zm \gCC^{\gix{k-1}}_{\rix{2} \Zo}
      {\,+\,}
       \gamma_{\Zm} \gCC^{\gix{k+1}}_{\rix{0} \Zo} 
       \,=\: 0
       \,,
     \\[0.4em]
      - 
       d\, \gCC^{\gix{k-1}}_{\rix{1} \Zo}
       =\: 0
       \,,
   \end{array}
   \quad
   \begin{array}{l@{\hspace{3pt}}c@{\hspace{5pt}}l@{\hspace{10pt}}|r@{\hspace{1pt}}}
       &
     \\[0.4em]
       &  {\small{\text{  $(\, 
                           2 \tgeq k \tgeq 1\,)$}}} \,,
     \\[0.4em]
      &  {\small{\text{ $(\, 
                         3 \tgeq k \tgeq 1 \,)$}}} \,.
   \end{array}
 \label{gf_equations_full_in_components_p=3} 
 \end{eqnarray}
One can note that these constraints fit the gauge-fixing chains, depicted in Fig.~\ref{fig:BVp=3_Triangular_diagram} by left-downward lines. The first equation, the second one for $k{\,=\,}2$ and the third one for $k{\,=\,}3$ originate from the longest, left $l{\,-\,}k{\,=\,}0$ chain which starts from the top. The second $l-k{\,=\,}1$ chain generate the second equation for $k{\,=\,}1$ and the third one for $k{\,=\,}2$. The shortest, right chain gives $\gCC^{\gix{0}}_{\rix{1} \Zo}  {\,=\,} 0$ for $k{\,=\,}1$ from the third equation, forcing the subleading degree-$0$ component of the minimal ghost $\gC^{\gix{0}}_{\Zm}$ to vanish.
For Dirac conjugated fields one gets analogous equations modulo additional sign factors for various components.
  Equating to zero combinations of irreducible fields at independent matrices from antisymmetric gamma-matrix basis 
in the first equation and expressing irreducible fields from bottom to top one finally gets
 \begin{eqnarray}
  \hspace{0mm}
   \begin{array}{|c@{\hspace{8pt}}lll}
       &
       \CPsi_{\Zm\Zn}
       = 0
       ,\;
       &
       \gCC^{\gix{2}}_{\rix{1} \Zm}
       =
       \HIDE{+} 
       \nfrac23 (d{-}3) \CPsi_{\Zm}
       ,\;
       &
        \CPsi_{\Zo}
       =
       \HIDE{+} 
        \nfrac1{(d{-}2)}
       \gCC^{\gix{2}}_{\rix{1} \Zo}
       = 0
       \,,
     \\[0.4em]
       &
       \gCC^{\gix{k-1}}_{\rix{2} \Zm}
       = 0
       ,\;
       &
        \gCC^{\gix{k+1}}_{\rix{0} \Zo}
        =
       - 
       (d{-}1) \gCC^{\gix{k-1}}_{\rix{2} \Zo}
       \,,
     \\[0.4em]
       &
       \gCC^{\gix{k-1}}_{\rix{1} \Zo}
       =\: 0
       \,,
   \end{array}
   \quad
   \begin{array}{l@{\hspace{3pt}}c@{\hspace{5pt}}l@{\hspace{10pt}}|r@{\hspace{1pt}}}
       &
     \\[0.4em]
      &  {\small{\text{ $(\, 
                         2 \tgeq k \tgeq 1 \,)$}}} \,,
     \\[0.4em]
      &  {\small{\text{ $(\, 
                         3 \tgeq k \tgeq 1 \,)$}}} \,,
   \end{array}
 \label{gf_solutions_full_in_components_p=3} 
 \end{eqnarray}
together with analogous relation for Dirac conjugated fields.

Only original fields, ghosts and antighosts ($k{\,=\,}0,1$)\HIDE{ though partially fixed,} still have unfixed components:
 \begin{equation}
  \begin{array}{c}
   \text{Fixed}
    \\[0.2em]
   \hline \vphantom{\Big|}
     \CPsi_{\Zm\Zn} {\,=\,} \CPsi_{\Zo}
     {\,=\,} 0
     \,,\quad
     \dcCPsi_{\Zm\Zn} {\,=\,} \dcCPsi_{\Zo}
     {\,=\,} 0
     \,,
    \\[0.4em]
     \gCC^{\gix{k}}_{\rix{2} \Zm}
     {\,=\, 0}
     \,,\quad
     \dcgCC^{\gix{k}}_{\rix{2} \Zm}
     {\,=\, 0}
     \,,
    \\[0.4em]
     \gCC^{\gix{k}}_{\rix{1} \Zo}
     {\,=\, 0}
     \,,\quad
     \dcgCC^{\gix{k}}_{\rix{1} \Zo}
     {\,=\, 0}
     \,,
    \\[0.4em]
     {}
    \\
   \end{array}
   \!\!\!
  \begin{array}{cc}
   & \vphantom{\text{F}}
    \\[0.2em]
   \vphantom{\Big|}
     & \text{\small $\begin{array}{l}  (\,k=0\,;\; l=0\,;\; r=3\,)\,,\end{array}$}
    \\[0.4em]
     &  \text{\small $\begin{array}{l} (\,k = 0,\,1\, ;\;\; l=1\,;\;  r=2\,)\,,\end{array}$}
    \\[0.4em]
     &  \text{\small $\begin{array}{l} (\,k = 0,\,1\, ;\;\; l=2\,;\;  r=1\,)\,,\end{array}$}
    \\[0.4em]
     &  \text{\small $\begin{array}{l} (\,k = 0,\,1\, ;\;\; l=3\,;\;  r=0\,)\,,\end{array}$}
     \\
   \end{array}
   \begin{array}{c}
   \text{Free} \; (\breve{\CPhi})
    \\[0.2em]
   \hline \vphantom{\Big|}
     \CPsi_{\Zm\Zn\Zl},\: \CPsi_{\Zm}
     ;\quad
     \dcCPsi_{\Zm\Zn\Zl},\: \dcCPsi_{\Zm}
     \,,
    \\[0.4em]
     \gCC^{\gix{k}}_{\rix{2} \Zm\Zn},\: \gCC^{\gix{k}}_{\rix{2} \Zo}
     \,,\,\;
     \dcgCC^{\gix{k}}_{\rix{2} \Zm\Zn},\: \dcgCC^{\gix{k}}_{\rix{2} \Zo}
     \,,
    \\[0.4em]
     \gCC^{\gix{k}}_{\rix{1} \Zm}
     \,,\,\;
     \dcgCC^{\gix{k}}_{\rix{1} \Zm}
     \,,
    \\[0.4em]
     \gCC^{\gix{k}}_{\rix{0} \Zo}
     \,,\,\;
     \dcgCC^{\gix{k}}_{\rix{0} \Zo}
     \,,
     \\
    \end{array}
 \label{gf_full-rank_reduction_surface_p=3} 
 \end{equation}
where in the table we also restored the level number $l$ for more direct matching with triangular diagram in Fig.~\ref{fig:BVp=3_Triangular_diagram}.
All extraghosts ($k\tgeq2$) are fixed by constraints (\ref{gf_equations_full_p=3}):
 \begin{equation}
  \begin{array}{r@{\hspace{5pt}}l}
   \text{Fixed}
    \\[0.2em]
   \hline \vphantom{\Big|}
     \gCC^{\gix{2}}_{\rix{1} \Zm}
       {\,=\,}
     \HIDE{+} 
       \nfrac23 (d{-}3) \CPsi_{\Zm}
     \,,\;
     \gCC^{\gix{2}}_{\rix{1} \Zo}
     {\,=\, 0}
     \,;
   &  \dcgCC^{\gix{2}}_{\rix{1} \Zm}
       {\,=\,}
      - 
       \nfrac23 (d{-}3) \dcCPsi_{\Zm}
     \,,\;
     \dcgCC^{\gix{2}}_{\rix{1} \Zo}
     {\,=\, 0}
     \,,
   \\[0.4em]
     \gCC^{\gix{k}}_{\rix{0} \Zo}
        =
      - 
      (d{-}1) \gCC^{\gix{k-2}}_{\rix{2} \Zo}
     \,;
   &  \dcgCC^{\gix{k}}_{\rix{0} \Zo}
        =
       \HIDE{+} 
       (d{-}1) \dcgCC^{\gix{k-2}}_{\rix{2} \Zo}
        \,
     \\
  \end{array}
   \!\!\!\!\!\!\!
  \begin{array}{cc}
   & \vphantom{\text{F}}
    \\[0.2em]
   \vphantom{\Big|}
     & \text{\small $\begin{array}{l} (\,k=2\,;\; l=2\,;\;  r = 1\,)\,,\end{array}$}
    \\[0.4em]
     & \text{\small $\begin{array}{l} (\,k=2,\,3\,;\; l=3\,;\;  r = 0\,)\,.\end{array}$}
     \\
   \end{array}
 \label{gf_full-rank_reduction_surface_p=3_extraghosts}
 \end{equation}

For original fields, minimal ghosts and antighosts, their leading-degree irreducible component (denote it by $r$) and components with degrees of the same parity: $r{\,-\,}2n$,\, $n{\,\in\,}\mathbb{N}_0$, remain unfixed by the constraints (\ref{gf_equations_full_p=3}). At the same time, the complementary components with degrees $r{\,-\,}1{\,-\,}2n$ (\emph{subleading-parity} components) are fixed to zero.
Moreover, for extraghosts ($k \tgeq 2$), who are all fixed by (\ref{gf_equations_full_p=3}), leading-degree components are expressed in terms of unfixed (free) fields, whereas the subleading component (for degree-$1$ extraghost) vanish.
In Section \ref{Sect:p=p} for general $p$ case we show that these facts are not accidental for the considered type of gauges.


  \subsection{Reduced gauge-fixed action and effective action}
   \label{}
    \hspace{\parindent}
After integrating out Lagrange multiplier fields and resolving delta-functions, only the leading-parity\footnote{For a field $\fA_{\ZZm{r}}$ of degree $r$ we refer its irreducible components $\fCA_{\rix{r} \ZZm{r-2n}}$ of degrees $r-2n$, $n\in\mathbb{N}_0$ as \emph{leading-parity} components, whereas that of degrees $r{-}2n{-}1$ as \emph{leading-parity} ones.} components of the original fields, minimal ghosts and antighosts
 \begin{eqnarray}
  \breve{\CPhi}\,:\quad\!
  \begin{array}{|llll}
   & \CPsi_{\Zm\Zn\Zl},\, \CPsi_{\Zm}\,;
   &  \gCC^{\gix{0}}_{\rix{2} \Zm\Zn},\, \gCC^{\gix{0}}_{\rix{2} \Zo},\,
      \gCC^{\gix{0}}_{\rix{1} \Zm},\, \gCC^{\gix{0}}_{\rix{0} \Zo}\,;
   &  \gCC^{\gix{1}}_{\rix{2} \Zm\Zn},\, \gCC^{\gix{1}}_{\rix{2} \Zo},\,
      \gCC^{\gix{1}}_{\rix{1} \Zm},\, \gCC^{\gix{1}}_{\rix{0} \Zo}\,;
      \\[0.4em]
   & \dcCPsi_{\Zm\Zn\Zl},\, \dcCPsi_{\Zm}\,;
   &  \dcgCC^{\gix{0}}_{\rix{2} \Zm\Zn},\, \dcgCC^{\gix{0}}_{\rix{2} \Zo},\,
      \dcgCC^{\gix{0}}_{\rix{1} \Zm},\, \dcgCC^{\gix{0}}_{\rix{0} \Zo}\,;
   &  \dcgCC^{\gix{1}}_{\rix{2} \Zm\Zn},\, \dcgCC^{\gix{1}}_{\rix{2} \Zo},\,
      \dcgCC^{\gix{1}}_{\rix{1} \Zm},\, \dcgCC^{\gix{1}}_{\rix{0} \Zo}\,,
  \end{array}
 \label{gf_min-rank_survive}
 \end{eqnarray}
survive as independent fields on the reduction surface.
As seen from (\ref{gf_min-rank_reduction_surface_p=3}) and (\ref{gf_solutions_full_in_components_p=3}) this is true for \emph{both} considered \emph{gauges}.

The reduced gauge-fixed action (\ref{gf_action_p=3}), which remains in the exponent of the generating functional considerably simplifies.
The auxiliary part of the gauge-fixed action (\ref{gf_aux_action_p=3}) vanishes, being proportional to constraints (\ref{gf_equations_full_p=3}).
The original action part (\ref{So_in_components_p=3}) is the functional of the original fields and the ghost part (\ref{gf_gh action_p=3}) is the functional of only minimal ghosts and antighosts. The resolution of constraints (\ref{gf_conditions_aux_p=3}) for minimal-rank gauge, (\ref{gf_min-rank_reduction_surface_p=3}), is identical to resolution (\ref{gf_full-rank_reduction_surface_p=3}) for the full-rank gauge in these sectors, consisting in zeroing terms with subleading-parity components. Moreover, though ghost actions (\ref{gf_gh action_p=3_min-rank_in_components}) and (\ref{gf_gh action_p=3_full-rank_in_components}) depend on gauge-fixing matrices and initially differ, restricting onto the reduction surface make them identical.
Also note, that in the original and ghost actions  (\ref{So_in_components_p=3}),(\ref{gf_gh action_p=3_min-rank_in_components}),(\ref{gf_gh action_p=3_full-rank_in_components}), off-diagonal contributions completely vanish, because their terms are from sub- and super-diagonals and couple an irreducible field of neighboring degrees. One of these is a subleading-parity degree and the field vanishes in course of reduction.

Finally, the reduced gauge-fixed BV action
 $
   \breve{S}^{\scriptscriptstyle(3)}_{\delta}[\breve{\CPhi}]
   \defeq  {S}^{\scriptscriptstyle(3)}_{\delta}[\varPhi(\breve{\CPhi})]
 $
for $p=3$ spin-form theory (\ref{S0_p=3}) in both gauges takes the block-diagonal\footnote{The reduced gauge-fixed action is the sum of terms, bilinear in irreducible fields of the same ranks. So its Hessian is the block-diagonal quadratic form.} form
 \begin{equation}
  \begin{array}{lllllll}
    \breve{S}^{\scriptscriptstyle(3)}_{\delta}[\breve{\CPhi}]
    &\!\!=\!\!\!&   \displaystyle
     \int \!d^dx\,
     \Bigl\{\,
     \HIDE{+} i 
     \sqrt{|\GG|}
     \, 6
     \,\dcCPsi^{\Zm\Zn\Zk}
     \,\slashed{\nabla}_{\!\oix{3}}
     \CPsi_{\Zm\Zn\Zk}
    + i 
     \sqrt{|\GG|}
    \, \nfrac{(d{-}3)!}{(d{-}7)!}
     \,\dcCPsi^{\Zm}
     \,\slashed{\nabla}_{\!\oix{1}}
     \CPsi_{\Zm}
    \\[0.7ex]
      && \hspace{12mm} \displaystyle
      +
       i \dcsc[2]{2} \nfrac13
      \big(
        \dcgCC^{\gix{1} \Zm\Zn}_\rix{2} \slashed\nabla_{\!\oix{2}}  \gCC^{\gix{0}}_{\rix{2} \Zm\Zn}
       -
        \dcgCC^{\gix{0} \Zm\Zn}_\rix{2} \dc{\slashed{\nabla}}_{\!\oix{2}}  \gCC^{\gix{1}}_{\rix{2} \Zm\Zn}
      \big)
      - 
      i \dcsc[2]{0} \, \nfrac{(d{-}1)!}{(d{-}3)!}  
      \big( \dcgCC^{\gix{1} \Zo}_\rix{2} \slashed\nabla_{\!\oix{0}} \gCC^{\gix{0}}_{\rix{2} \Zo}
         -
          \dcgCC^{\gix{0} \Zo}_\rix{2} \dc{\slashed{\nabla}}_{\!\oix{0}} \gCC^{\gix{1}}_{\rix{2} \Zo}
      \big)
   \\[0.7ex]
    && \hspace{12mm} \displaystyle 
     -\,
      i \dcsc[1]{1} \nfrac12 \big(
          \dcgCC^{\gix{1} \Zm}_\rix{1} \slashed\nabla_{\!\oix{1}} \gCC^{\gix{0}}_{\rix{1} \Zm}
         -
          \dcgCC^{\gix{0} \Zm}_\rix{1} \dc{\slashed{\nabla}}_{\!\oix{1}}  \gCC^{\gix{1}}_{\rix{1}\Zm} \big)
      +\,
      i \dcsc[0]{0} \big(
          \dcgCC^{\gix{1} \Zo}_\rix{0} \slashed\nabla_{\!\oix{0}}  \gCC^{\gix{0}}_{\rix{0} \Zo}
         -
          \dcgCC^{\gix{0} \Zo}_\rix{0} \dc{\slashed{\nabla}}_{\!\oix{0}} \gCC^{\gix{1}}_{\rix{0}\Zo} \big)
     \Bigr\}
     \,,
  \end{array}
 \label{S_red_p=3} 
 \end{equation}
where in Dirac-like massive operators
 \begin{equation}
  \slashed{\nabla}_{\!\oix{s}}
  \,\defeq\,
  \slashed{\nabla}
   \mp 
   \nfrac{i}{2}\, \sqrr\,(d{-}2s)
  \,,
  \qquad
  \dc{\slashed{\nabla}}_{\!\oix{s}}
  \,\defeq\,
  \dc{\slashed{\nabla}}
   \pm 
  \nfrac{i}{2}\, \sqrr\,(d{-}2s)
  \,,
 \label{nabla-s_p=3} 
 \end{equation}
$\slashed{\nabla}$ acts to the right and $\dc{\slashed{\nabla}}$ acts to the left.
The spectrum of masses depends entirely on the degree of fields on which $\slashed{\nabla}_{\!\oix{s}}$ acts: relative coefficient at mass term is proportional to $d{\,-\,}2s$, where $s$ is the degree of irreducible fields in the term.


After reduction, the generating functional (\ref{Z_with_Sdelta_p=3}) becomes
 \begin{eqnarray}
   Z^{\scriptscriptstyle(3)}
   \:=\:
   \!\int\! D \breve{\CPhi} \, \exp \Bigl\{ i \breve{S}^{\scriptscriptstyle(3)}_{\delta}[\breve{\CPhi}] \Bigr\}
   \,.
 \label{Z_reduced_p=3}
 \end{eqnarray}
The reduced action (\ref{S_red_p=3}) in the exponent is quadratic in reduced set of fields and block-diagonal, so integration over the remaining set of fields yield the number of determinants of the form
 \begin{equation}
   \Delta_{\oix{s}}
   \,\defeq\,
   \mathrm{Det} \,
     \slashed{\nabla}_{\!\oix{s}}
     \,,
 \end{equation}
generated by pairs of degree $s$ irreducible fields.\footnote{
 We do not additionally mark the degree of the fields which generate a determinant because the parameter $s$ of the mass coincides with the degree of the fields.}
There is one determinant of $\slashed{\nabla}_{\!\oix{3}}$, two determinants of $\slashed{\nabla}_{\!\oix{2}}$, three determinants of $\slashed{\nabla}_{\!\oix{1}}$, and four determinants of $\slashed{\nabla}_{\!\oix{0}}$. Irreducible components of $\psi_{\ZZm{3}}$ , $\gC^{\gix{}}_{ \Zm}$ and their Dirac conjugates are Grassmann odd fields, whereas components of $\gC^{\gix{}}_{ \ZZm{2}}$, $\gC^{\gix{}}_{ \Zo}$ and their Dirac conjugates are even. Thus finally, the generating functional can be expressed in the form
\begin{eqnarray}
  Z^{\scriptscriptstyle(3)}
  &\!=\!&
  \frac{\Delta_{\oix{3}}\;(\Delta_{\oix{1}})^3\;}
  {(\Delta_{\oix{2}})^2\;(\Delta_{\oix{0}})^4}
  \,.
 \label{result_EA_p=3}
\end{eqnarray}
which fits the general result predicted in \cite{BBKN4:2025}.


\section{Effective action of fermionic $p$-form theory of arbitrary degree}
 \label{Sect:p=p} 
  \hspace{\parindent}
In this section, we generalize the BV quantization procedure to the totally antisymmetric tensor-spinor $p$-form gauge theory of arbitrary degree $p$. The initial classical theory is defined on $AdS_d$ space by the action (\ref{def_S0})
 \begin{equation}
  S_0[\psi, \dcpsi]
  \:=\:
   \int\!d^dx
    \,i\sqrt{|\GG|}
    \,\dcpsi_{\ZZm{p}} \gamma^{\ZZm{2p+1}}\, D_{\Zm}\, \psi_{\ZZm{p}}
   \,,
 \label{def_S0_p=p}  
 \end{equation}
where $\psi_{\ZZm{p}}$ and its Dirac conjugate are spinor-valued tensor fields with $p$ antisymmetrized cotangent indices, and the multigamma structure $\gamma^{\ZZm{2p+1}}$ denotes the completely antisymmetrized product of $2p{+}1$ gamma matrices (\ref{def_mmg}). The operator
 $
   D_{\Zm}
   {\:=\:}
   \nabla_{\!\Zm} \pm \nfrac{i}{2}\, \sqrr\,\gamma_{\Zm}
 $,
(\ref{Dmu}), with a curvature-related mass (\ref{AdS_d}) satisfies the nilpotency property
 $
   D_{\Zm} D_{\Zm} f_{\ZZm{k}}
   {\:=\:} 0
   \,,
 $
which gives rise to the $(p{-}1)$-stage reducible gauge structure (\ref{gauge_transfs_p=p}):
 \begin{eqnarray}
  \begin{array}{llr@{\hspace{6pt}}c@{\hspace{6pt}}llr@{\hspace{6pt}}c@{\hspace{6pt}}llllllll}
  &&
  \gvar \psi_{\ZZm{p}}
  &=&
  D_{\Zm}  \glambda_{\ZZm{p-1}}
  \,,
   &&
  \gvar \dcpsi_{\ZZm{p}}
  &=&
    \dc{\glambda}_{\ZZm{p-1}} \dc{D}_{\Zm}
  \,,
  \quad
  \\[1ex]
  &&
   \gvar \glambda_{\ZZm{p-l}}
  &=&
  D_{\Zm} \glambda_{\ZZm{p-l-1}}
  \,,
  &&
   \gvar \dc{\glambda}_{\ZZm{p-l}}
  &=&
   \dc{\glambda}_{\ZZm{p-l-1}}   \dc{D}_{\Zm}
  \,,
  &   \quad
      \text{\small(\,$1 \tleq l \tleq p{-}1$\,)}
      \,,
  \\[1ex]
  &&
   \gvar \glambda_{\Zo}
  &=&
  0
  \,,
  &&
   \gvar \dc{\glambda}_{\Zo}
  &=&
  0
  \,,
  &   \quad
      \text{\small(\,$l = p$\,)}
      \,,
  \end{array}
 \label{gauge_transfs_p=p} 
 \end{eqnarray}
where the right-acting conjugate operator
 $
  \dc{D}_{\Zm}
  \defeq  \dc{\nabla}_{\!\Zm} \mp \nfrac{i}{2}\, \sqrr\,\gamma_{\Zm}
 $,
(\ref{dcDmu}), is also nilpotent:
 $
  \dc{f}_{\Al\Zm\ldots\Zn} \dc{D}_{\Zk} \dc{D}_{\Zl\Ar}
   = 0
 $.

Reducible gauge symmetry (\ref{gauge_transfs_p=p}) requires the introduction of $p$ levels of gauge parameters --- spinor-valued forms $\glambda_{\ZZn{p-l}}$,\, $1 \tleq l \tleq p$,\, of degrees $p{\,-\,}l$.
Gauge transformations of the original fields correspond to $\gvar \psi^i {\,=\,} R^i_{\zza{1}} \glambda^{\zza{1}}$, and reducibility gauge transformations at level $l$ correspond to $\gvar \glambda^{\zza{l}} {\,=\,} \,Z^{\zza{l}}_{\,\zza{l+1}} \glambda^{\zza{l+1}}$ in the general Batalin-Vilkovisky (BV) scheme of Appendix~\ref{Sect:BVst}. This yields the following gauge setup for applying the BV scheme:
 \begin{equation}
  \begin{array}{lcr@{\hspace{14pt}}l}
   R^i_{\,\zza{1}} &\mapsto&
   R_{\ZZm{p}}^{\,\ZZn{p-1}}
   \defeq
   D_{\Zm} \delta_{\ZZm{p-1}}^{\,\ZZn{p-1}}
   \,,&
   \dc{R}^{\ZZn{p-1}}_{\,\ZZm{p}}
   \defeq
   \delta^{\ZZn{p-1}}_{\,\ZZm{p-1}} \dc{D}_{\Zm}
   \,,
  \\[1.6ex]
   Z^{\zza{l-1}}_{\,\zza{l}}
   &\mapsto&
   Z_{\ZZm{p-l+1}}^{\,\ZZn{p-l}}
   \defeq
   D_{\Zm} \delta_{\ZZm{p-l}}^{\,\ZZn{p-l}}
   \,,&
   \dc{Z}^{\ZZn{p-l}}_{\,\ZZm{p-l+1}}
   \defeq
   \delta^{\ZZn{p-l}}_{\,\ZZm{p-l}} \dc{D}_{\Zm}
   \,,
   \quad\;
   \text{\small(\,$2 \tleq l \tleq p$\,)},
  \end{array}
 \label{gauge_generators_p=p} 
 \end{equation}
where the Dirac-conjugated operator is defined in (\ref{dcDmu}) as the left-acting operator. This gauge structure is \emph{abelian} because the gauge and reducibility generators are field-independent.

\newpar

At level $l$, a spinor-form $\glambda_{\ZZn{p-l}}$ of degree $r=p{\,-\,}l$ carries $\nnn_l$ independent spinor components, equal to the binomial coefficient $\binom{d}{r}\equiv \binom{d}{p-l}$. Recursively counting the number $\mmm_l = \nnn_{l} {\,-\,} \mmm_{l+1}$ of non-gauge degrees of freedom in $\glambda_{\ZZn{p-l}}$ yields
 $
   \binom{d{-}1}{p{-}l} = \binom{d}{p{-}l} - \binom{d{-}1}{p{-}l{-}1}
 $
non-gauge spinor components,\footnote{
 In the context of spin-form models, we systematically count the degrees of freedom as independent tensor index combinations, each of which is a spinor-valued function. Thus, for spin-form fields, these numbers always represent the number of independent spinor components.
}
which is the same for $\dc{\glambda}_{\ZZn{p-l}}$. For each of the gauge fields $\psi_{\ZZm{p}}$ and $\dcpsi_{\ZZm{p}}$, the symmetry leaves
 \begin{equation}
  \textstyle \binom{d}{p} - \binom{d}{p-1}+ \binom{d}{p-2} - \ldots + (-1)^{p-1} d +(-1)^p 1
  \:=\:  {\binom{d}{p} - \binom{d{-}1}{p{-}1} }
  \:=\: \binom{d{-}1}{p}
  \:\equiv\: \nfrac{(d{-}1)!}{(d{-}1{-}p)!\,p!}
 \label{mmm_0_p=p}
 \end{equation}
non-gauge spinor components \cite{FVP}, which corresponds to  $\nnn_0-\nnn_1+\nnn_2- \ldots + (-1)^{p}\nnn_{p} = \mmm_0$\HIDE{ in general BV scheme}.


  \subsection{BV construction for arbitrary $p$}
   \label{SSect:BV extension_p=p}
    \hspace{\parindent}
The BV extension (\ref{BV_master_action}--\ref{BV_aux_action})
implies introducing the set of (minimal) ghosts $\gC^{\gix{0}}_{ \ZZn{p-l}},\,\dcgC^{\gix{0}}_{ \ZZm{p-l}}$ of $p$ levels: $1 \tleq l \tleq p$.
For the gauge-fixing purposes we also introduce the set of auxiliary fields --- pairs of auxiliary ghost fields $\gC^{\gix{k}}_{ \ZZm{p-l}} ,\, \dcgC^{\gix{k}}_{ \ZZm{p-l}}$ and\HIDE{ so-called} Lagrange-multiplier fields $\gB^{\gix{k}}_{ \ZZm{p-l}} ,\, \dcgB^{\gix{k}}_{ \ZZm{p-l}}$. At each level $l$ there are $l$ generations of the auxiliary ghost pairs: $1\tleq k \tleq l$. All level-$l$ ghost fields are spinor-valued forms of degree $r = p{\,-\,}l$.
Grassmann parities of the fields in this theory are \emph{opposite} to the numerical parities of their levels $l$.

Field correspondence with the BV method of Appendix \ref{Sect:BVst} is given by
 \begin{equation}
   \begin{array}{llll}
    \psi^i & \mapsto &  \psi_{\ZZm{p}} , \dcpsi_{\ZZm{p}}
   \,,\\[0.3em]
    \gC_\gix{k}^{\zza{l}} & \mapsto & \gC^{\gix{k}}_{ \ZZm{p-l}} ,\, \dcgC^{\gix{k}}_{ \ZZm{p-l}}
   \,,
   \quad
    0\tleq k \tleq l\,,
   \\[0.3em]
    \gB_\gix{k}^{\zza{l}} & \mapsto & \gB^{\gix{k}}_{ \ZZm{p-l}} ,\, \dcgB^{\gix{k}}_{ \ZZm{p-l}}
   \,,
   \quad
    1\tleq k \tleq l \,.
   \end{array}
 \end{equation}
which are assigned the following ghost numbers:
\def \TblSpinpFormGhNumDNIB
{
\vspace{-2mm}
  \begin{center}
    \begin{tabular}{c@{\hspace{2pt}}c|c|c@{\hspace{2pt}} c@{\hspace{2pt}}||c|c@{\hspace{2pt}} c@{\hspace{2pt}}|}
    \cline{3-8}
    General BV
    && \quad$\varPhi^I$\quad  & $\gh{\varPhi^I}$
    && \quad$\varPhi^\af_I$\!\quad  & $\gh{\varPhi^\af_I}$ &
    \vphantom{\Big|}
    \\[0.1em]
    \cline{3-8}
    \multirow{10}{0cm}{}
    &
    & \multicolumn{6}{c|}{\textrm{\small minimal sector:}}
    \\
    \multirow{10}{0cm}{}
    &
    & \multicolumn{6}{c|}{\small\textrm{original fields and minimal ghosts of levels $1 \tleq l \tleq 3$}}
    \\
    \cline{3-8}
     $\;\;\psi^i \;|\!|\; \psi^\af_i \,\;\; $
    & $ \;\; \mapsto \;\; $
    \vphantom{\Big|}
    &  $\psi_{\ZZm{p}}
        ,\;
        \dc{\psi}_{\ZZm{p}}
        $
       & $0$  &
    &  $\psi_{\af}^{\ZZm{p}}
        ,\;
        \dc{\psi}_{\af}^{\ZZm{p}}
        $
       & $-1$ &
    \\
     $\gC_{\gix{0}}^{\zza{l}} \;|\!|\; \gC^\af_{\gix{0}\zza{l}} $
    & $\;\; \mapsto \;\; $
    \vphantom{\Big|} 
    &  $\gC^{\gix{0}}_{\ZZm{p-l}}
        ,\,
        \dcgC^{\gix{0}}_{\ZZm{p-l}}
        $
      & $l$  &
    & $\gC_{\af}^{\gix{0} \ZZm{p-l}}
       ,\;
       \dcgC_{\af}^{\gix{0} \ZZm{p-l}}
      $
      & $-l{-}1$  &
    \\[0.5ex]
    \cline{3-8}
    \multirow{10}{0cm}{}
    &
    & \multicolumn{6}{c|}{\textrm{\small auxiliary sector:}}
    \\
    \multirow{10}{0cm}{}
    &
    & \multicolumn{6}{c|}{\small\textrm{auxiliary ghosts and Lagrange multiplier fields}}
    \\
    &
    \multirow{10}{0cm}{}
    & \multicolumn{6}{c|}{\small\textrm{of levels $1 \tleq l \tleq p$, \,generations $k=1,...,\,l$}}
    \\
    \cline{3-8}
     $\gC_{\gix{k}}^{\zza{l}} \;|\!|\; \gC^\af_{\gix{k}\zza{l}} $
    & $ \;\; \mapsto \;\; $
    \vphantom{\Big|} 
    & \;\;$\gC^{\gix{k}}_{\ZZm{p-l}},\, \dcgC^{\gix{k}}_{\ZZm{p-l}}$\;\;
      & $ \msp^k(l{-}k{+}\tfrac12) - \tfrac12 $  &
    &  $\gC_{\af}^{\gix{k} \ZZm{p-l}} ,\, \dcgC_{\af}^{\gix{k} \ZZm{p-l}}$
      & $ -\,\msp^k(l{-}k{+}\tfrac12) - \tfrac12 $   &
    \\[0.5em]
     $\gB_{\gix{k}}^{\zza{l}} \;|\!|\; \gB^\af_{\gix{k}\zza{l}} $
    & $ \;\; \mapsto \;\; $
    \vphantom{\Big|} 
    & $\gB^{\gix{k}}_{\ZZm{p-l}},\,
       \dcgB^{\gix{k}}_{\ZZm{p-l}}$
      & $ \msp^k(l{-}k{+}\tfrac12) + \tfrac12 $   &
    &    &   &
    \\[0.5ex]
    \cline{3-8}
  \end{tabular}
 \end{center}
 \vspace{-5mm}
}
\begin{table}[h!]
 \centering
  \TblSpinpFormGhNumDNIB
 \caption{BV-extended field set of the $(p{-}1)$-stage reducible gauge theory of a $p$-form spinor field.}
 \label{Table:p-Improved}
\end{table}

\if{
Dimensions
 \begin{equation}
  \begin{array}{lllllll}
  & \maxL & \mapsto & p
  \,,\\[0.3em]
  & \nnn_{0} & \mapsto & 2 \binom{d}{p}
  \,,\quad
  & \mmm_{0} & \mapsto & 2 \binom{d-1}{p}
  \,,\\[0.3em]
  & \nnn_{l} & \mapsto & 2 \binom{d}{p-l}
  \,,\quad
  & \mmm_{l} & \mapsto & 2 \binom{d-1}{p-l}
  \,,\\[0.3em]
  \end{array}
 \end{equation}
}\fi

\vspace{-1mm}
Since the gauge group is \emph{abelian}, the minimal master sector extends the original action with a simple bilinear structure of antighosts and differentiated ghosts.
Thus the nonminimal BV master action for the $p$-form spinor field theory has the form (\ref{BV_master_action}):
 \begin{eqnarray}
  \hspace{-5mm}
   &&S_{\iBV} [\varPhi,\varPhi^\af]
    \;=\;
     S_{0} [\psi,\dcpsi]
     +
     S_{\text{gh}} [\varPhi_{\text{min}},\varPhi^\af_{\text{min}}]
    +
    S_{\text{aux}} [\varPhi_{aux},\varPhi^\af_{aux}]
    \,,
 \label{master_action_p=p}
 \end{eqnarray}
where $S_{0} [\psi,\dcpsi]$ is the original gauge action (\ref{def_S0_p=p}), minimal ghost part is
 \begin{eqnarray}
   \begin{array}{lllllll}
   &S_{\text{gh}}[\varPhi_{min},\varPhi_{min}^\af]
    &\!=\!&
    \displaystyle
     \! \int\!d^dx
     \,\Bigl\{
      %
       \,
       i
        \big(
          \,\psi_{\af}^{\ZZm{p}} D_{\Zm} \gC^{\gix{0}}_{ \ZZm{p-1}}
          - \dcgC^{\gix{0}}_{ \ZZm{p-1}} \dc{D}_{\Zm} \, \dcpsi_{\af}^{\ZZm{p}}
        \big)
    \\[.5em]
    &&&
    \displaystyle
     \qquad \quad
      +\, i \sum_{l=2}^{p}
      \big( \gC_{\af}^{\gix{0} \ZZm{p-l+1}} D_{\Zm} \gC^{\gix{0}}_{ \ZZm{p-l}}
      - 
       \dcgC^{\gix{0}}_{ \ZZm{p-l}} \dc{D}_{\Zm} \, \dcgC_{\af}^{\gix{0} \ZZm{p-l+1}} \big)
     \Bigr\}
     \,,
    \\[.5em]
   \end{array}
 \end{eqnarray}
where $D_{\mu}$ act to the right while $\dc{D}_{\mu}$ act to the left, and the contribution from the auxiliary fields is
 \begin{eqnarray}
   \begin{array}{lllllll}
    &S_{\text{aux}} [\varPhi_{aux},\varPhi^\af_{aux}]
     &\!=\!&
      \displaystyle
     \! \int\!d^dx
      \;\sum_{l=1}^{p} \sum_{k=1}^{l}
      \big(
      \gC_{\af}^{\gix{k} \ZZm{p-l}} \gB^{\gix{k}}_{ \ZZm{p-l}}
      \!+ \dcgB^{\gix{k}}_{\ZZm{p-l}} \dcgC_{\af}^{\gix{k} \ZZm{p-l}}
      \big)
    \,.
   \end{array}
 \end{eqnarray}

The sum of the first two terms in (\ref{master_action_p=p}), $S_{0} [\psi,\dcpsi]  +   S_{\text{gh}} [\varPhi_{\text{min}},\varPhi^\af_{\text{min}}]$, constitute the \emph{minimal} BV action  $S_{\text{min}}[\varPhi_{min},\varPhi_{min}^\af]$, which is the solution of master equation on the minimal BV set of fields.


\def \TriangleDNp
{
\hspace{-2.5cm}
\begin{picture}(300,250)
\put(171,223){\text{\small $\psi_\ZZm{p}$}}
\put(151,191.3){\vector(3,4){15}}
\put(171,218){\line(-3,-4){15}}
 \put(200,190){\line(-3,4){21}}
 \put(200.3,190){\line(-3,4){21}}
%
\put(121,188.5)
  {\text{\small $\gC^{\gix{1}}_{ \ZZm{p-1}}$ \phantom{,$\dcgB^{\gix{1}}_{ \ZZm{p-1}}$}}}
\put(108,174.5)
  {\text{\small \phantom{$\gC^{\gix{1}}_{ \ZZm{p-1}}$,} $\dcgB^{\gix{1}}_{ \ZZm{p-1}}$}}
\put(199,181)
  {\text{\small $\gC^{\gix{0}}_{ \ZZm{p-1}}$}}
\put(123,154){\vector(3,4){15}}
\put(141,178){\vector(-3,-4){15}}
\put(183,154){\vector(3,4){15}}
\put(201,178){\line(-3,-4){15}}
 \put(229,151){\line(-3,4){20}}
 \put(229.3,151){\line(-3,4){20}}
%
\put(94,149.5)
  {\text{\small $\gC^{\gix{2}}_{ \ZZm{p-2}}$ \phantom{,$\dcgB^{\gix{2}}_{ \ZZm{p-2}}$}}}
\put(82,135.5)
  {\text{\small \phantom{$\gC^{\gix{2}}_{ \ZZm{p-2}}$,}$\dcgB^{\gix{2}}_{ \ZZm{p-2}}$}}
\put(154,149.5)
  {\text{\small $\gC^{\gix{1}}_{ \ZZm{p-2}}$ \phantom{,$\dcgB^{\gix{1}}_{ \ZZm{p-2}}$}}}
\put(142,135.5)
  {\text{\small \phantom{$\gC^{\gix{1}}_{ \ZZm{p-2}}$,}$\dcgB^{\gix{1}}_{ \ZZm{p-2}}$}}
\put(228,142)
  {\text{\small $\gC^{\gix{0}}_{ \ZZm{p-2}}$}}
\put(93,114){\vector(3,4){15}}
\put(111,138){\vector(-3,-4){15}}
\put(153,114){\vector(3,4){15}}
\put(171,138){\vector(-3,-4){15}}
\put(213,114){\vector(3,4){15}}
\put(231,138){\line(-3,-4){15}}
 \put(259,111){\line(-3,4){20}}
 \put(259.3,111){\line(-3,4){20}}
%
\put(64,110) {\text{\small $\gC^{\gix{3}}_{ \ZZm{p-3}}$ \phantom{,$\dcgB^{\gix{3}}_{ \ZZm{p-3}}$}}}
\put(52,96) {\text{\small \phantom{$\gC^{\gix{3}}_{ \ZZm{p-3}}$,}$\dcgB^{\gix{3}}_{ \ZZm{p-3}}$}}
\put(124,110) {\text{\small $\gC^{\gix{2}}_{ \ZZm{p-3}}$ \phantom{,$\dcgB^{\gix{2}}_{ \ZZm{p-3}}$}}}
\put(112,96) {\text{\small \phantom{$\gC^{\gix{2}}_{ \ZZm{p-3}}$,}$\dcgB^{\gix{2}}_{ \ZZm{p-3}}$}}
\put(184,110) {\text{\small $\gC^{\gix{1}}_{ \ZZm{p-3}}$ \phantom{,$\dcgB^{\gix{1}}_{ \ZZm{p-3}}$}}}
\put(172,96) {\text{\small \phantom{$\gC^{\gix{1}}_{ \ZZm{p-3}}$,}$\dcgB^{\gix{1}}_{ \ZZm{p-3}}$}}
\put(258,102)  {\text{\small $\gC^{\gix{0}}_{ \ZZm{p-3}}$}}
%
\put(63,74){\vector(3,4){15}}
\put(81,98){\vector(-3,-4){28}}
\put(81,98){\line(-3,-4){31.5}}
\put(123,74){\vector(3,4){15}}
\put(141,98){\vector(-3,-4){28}}
\put(141,98){\line(-3,-4){28}}
\put(141,98){\line(-3,-4){31.5}}
\put(188.5,81.5){\vector(3,4){9}}
\put(201,98){\line(-3,-4){9}}
\put(243,74){\vector(3,4){15}}
\put(261,98){\line(-3,-4){22}}
%
\put(268,80){\line(-3,-4){18}}
\put(268,80){\vector(-3,-4){14}}
 \put(268.5,97){\line(3,-4){32}}
 \put(268.8,97){\line(3,-4){32}}

\put(28,44) 
  {\text{\small $\gC^\gix{p}_{\Zo}$,\,$\dcgB^\gix{p}_{\Zo}$}}
\put(85,44)
  {\text{\small $\gC^\gix{p-1}_{\Zo}$,\,$\dcgB^\gix{p-1}_{\Zo}$}}
\put(157,46)
  {\text{\small $\ldots$}}
\put(198,46)
  {\text{\small $\ldots$}}
\put(233,44)
  {\text{\small $\gC^\gix{1}_{\Zo}$,\,$\dcgB^\gix{1}_{\Zo}$}}
\put(300,44)
  {\text{\small $\gC^\gix{0}_{\Zo}$}}
%
    \put(350,242) {\text{level \quad \;degree}}
    \multiput(350,220)(-5,0){22}{\line(1,0){3}}
    \put(360,220) {\text{\small $0$ \qquad $\;\,p$}}
    \multiput(350,181)(-5,0){17}{\line(1,0){3}}
    \put(360,181) {\text{\small $1$ \qquad\; $p{\,-\,}1$}}
    \multiput(350,142)(-5,0){12}{\line(1,0){3}}
    \put(360,142) {\text{\small $2$ \qquad\; $p{\,-\,}2$}}
    \multiput(350,102)(-5,0){7}{\line(1,0){3}}
    \put(360,102) {\text{\small $3$ \qquad\; $p{\,-\,}3$}}
    \multiput(350,44)(-5,0){2}{\line(1,0){3}}
    \put(361,44) {\text{\small $p$ \qquad\; $0$}}
 {\color{white}
  \linethickness{14pt}
  \put(40,75){\line(1,0){300}}
  \linethickness{1pt}
 }
\put(60,74) {\text{\small $...$}}
\put(120,74) {\text{\small $...$}}
\put(180,74) {\text{\small $...$}}
\put(240,74) {\text{\small $...$}}
\put(258,74) {\text{\small $...$}}
\put(281,74) {\text{\small $...$}}
\put(358,74) {\text{\small $...$}}
\put(396,74) {\text{\small $...$}}
%
\end{picture}
\vspace{-40pt}
}


  \subsection{Gauge fixing and quantization} 
   \label{SSSect:Gauge-fixing_p=p}
    \hspace{\parindent}
The BV master action possesses an extended local symmetry (\ref{BV_NoetherIdmaS}), which must be gauge-fixed before quantization. The standard gauge fixing (\ref{BV_gauge_fixing}) excludes antifields via
 \begin{eqnarray}
    \varPhi^\af_I {\,=\,} \frac{\var^{\HIDE{r}} \varPsi[\varPhi]}{\,\var \,\varPhi^I}
    \,,\qquad
    \dc{\varPhi}^\af_I {\,=\,} \frac{\var^{\HIDE{l}} \varPsi[\varPhi]}{\,\var \,\dc{\varPhi}^I \vphantom{|^{l}}}
    \,,
 \label{gauge_fixing_p=p} 
 \end{eqnarray}
with an appropriate choice of the gauge fermion $\varPsi[\varPhi]$ --- a function of only the BV fields $\varPhi^{I}$.
We consider a gauge fermion (\ref{BV_gauge_fermion}) independent of the Lagrange-multiplier fields, which imposes a delta-function type of gauge fixing. We impose the linear gauge conditions, parameterized by field-independent operators so that the gauge fermion takes the form
 \begin{equation}
  \begin{array}{lll}
   \varPsi_{\delta}[\varPhi] &=&
   \displaystyle
   \!\!\int \!d^dx\,
   \Bigl\{
    \dcgC^{\gix{1}}_{ \ZZn{p-1}} X^{\ZZn{p-1}\ZZm{p}} \psi_{\ZZm{p}}
    +
    \dcpsi_\ZZm{p} \dc{X}^{\ZZm{p}\ZZn{p-1}}  \gC^{\gix{1}}_{ \ZZn{p-1}}
   \\[0.2em]
   &&
   \displaystyle
    + \sum_{l=2}^{p} \sum_{k=1}^{l}
    \big(
    \dcgC^{\gix{k}}_{ \ZZn{p-l}} X^{\ZZn{p-l}\ZZm{p-l+1}} \gC^{\gix{k-1}}_{ \ZZm{p-l+1}}
    +
    \dcgC^{\gix{k-1}}_{ \ZZm{p-l+1}} \dc{X}^{\ZZm{p-l+1}\ZZn{p-l}} \gC^{\gix{k}}_{ \ZZn{p-l}}
    \big)
    \Bigr\}
    \,.
   \\[1em]
  \end{array}
 \label{gauge_fermion_gen_p=p} 
 \end{equation}
The first line with $\chi^{\ZZn{p-1}}(\psi) = X^{\ZZn{p-1}\ZZm{p}} \psi_{\ZZm{p}}$  and $\dc{\chi}^{\ZZn{p-1}}(\dcpsi) = \dcpsi_\ZZm{p} \dc{X}^{\ZZm{p}\ZZn{p-1}}$, represents the $l{\,=\,}1$ case of the structure in the double sum upon identification  $\psi_{\ZZn{p}} \leftrightarrow \gC^{\gix{0}}_{ \ZZn{p}}, \; \dcpsi_{\ZZn{p}} \leftrightarrow \dcgC^{\gix{0}}_{ \ZZn{p}}$.

For fixed $l$ but different generation numbers $k$, the gauge-fixing operators $X_{\gix{k}}^{\ZZn{p-l}\ZZm{p-l+1}}$ and $\dc{X}_{\gix{k}}^{\ZZm{p-l+1}\ZZn{p-l}}$ can be chosen independently. We choose these operator pairs to be identical for all $k$ and omit the generation index in the operators.

\newpar

The gauge fixing and related structures are often illustrated by a triangular diagram. In Fig.~\ref{BVp_Triangular_diagram} for each spinor form field depicted, the presence of a Dirac conjugate partner is also assumed.

\begin{figure}[htbp]
  \centering
  \TriangleDNp  
  \caption{\hspace{12.2ex} Triangular diagram for fermionic $p$-form model
  \\ \phantom{.} \\
  \small \it
  At each vertex, Dirac-conjugated partners $\dcpsi,\dcgC{}{},\gB{}{}$ of the depicted fields are assumed. The degree $r$ of a field corresponds to its level as $r=p{\,-\,}l$. The gauge fermion (\ref{gauge_fermion_gen_p=p}) couples the fields along gauge-fixing chains. Each chain originates from its own minimal-ghost vertex at the right slope and extends left-downward until the maximal level $l{\,=\,}p$, with spinor fields having no tensor indices. Edges in these chains are associated with operators $X_{\gix{k}}^{\ZZn{p-l}\ZZm{p-l+1}}$ and $\dc{X}_{\gix{k}}^{\ZZm{p-l+1}\ZZn{p-l}}$, which, in the gauge fermion, couple ghosts from adjacent vertices.
  }
 \label{BVp_Triangular_diagram} 
\end{figure} 

Gauge fixing (\ref{gauge_fixing_p=p}) for the gauge fermion (\ref{gauge_fermion_gen_p=p}) implies the following substitution for antifields:
 \begin{equation}
   \begin{array}{|lcl@{\hspace{1pt}}c@{\hspace{1pt}}}
      \gC_{\af}^{\gix{k} \ZZm{p-l}}
      &\mapsto&
       \dcgC^{\gix{k-1}}_{ \ZZn{p-l+1}} \dc{X}^{\ZZn{p-l+1}\ZZm{p-l}}
       +
       \dcgC^{\gix{k+1}}_{ \ZZn{p-l-1}} X^{\ZZn{p-l-1}\ZZm{p-l}}
       \,,
      &
     \\[0.8em]
      %
      \dcgC_{\af}^{\gix{k} \ZZm{p-l}}
      &\mapsto&
       X^{\ZZm{p-l}\ZZn{p-l+1}} \gC^{\gix{k-1}}_{ \ZZn{p-l+1}}
       +
       \dc{X}^{\ZZm{p-l}\ZZn{p-l-1}} \gC^{\gix{k+1}}_{ \ZZn{p-l-1}}
       \,,
      &
   \end{array}
   \quad {\small{ \text{(\,$0\tleq k\tleq l$,\,\; $0 \tleq l\tleq p$\,)}}}
   .
 \label{gf_antifields_gen_p=p} 
 \end{equation}
Boundary cases\footnote{
 Boundary cases of (\ref{gf_antifields_gen_p=p}):
 \begin{equation}
  \hspace{-5mm}
   \begin{array}{l@{\hspace{1pt}}c@{\hspace{5pt}}l@{\hspace{1pt}}c@{\hspace{1pt}}}
      \psi_{\af}^{\ZZm{p}}
      &\mapsto&
       \dcgC^{\gix{1}}_{ \ZZn{p-1}} X^{\ZZn{p-1} \ZZm{p}}
       \vphII\,,
     \\[0.2em]
      \gC_{\af}^{\gix{0} \ZZm{p-l}}
      &\mapsto&
       \dcgC^{\gix{1}}_{ \ZZn{p-l-1}} X^{\ZZn{p-l-1}\ZZm{p-l}} \,,
     \\[0.2em]
      \gC_{\af}^{\gix{0} \Zo}
      &\mapsto&
       0  \,,
     \\[0.2em]
      &&&
     \\[-0.8em]
      \gC_{\af}^{\gix{1}  \ZZm{p-1}}
      &\mapsto&
       \dcpsi_\ZZn{p} \dc{X}^{\ZZn{p}\ZZm{p-1}}
       {\,+\,}
       \dcgC^{\gix{2}}_{ \ZZn{p-2}} X^{\ZZn{p-2}\ZZm{p-1}}
       ,
     \\[0.3em]
      %
      \gC_{\af}^{\gix{k} \Zo}
      &\mapsto&
       \dcgC^{\gix{k-1}}_{\Zn} \dc{X}^{\Zn}
       \,,
     \\[0.2em]
    \end{array}
    \begin{array}{l@{\hspace{1pt}}c@{\hspace{5pt}}l@{\hspace{1pt}}r@{\hspace{1pt}}}
      \dcpsi_{\af}^{\ZZm{p}}
      &\mapsto&
       \dc{X}^{\ZZm{p} \ZZn{p-1}} \gC^{\gix{1}}_{ \ZZn{p-1}} \,,
      &  {\small{\text{   $(\,l=0,\; k=0\,)$   }}}
       \vphII
     \\[0.2em]
       \dcgC_{\af}^{\gix{0} \ZZm{p-l}}
      &\mapsto&
         \dc{X}^{\ZZm{p-l}\ZZn{p-l-1}} \gC^{\gix{1}}_{ \ZZn{p-l-1}} \,,
       &  {\small{\text{   $(\,1 \tleq l \tleq p{\,-\,}1,\, k{\,=\,}0\,)$   }}}
     \\[0.2em]
      \dcgC_{\af}^{\gix{0} \Zo}
      &\mapsto&
       0 \,,
      &  {\small{\text{   $(\,l=p,\; k=0\,)$   }}}
     \\[0.2em]
      &&&
     \\[-0.8em]
      \dcgC_{\af}^{\gix{1} \ZZm{p-1}}
      &\mapsto&
       X^{\ZZm{p-1}\ZZn{p}} \psi_{\ZZn{p}}
       {\,+\,}
       \dc{X}^{\ZZm{p-1}\ZZn{p-2}} \gC^{\gix{2}}_{ \ZZn{p-2}} 
       ,
    \hspace{-15mm}
       &  {\small{\text{  $(\,l{\,=\,}1,\, k{\,=\,}1\,)$   }}}
     \\[0.3em]
      \dcgC_{\af}^{\gix{k} \Zo}
      &\mapsto&
       X^{\Zn} \gC^{\gix{k-1}}_{\Zn}
      \,,
      &  {\small{ \text{$(\, l{\,=\,}p,\; 1\tleq k\tleq p \,)$ }}}
      \\[0.2em]
   \end{array}
 \nonumber
 \end{equation}
}
imply that the level $0$ ghosts, $\gC^{\gix{0}}_{ \ZZn{p}},\,\dcgC^{\gix{0}}_{ \ZZn{p}}$, are the original fields $\psi_{\ZZn{p}},\,\dcpsi_{\ZZn{p}}$, and the ghosts $\gC^{\gix{k}}_{ \ZZn{p-l}},\,\dcgC^{\gix{k}}_{ \ZZn{p-l}}$ of levels $l<0$,\, $l>p$, and generations $k<0$,\, $k>l$ are zero.

Reducing antifields via (\ref{gf_antifields_gen_p=p}) leads to the \emph{gauge-fixed action}  (\ref{BV_gf_action})
 \begin{eqnarray}
  &&S_{\delta} [\varPhi]
   \;=\;
    S_{0} [\psi,\dcpsi]
    +
    S_{\delta\,\text{gh}} [\varPhi]
    +
    S_{\delta\,\text{aux}} [\varPhi]
  \,,
 \label{gf_action_p=p} 
 \end{eqnarray}
where the original action term $S_{0}$, (\ref{def_S0_p=p}), does not change, the gauge-fixed representation of the minimal ghost part
 \begin{eqnarray}
  \begin{array}{lllllll}
   & S^{}_{\delta\,\text{gh}} [\varPhi]
    &\!=\!&
    \displaystyle
    \!\int \!d^dx\,
       \,i \sum_{r=0}^{p-1}  
      \big( \dcgC^{\gix{1}}_{ \ZZn{r}} X^{\ZZn{r}\ZZm{r+1}}
           D_{\Zm} \gC^{\gix{0}}_{ \ZZm{r}}
         {\,-\,}
          \dcgC^{\gix{0}}{}_{\ZZm{r}} \dc{D}_{\Zm}
           \dc{X}^{\ZZm{r+1}\ZZn{r}} \gC^{\gix{1}}_{ \ZZn{r}} \big)
     \,,
  \end{array}
 \label{gf_gh action_p=p} 
 \end{eqnarray}
becomes a functional of only the minimal ghosts and antighosts,
$S_{\delta\,\text{gh}} [\varPhi] =  S_{\delta\,\text{gh}} [\gC^\gix{0},\gC^\gix{1},\dcgC^\gix{0},\dcgC^\gix{1}]$,
and the gauge-fixed auxiliary action takes the form
 \begin{eqnarray}
  \begin{array}{lllllll}
   & S^{}_{\delta\,\text{aux}} [\varPhi]
   &\!=\!& \displaystyle
    \!\int \!d^dx\,
    \sum_{r=0}^{p-1}  \sum_{k=1}^{p-r}
    \Bigl\{
    \big(
    \dcgC^{\gix{k-1}}_{ \ZZn{r+1}} \dc{X}^{\ZZn{r+1}\ZZm{r}}
     {\,+\,}
     \dcgC^{\gix{k+1}}_{ \ZZn{r-1}} X^{\ZZn{r-1}\ZZm{r}}
     \big)
    \gB^{\gix{k}}_{ \ZZm{r}}
    \\[.3em]
   &
   && \qquad\qquad\qquad \displaystyle
    +\,
    \dcgB^{\gix{k}}_{ \ZZm{r}}
    \big( X^{\ZZm{r}\ZZn{r+1}} \gC^{\gix{k-1}}_{ \ZZn{r+1}}
     {\,+\,}
     \dc{X}^{\ZZm{r}\ZZn{r-1}} \gC^{\gix{k+1}}_{ \ZZn{r-1}}
    \big)
    \Bigr\}
    \,.
  \end{array}
 \label{gf_aux_action_p=p} 
 \end{eqnarray}
We sum over the degree parameters $r$ of spinor form fields, which are uniquely related to level numbers $l$ as $r=p{\,-\,}l$.
Gauge fixing entangles the minimal and auxiliary fields. In particular, the original fields ($ \psi_{\ZZn{p}},\,\dcpsi_{\ZZn{p}}  \leftrightarrow \gC^{\gix{0}}_{ \ZZn{p}},\,\dcgC^{\gix{0}}_{ \ZZn{p}}$) enter the $S^{}_{\delta\,\text{aux}}$ via the $r {\,=\,} p{\,-\,}1$ term, which breaks the gauge symmetry of the original gauge action (\ref{def_S0_p=p}).


\newpar

With proper gauge fixing, the gauge-fixed action (\ref{gf_action_p=p}) is \emph{nondegenerate} and is suitable for quantization with the generating functional
 \begin{eqnarray}
   Z^{\scriptscriptstyle(p)}
   \:=\: \!\int\! D\varPhi\, \exp \Bigl\{ i S_{\delta}[\varPhi] \Bigr\}
   \,.
 \label{Z_with_Sdelta_p=p} 
 \end{eqnarray}
The model under consideration is a free theory, for which the part or all fields can be integrated out producing the effective action.
A gauge fixing with \emph{algebraic} gauge-fixing operators gives rise to a huge subset of nondynamical fields, which can be eliminated directly in the classical gauge-fixed action before quantization. Or, equivalently, can be integrated out in (\ref{Z_with_Sdelta_p=p}) with the ultralocal contribution to the measure.  The reduced representation of the generating functional is often more convenient to describe the quantum properties of the theory and calculating the effective action.

\newpar

In what follows, we consider algebraic gauge fixing and derive the reduced representation of the gauge-fixed action $S^{\scriptscriptstyle(p)}_{\delta}[\varPhi] \to \breve{S}^{\scriptscriptstyle(p)}_{\delta}[\breve{\varPhi}]$ by performing a reduction for two gauges: one defined by minimal-rank gauge-fixing operators\HIDE{ allowable by (\ref{GF_rank_conds_p=p})} with $\rank X_{\gix{k}}^{\ZZm{r}\ZZn{r+1}}\! = \binom{d-1}{r}$, and another with full-rank operators having $\rank X_{\gix{k}}^{\ZZm{r}\ZZn{r+1}} \!= \binom{d}{r}$. To define these operators and solve the reduction constraints on the fields
\if{
 $
  \frac{\var S_{\delta}}{\var{\gB^{\gix{k}}_{ \ZZm{r}}}} {\,=\,} 0
   ,\;
  \frac{\var S_{\delta}}{\var{\dcgB^{\gix{k}}_{ \ZZm{r}}}} {\,=\,} 0
   ,\,
 $
}\fi
we begin with the decomposition of the fields into algebraically irreducible components.


  \subsection{Irreducible fields for arbitrary $p$}
   \label{SSect:Irreducible_basis_p=p}
    \hspace{\parindent}
The spinor-valued form of degree $r$ and its Dirac conjugate can be algebraically decomposed into $\gamma$-irreducible fields of degrees $s$ ($0\tleq s\tleq r$) as
 \begin{equation} 
  \hspace{-10mm}
   \begin{array}{|llllll}
    &   \displaystyle
    \fA_{\ZZm{r}}  \;=\;
     \sum_{s=0}^{r} 
     \gamma_{\ZZm{r-s}} \fCA_{\rix{r} \ZZm{s}}
    \,,
    &&
    \gamma^{\Zn} \fCA_{\rix{r} \ZZn{s}} = 0
    \,,
    \\[0.5em]
    &   \displaystyle
    \dc{\fA}_{\ZZm{r}}  \;=\;
     \sum_{s=0}^{r}
     \dcsc[r]{s} \dc{\fCA}_{\rix{r} \ZZm{s}} \gamma_{\ZZm{r-s}}
    \,,
    &&
    \dc{\fCA}_{\rix{r} \ZZn{s}}  \gamma^{\Zn}=0
    \,,
    \\[0.5em]
   \end{array}
 \label{A_decomp} 
 \end{equation}
where  $\fCA_{\rix{r} \ZZm{s}}$ are $\gamma$-irreducible components of degree $s$ of the field $\fA_{\ZZm{r}}$, being the coefficients at independent matrix basis elements of the Clifford algebra. Sign factors $\dcsc[r]{s}$ are a part of the definition of the conjugated components. To ensure $\dcfCA_{\rix{r} \ZZm{s}}= \dcl{\fCA_{\rix{r} \ZZm{s}}}$ for $\dcfA_{\ZZm{r}}= \dcl{\fA_{\ZZm{r}}}$ one must choose
 \begin{equation}
  \dcsc[r]{s} = \msp^{\frac{r(r-1)}{2} - \frac{s(s-1)}{2}}.
 \label{def_dcsc}
 \end{equation}

The transition of the configuration space of BV fields $\Phi^I$ to the basis of irreducible fields
 \begin{equation}
  \Phi \to \CPhi\;:
  \qquad
  \psi, \dcpsi \,\to\, \CPsi{}, \dcCPsi{}
  \,;\;\;
  \gC, \dcgC \,\to\, \gCC, \dcgCC
  \,;\;\;
  \gB, \dcgB \,\to\, \gCB, \dcgCB
  \,;
 \end{equation}
proceeds the same way
 \begin{equation}
  \begin{array}{|@{\hspace{4pt}}lll@{\hspace{15pt}}|@{\hspace{4pt}}lll}  
   & \displaystyle
    \psi_{\ZZm{p}} \;=\;
     \sum_{s=0}^{p}
     \gamma_{\ZZm{p-s}} \CPsi_{\ZZm{s}}
    \,,
    & \gamma^{\Zn} \CPsi_{\ZZn{s}}=0
    \,,
    &&   \displaystyle
    \dcpsi_{\ZZm{p}} \;=\;
     \sum_{s=0}^{p}
    \dcsc[p]{s}  \dcCPsi_{\ZZm{s}}\gamma_{\ZZm{p-s}}
    \,,
    &
    \dcCPsi_{\ZZn{s}}\gamma^{\Zn}=0
    \,,
  \\[0.5em]
   &   \displaystyle
    \gC^{\gix{k}}_{ \ZZm{r}}  \;=\;
     \sum_{s=0}^{r}
     \gamma_{\ZZm{r-s}} \gCC^{\gix{k}}_{\rix{r} \ZZm{s}}
    \,,
    &
    \gamma^{\Zn} \gCC^{\gix{k}}_{\rix{r} \ZZn{s}} = 0
    \,,
    &&   \displaystyle
    \dcgC^{\gix{k}}_{ \ZZm{r}}  \;=\;
     \sum_{s=0}^{r}
     \dcsc[r]{s} \dcgCC^{\gix{k}}_{\rix{r} \ZZm{s}} \gamma_{\ZZm{r-s}}
    \,,
    &
    \dcgCC^{\gix{k}}_{\rix{r} \ZZn{s}}  \gamma^{\Zn}=0
    \,,
  \\[0.5em]
   &   \displaystyle
    \gB^{\gix{k}}_{\ZZm{r}}  \;=\;
     \sum_{s=0}^{r}
     \gamma_{\ZZm{r-s}} \gCB^{\gix{k}}_{\rix{r} \ZZm{s}}
     \,,
    &
    \gamma^{\Zn} \gCC^{\gix{k}}_{\rix{r} \ZZn{s}} = 0
    \,,
    &&   \displaystyle
    \dcgB^{\gix{k}}_{\ZZm{r}}  \;=\;
     \sum_{s=0}^{r}
     \dcsc[r]{s} \dcgCB^{\gix{k}}_{\rix{r} \ZZm{s}} \gamma_{\ZZm{r-s}}
    \,,
    &
    \dcgCB^{\gix{k}}_{\rix{r} \ZZn{s}}  \gamma^{\Zn}=0
    \,,
  \\[0.5em]
  \end{array}
 \label{decomp_p=p} 
 \end{equation}
For mostly presentational reasons we keep the coefficients $\dcsc[r]{s}$ unspecified.\footnote{The presence of unspecified $\dcsc[r]{s}$ keeps signs in general formulas more symmetric.
In any case, the field spectrum and the corresponding masses in the irreducible basis do not depend on these coefficients due to the diagonalization of the reduced gauge-fixed action in the basis of irreducible fields.}

\newpar

The initial action $ S_0[\psi, \dcpsi]$, (\ref{def_S0_p=p}), as the part of the BV action (\ref{master_action_p=p}),
does not acquire dependence on the gauge parameters during gauge fixing. Therefore, we can derive the representation in irreducible fields for this term before specifying a particular gauge. The result reads
 \begin{eqnarray}
  \begin{array}{@{\hspace{2pt}}llll}
   & S^{}_{0} [\CPsi,\dcCPsi]
    &\!\!=\!\!&   \displaystyle
     \int \!d^dx\,
     i\sqrt{|\GG|}
      \sum_{s=0}^{p}
     \bnc{s}\,
    \dcCPsi^{\ZZm{s}}
     \big( \slashed{\nabla}
     -i 
     \mAdS 
      \msp^{p-s} (d{-}2s)\big)
    \, \CPsi_{\ZZm{s}}
  \\
    &&&
    \displaystyle
     +\, 
     \int \!d^dx\,
     i\sqrt{|\GG|}
     \sum_{s=1}^{p}
     \bnc{s}
     \, (d{-}2s)
      \,\big(
       \,\dcCPsi^{\ZZm{s}} \,{\nabla}_{\!\Zm} \,\CPsi_{\ZZm{s-1}}
      +
       \dcsc[s]{s-1} 
       \,\dcCPsi^{\ZZm{s-1}} \,{\nabla}^{\Zm} \,\CPsi_{\ZZm{s}}
      \big)
    \,,
  \end{array}
 \label{So_in_components_p=p} 
 \end{eqnarray}
where $\mAdS \equiv \pm \frac{1}{2}\, \sqrr$.
 %
The structure of the original action comprises a sum of diagonal terms with Dirac-like massive operators, characterized by the degree-dependent masses,
and sub- and super-diagonal kinetic contributions.
The overall coefficients on the right-hand side of (\ref{So_in_components_p=p}) are:
 \begin{equation}
   \bnc{s}
   \:=\:
   \msp^{\frac{s(s-1)}{2} - s} 
    \, \dcsc[p]{s}
    \, s!\, \nfrac{(d{-}2s{-}1)!}{(d{-}2p{-}1)!}
    \,,
 \label{So_in_components_p=p_coef} 
 \end{equation}
where $\dcsc[p]{s}$ are sign factors (\ref{def_dcsc}), and the relative sign factors $\dcsc[s]{s-1}$\HIDE{ in the nondiagonal terms} originate from ${\dcsc[p]{s-1}}/{\dcsc[p]{s}}$.

\newpar

For a generic field $\fA_{\ZZm{r}}$ of degree $r$, (\ref{A_decomp}), we introduce the corresponding projectors $P_\pix{s} {}_{\ZZn{r}}^{\,\ZZm{r}}$ onto basis terms with irreducible components $\fCA_{\rix{r} \ZZn{s}}$ of particular degrees $s$,\, $0 \tleq s \tleq r$:
 \begin{eqnarray}
  P_\pix{s} {}_{\ZZn{r}}^{\,\ZZm{r}} \fA_{\ZZm{r}}
    \:=\:
   \gamma_{\ZZn{r-s}} \fCA_{\rix{r} \ZZn{s}}
   \,,
 \label{Projector_gen_s_r_COPY} 
 \end{eqnarray}
whose explicit representations is given by
 \begin{equation}
  P_\pix{s} {}_{\ZZn{r}}^{\:\ZZm{r}}
  \;=\;
  \msp^{\frac{r(r-1)}{2}}
  \tfrac{r!}{s!(r{-}s)!} \tfrac{(d{-}r{-}s)!}{(d{-}2s)!}
  \sum_{u=0}^{{\color{Navy}s}}
  \msp^{r-s} 
  \msp^{\frac{u(u-1)}{2}}
  \tfrac{s!}{u!(s{-}u)!} \tfrac{(d{+}1{-}2s)!}{(d{+}1{-}s{-}u)!}
  \cdot
  \gamma_{\ZZn{r-u}} \delta_{\ZZn{u}}^{\,\ZZm{u}} \gamma^{\ZZm{r-u}}
  \,.
 \label{Projector_gen_s_r} 
 \end{equation}

For the Dirac-conjugated field $\dc{\fA}_{\ZZm{r}}$ with decomposition
(\ref{A_decomp}), the projectors $P_\pix{s} {}^{\ZZn{r}}_{\,\ZZm{r}}$ onto the basis elements are:
 \begin{equation}
  P_\pix{s} {}^{\ZZn{r}}_{\,\ZZm{r}}
   \equiv
  \GG^{\ZZn{r}\ZZk{r}} P_\pix{s} {}_{\ZZk{r}}^{\:\ZZl{r}}\, \GG_{\ZZl{r}\ZZm{r}}
    \,.
 \label{dcProj_p=p}
 \end{equation}

\newpar

Each field of level $l$ has degree $r=p{\,-\,}l$ and describes $\binom{d}{r}\equiv \binom{d}{p-l}$ independent spinor components. Its irreducible algebraic components are spinor-valued forms of degrees $s$ subject to $\binom{d}{s-1}$ conditions of the type $\gamma^{\Zm}\fCA^{\gix{k}}_{\rix{r} \ZZm{s}}{\,=\,}0$, ($s \tgeq 1$), and thus contain
$\binom{d}{s}-\binom{d}{s-1} = \frac{d-2s+1}{d-s+1}\binom{d}{s} = \frac{d-2s+1}{s}\binom{d}{s-1}$ independent spinor components\HIDE{(we assume $\binom{d}{-1}=0$)}.

\newpar

It is convenient to introduce two complementary projectors
 \begin{equation}
  \begin{array}{ll}
    \displaystyle
    P_{\lead} {}_{\ZZm{r}}^{\ZZn{r}} \fA_{ \ZZn{r}}
    =
     \sum_{n=0}^{[r/2]} \gamma_{\ZZm{2n}} \fCA_{\rix{r} \ZZm{r-2n}}
   \,,
   \qquad
    &    \displaystyle
    P_{\lead} {}_{\ZZm{r}}^{\ZZn{r}}
    \,\equiv
    \! \sum_{n=0}^{[r/2]} P_\pix{r-2n} {}_{\ZZm{r}}^{\ZZn{r}}
    \,,
  \\[1ex]
    \displaystyle
    P_{\sublead} {}_{\ZZm{r}}^{\:\ZZn{r}} \gC^{\gix{k}}_{ \ZZn{r}}
    =
    \! \sum_{n=0}^{[(r-1)/2]} \gamma_{\ZZm{2n+1}} \gCC^{\gix{k}}_{\rix{r} \ZZm{r-2n-1}}
    \,,
    \qquad
    &    \displaystyle
    P_{\sublead} {}_{\ZZm{r}}^{\ZZn{r}}
    \,\equiv
    \! \sum_{n=0}^{[(r-1)/2]} \!\!\!\! P_\pix{r-2n-1} {}_{\ZZm{r}}^{\ZZn{r}}
    \,,
 \label{gf_Projectors_p=p} 
  \end{array}
 \end{equation}
where square brackets in the sums' upper limits denote the integer part of a number.
The first --- \emph{leading-parity} projector ---  when acting on spinor forms of degree $r$ preserves basis terms with irreducible components of degrees $r{\,-\,}2n$,\,  $n\in 0,1,2,...\,,[r/2]$,\, forcing irreducible contributions of degrees $r{\,-\,}2n{\,-\,}1$,\, $n\in 0,1,2,...\,,[(r{-}1)/2]$ to vanish.
The second --- \emph{subleading-parity} projector --- makes the opposite: it projects out basis terms with irreducible components of degrees $r{\,-\,}2n$, while preserving irreducible contributions of degrees $r{\,-\,}2n{\,-\,}1$.
The projector $P_{\lead} {}_{\ZZm{r}}^{\ZZn{r}}$ selects $\binom{d-1}{r}$ elementary spinor components and has that rank, which coincides with $\mmm_{p-r}$ --- the number of non-gauge degrees of freedom at level $l{\:=\:}p{\,-\,}r$. Whereas the projector $P_{\sublead} {}_{\ZZm{r}}^{\ZZn{r}}$ selects complementary $\binom{d-1}{r-1}$ elementary spinor components and has the corresponding rank, which coincides with $\mmm_{p-r+1}$ --- the number of non-gauge degrees of freedom at level $l{\,+\,}1{\:=\:}p{\,-\,}r{\,+\,}1$.\footnote{
 Recall that the degrees of freedom of spinor-valued form fields (or ranks of spin-tensor operators) we count as the number of elementary spinors (or, correspondingly, the rank of a matrix operator, whose elements are spinorial endomorphisms). In other words, this counting reflects the tensor properties of the spin-tensor objects.
}


  \subsection{Minimal-rank gauge fixing and physical space in general case}
   \label{SSect:Gauge_fixing_min_p=p}
    \hspace{\parindent}
The rank properties of projectors (\ref{gf_Projectors_p=p}) suggest that an algebraic gauge fixing, along the lines of the the original prescription \cite{Batalin:1983ggl}, can be defined by
 \begin{equation}
  \begin{array}{lllllll}
    &X^{\ZZm{r}\ZZn{r+1}}
    &\!\!=\!\!&
   \displaystyle
    \gamma_{\Zm} P_{\sublead}^{\ZZm{r+1}\ZZn{r+1}}
    =  P_{\lead}^{\ZZm{r}\ZZn{r}} \gamma^{\Zn}
    \,, \;\;\quad
    &\dc{X}^{\ZZn{r+1}\ZZm{r}}
    &\!\!=\!\!&
   \displaystyle
    P_{\sublead}^{\ZZn{r+1}\ZZm{r+1}}\gamma_{\Zm}
    = \gamma^{\Zn} P_{\lead}^{\ZZn{r}\ZZm{r}}
    \,.
  \end{array}
 \label{min-rank_gf_p=p} 
 \end{equation}
with lower indices raised by the inverse metric tensors.\footnote{
 The possibility of two representations for the gauge-fixing matrices, which can be expressed in terms of both projectors (\ref{gf_Projectors_p=p}), directly follow from the defining properties (\ref{Projector_gen_s_r_COPY}).
}
These gauge-fixing matrices (\ref{min-rank_gf_p=p}) are of minimal possible rank, and which imply kernel complementarity conditions
 $ 
  \rker X^{\ZZn{r-1}\ZZm{r}}  \cap \rker \dc{X}^{\ZZn{r+1}\ZZm{r}} = \emptyset\,,
  \;\; 
  \rker X^{\ZZn{r-1}\ZZm{r}}  \cup \rker \dc{X}^{\ZZn{r+1}\ZZm{r}} = V^{\rix{r}}\,,
 $ 
where $V^{\rix{r}}$ is the vector space of spin-forms of degree $r$ and \,$\rker$\, denotes the space of right zero modes of an operator. Together, $X^{\ZZm{r-1}\ZZn{r}}$ and $\dc{X}^{\ZZn{r+1}\ZZm{r}}$ contain a complete set of projectors onto a field space of degree $r$ and form a full-rank block matrix.

\newpar

It is informative to look at the structure of the gauge-fixed action in the basis of irreducible fields. In the gauge (\ref{min-rank_gf_p=p}), the gauge-fixed ghost part (\ref{gf_gh action_p=p}) of the minimal action,
 \begin{equation}
  \begin{array}{lllllll}
   & S^{}_{\delta\,\text{gh}} [\varPhi]
    &\!\!=\!\!&
    \displaystyle
    \!\!\int \!d^dx\,
        i \sum_{r=0}^{p-1}  
      \big( \dcgC^{\gix{1}}_{ \ZZn{r}}\gamma_{\Zn} P_{\sublead}^{\ZZn{r+1}\ZZm{r+1}} D_{\Zm} \gC^{\gix{0}}_{ \ZZm{r}}
      -
       \dcgC^{\gix{0}}_{ \ZZn{r}}\dc{D}_{\Zn} P_{\sublead}^{\ZZn{r+1}\ZZm{r+1}} \gamma_{\Zm} \gC^{\gix{1}}_{ \ZZm{r}} \big)
     \,,
   \end{array}
 \label{gf_aux_action_min-rank_in_components_p=p}
 \end{equation}
is a sum of terms bilinear in minimal ghosts and antighosts of the same degree, and, in the basis of irreducible fields, reads
 \begin{equation}
  \begin{array}{@{\hspace{2pt}}l@{\hspace{2pt}}llllll}
   & S^{}_{\delta\,\text{gh}} [\CPhi]
  \if{
    &\!\!=\!\!&
    \displaystyle
    \!\!\int \!d^dx\,
        i \sum_{r=0}^{p-1}  
      \big( \dcgC^{\gix{1}}_{ \ZZn{r}}\gamma_{\Zn} P_{\sublead}^{\ZZn{r+1}\ZZm{r+1}} D_{\Zm} \gC^{\gix{0}}_{ \ZZm{r}}
      -
       \dcgC^{\gix{0}}_{ \ZZn{r}}\dc{D}_{\Zn} P_{\sublead}^{\ZZn{r+1}\ZZm{r+1}} \gamma_{\Zm} \gC^{\gix{1}}_{ \ZZm{r}} \big)
     \,,
    \\[.3em]
  }\fi
    &\!=\!&
    \displaystyle
    \!\int \!d^dx\,
        i
      \sum_{r=0}^{p-1} 
      \sum_{n=0}^{[r/2]}
        \cnc[r]{r-2n}
   \textstyle
     \Big(
       \dcgCC^{\gix{1} \ZZm{r-2n}}_{\rix{r}}
           \big( \slashed{\nabla}
           {\,-\,} i 
           \mAdS (d{-}2r{+}4n) \big)
       \gCC^{\gix{0}}_{\rix{r} \ZZm{r-2n}}
  \\
  &&& \hspace{30ex}
       -\,
       \dcgCC^{\gix{0} \ZZm{r-2n}}_{\rix{r}}
         \big( \dc{\slashed{\nabla}}
         {\,+\,} i 
         \mAdS (d{-}2r{+}4n) \big)
       \gCC^{\gix{1}}_{\rix{r} \ZZm{r-2n}}
     \Big)
   \\[.5em]
    &&\!\!\!\!&
    \displaystyle
    {\color{black}\!-\,} 
     \!\int \!d^dx\,
       i
     \sum_{r=0}^{p-1}  
     \sum_{n=1}^{[{r/2}]} 
    \textstyle
       \msp^{ r}
       \cnc[r]{r-2n}
       \nfrac{(r{-}2n{+}1)}{(d{-}2r{+}4n{-}2)}
     \Big(
       \dcgCC^{\gix{1} \ZZm{r-2n}}_{\rix{r}}
        {\nabla}^\Zm
       \gCC^{\gix{0}}_{\rix{r} \ZZm{r-2n+1}}
  \\
  &&& \hspace{40ex}
       -\,
       \dcsc[{r-2n}]{r-2n+1} 
       \dcgCC^{\gix{0} \ZZm{r-2n+1}}_{\rix{r}}
        \,\dc{\nabla}_{\!\Zm}
       \gCC^{\gix{1}}_{\rix{r} \ZZm{r-2n}}
       \Big)
    \\[.5em]
     &&\!\!\!\!&
    \displaystyle
    {+\,} 
    \int \!d^dx\,
       i
    \sum_{r=0}^{p-1}\!  
    \sum_{n=0}^{[(r-1)/2]} 
       \!\!\!
       \cnc[r]{r-2n} (d{-}2r{+}4n)
   \textstyle
     \Big(
       \dcgCC^{\gix{1} \ZZm{r-2n}}_{\rix{r}}
        {\nabla}_{\!\Zm}
       \gCC^{\gix{0}}_{\rix{r} \ZZm{r-2n-1}}
  \\
  &&& \hspace{40ex}
       -\,
       \dcsc[{r-2n}]{r-2n-1} 
       \dcgCC^{\gix{0} \ZZm{r-2n-1}}_{\rix{r}}
        \,\dc{\nabla}^\Zm
       \gCC^{\gix{1}}_{\rix{r} \ZZm{r-2n}}
       \Big)
     \,,
    \\[.3em]
  \end{array}
 \label{Sgh_gf_min-rank_p=p} 
 \end{equation}
where $\mAdS \equiv \pm \frac{1}{2}\, \sqrr$. The overall normalizing coefficients are given by
 \begin{equation}
   \cnc[r]{r-2n}
   \:=\:
    \HIDE{+} 
     \dcsc[r]{r-2n}
     \msp^{r-n}
     \nfrac{(2n{+}1)!(r{-}2n)!}{(r{+}1)!} \nfrac{(d{-}2r{+}4n{-}1)!}{(d{-}2r{+}2n{-}1)!}
    \,,
 \label{Sgh_gf_min-rank_p=p_coef} 
 \end{equation}
where $\dcsc[r]{s}$ are the sign factors from definitions of the Dirac-conjugated irreducible fields (\ref{decomp_p=p}). For the choice (\ref{def_dcsc}),  $\dcsc[r]{r-2n} {\,=\,} \msp^{n}$ and the relative coefficient $\dcsc[{r-2n}]{r-2n\pm1} {\,=\,} \pm \msp^{r}$.
 %
 %
Similar to the irreducible representation of the original action (\ref{So_in_components_p=p}), it comprises a sum of diagonal terms with Dirac-like massive operators featuring a degree-dependent spectrum of masses, and sub- and super-diagonal kinetic contributions.

\newpar

The auxiliary part of the gauge-fixed action (\ref{gf_aux_action_p=p}) for the gauge (\ref{min-rank_gf_p=p})
 \begin{eqnarray}
  \begin{array}{llll}
   & S^{}_{\delta\,\text{aux}} [\CPhi]
    &\!\!=\!\!&
    \displaystyle
   \int \!d^dx\,\!
    \sum_{r=0}^{p-1}
    \sum_{k=1}^{p-r}
      \Bigl\{
      \big(
        \dcgC^{\gix{k-1}}_{ \ZZn{r+1}} P_{\sublead}^{\ZZn{r+1}\ZZm{r+1}} \gamma_{\Zm}
        +
        \dcgC^{\gix{k+1}}_{ \ZZn{r-1}} \gamma_{\Zn} P_{\sublead}^{\ZZn{r}\ZZm{r}} \big)
       \gB^{\gix{k}}_{ \ZZm{r}}
    \\
     &&&
      \hspace{30mm}
      \!+\,
      \dcgB^{\gix{k}}_{ \ZZn{r}}
      \big(
        \gamma_{\Zn} P_{\sublead}^{\ZZn{r+1}\ZZm{r+1}} \gC^{\gix{k-1}}_{ \ZZm{r+1}}
        +
        P_{\sublead}^{\ZZn{r}\ZZm{r}} \gamma_{\Zm} \gC^{\gix{k+1}}_{ \ZZm{r-1}}
      \big)
      \Bigr\}
      ,
  \end{array}
 \label{gf_aux_action_min-rank_p=p} 
 \end{eqnarray}
in the basis of irreducible fields, is expressed as
 \begin{equation}
 \hspace{-1ex}
  \begin{array}{@{\hspace{0pt}}llll}
   & S^{}_{\delta\,\text{aux}} [\CPhi]
   &\!\!=\!\!&
    \displaystyle
    \!\int \!d^dx \,\!
    \sum_{r=0}^{p-1}
    \sum_{k=1}^{p-r}
   \textstyle
    \textstyle
     \Bigl\{\,
       { \displaystyle \sum_{n=0}^{[r/2]} }
        \dcgCC^{\gix{k-1} \ZZn{r-2n}}_{\rix{r+1}}
        \anc[r+1]{r-2n}
        \gCB^{\gix{k}}_{\rix{r} \ZZn{r-2n}}
      +
       { \!\!\!\! \displaystyle \sum_{n=0}^{[(r{-}1)/2]} \!\!\!}
        \dcsc[r-1]{r} 
        \anc[r]{r-1-2n}
        \dcgCC^{\gix{k+1} \ZZn{r-1-2n}}_{\rix{r-1}}
        \gCB^{\gix{k}}_{\rix{r} \ZZn{r-1-2n}}
     \\
     &&&
      \hspace{25.5mm}
      +
    \textstyle
      { \displaystyle \sum_{n=0}^{[r/2]} }
       \dcsc[r]{r+1} 
       \anc[r+1]{r-2n}
        \dcgCB^{\gix{k} \ZZn{r-2n}}_{\rix{r}}
        \gCC^{\gix{k-1}}_{\rix{r+1} \ZZn{r-2n}}
      \!+
       { \!\!\!\! \displaystyle \sum_{n=0}^{[(r{-}1)/2]} \!\!\!}
        \dcgCB^{\gix{k} \ZZn{r-1-2n}}_{\rix{r}}
        \anc[r]{r-1-2n}
        \gCC^{\gix{k+1}}_{\rix{r-1} \ZZn{r-1-2n}}
     \Bigr\}
      ,
  \end{array}
 \label{gf_aux_action_min-rank_in_components_p=p} 
 \end{equation}
where the normalizing coefficients are
 \begin{eqnarray}
  \anc[r]{s}
   &\!=\!& \dcsc[r]{s}
       \msp^{\frac{r(r-1)}{2}-\frac{s(s-1)}{2} }
       \msp^{r-s} 
       \nfrac{(r{-}s)!\,s!}{r!} \nfrac{(d{-}2s)!}{(d{-}r{-}s)!}
       \,.
 \label{Sgh_in_components_p=p_coef} 
 \end{eqnarray}
For the choice (\ref{def_dcsc}), the relative sign factors are $\dcsc[r]{r+1}=\msp^{r}$.
 %
\emph{All} irreducible components  $\gCB{}{}, \dcgCB{}{}$ of \emph{all} Lagrange-multiplier fields $\gB{}{}, \dcgB{}{}$ are present on the right-hand side of (\ref{gf_aux_action_min-rank_in_components_p=p}) with non-vanishing numerical factors. Moreover, each irreducible field in (\ref{gf_aux_action_min-rank_in_components_p=p}) appears \emph{once}.

Together with the decomposed representation of the original action (\ref{So_in_components_p=p}), the expressions (\ref{Sgh_gf_min-rank_p=p}) and (\ref{gf_aux_action_min-rank_in_components_p=p}) constitute the representation of the BV gauge-fixed action (\ref{gf_action_p=p}) in the basis or the irreducible fields.

    \subsubsection*{Physical space and reduced gauge-fixed fields}
     \hspace{\parindent}
As noted in Section \ref{SSSect:Gauge-fixing_p=p}, with proper algebraic gauge fixing a large number of the fields which can be reduced in the action at the classical level. This includes all Lagrange-multiplier fields $\gB^{\gix{k}}_{\ZZm{r}}, \dcgB^{\gix{k}}_{\ZZm{r}}$,\, $1 \tleq k \tleq p{\,-\,}r$,\, and an equivalent number of degrees of freedom from ghosts and original fields, which are coupled to them in the gauge-fixed action. Based on the specific structure of the auxiliary part of the gauge-fixed action (\ref{gf_aux_action_min-rank_in_components_p=p}), this dual set is evident: all irreducible components of ghosts  $\gCC^{\gix{k}}_{\rix{r} \ZZm{s}}, \dcgCC^{\gix{k}}_{\rix{r} \ZZm{s}}$ of generations ($2 \tleq k \tleq p{\,-\,}r$) (referred to as \emph{extraghosts}), and the subleading-parity components of antighosts $\gCC^{\gix{1}}_{\rix{r} \ZZm{s}}, \dcgCC^{\gix{1}}_{\rix{r} \ZZm{s}}$,  and minimal fields $\gCC^\gix{0}_{\rix{r} \ZZm{s}}, \dcgCC^\gix{0}_{\rix{r} \ZZm{s}}$, $\CPsi_{\ZZm{s}}, \dcCPsi_{\ZZm{s}}$.

Due to the simplicity of the gauge-fixed auxiliary action in the irreducible representation (\ref{gf_aux_action_min-rank_in_components_p=p}), we can immediately identify the residual fields: the leading-parity components of antighosts $\gCC^\gix{1}_{\rix{r} \ZZm{s}}, \dcgCC^\gix{1}_{\rix{r} \ZZm{s}}$  and minimal fields $\gCC^\gix{0}_{\rix{r} \ZZm{s}}, \dcgCC^\gix{0}_{\rix{r} \ZZm{s}}$, $\CPsi_{\ZZm{s}}, \dcCPsi_{\ZZm{s}}$.
The auxiliary action $S^{}_{\delta\,\text{aux}}$ is evidently nondegenerate on the set of fields being reduced. The properness of the minimal-rank gauge fixing (\ref{min-rank_gf_p=p}), which ensure nondegeneracy of the gauge-fixed action, thus reduce to the nondegeneracy of the action at the reduction surface, which can be checked directly after reduction.\footnote{
 The residual properness condition can be encoded in rank conditions
  $\rank (\dc{X}^{\ZZn{p-1}\ZZl{p}} R_{\ZZl{p}}^{\;\ZZm{p-1}}) = \mmm_{p-1}$,\,
  $\rank (\dc{X}^{\ZZn{r}\ZZl{r+1}} Z_{\ZZl{r+1}}^{\;\ZZm{r}}) = \mmm_{r}$,\, ($0 \tleq r \tleq p-2$),\, and
  $ \rank ( P_{\lead} {}_{\ZZn{p}}^{\ZZk{p}} \gamma^{\ZZn{p} \Zr \ZZm{p}} D_{\Zr} P_{\lead} {}_{\ZZm{p}}^{\:\ZZl{p}} ) = \mmm_{p}$.
}

\newpar

For further comparison with the alternative full-rank gauge case, it is instructive to validate these conclusions from a perspective of solving the reduction equations (cf. \ref{BV_gf_constraints}):
 \begin{equation}
  \begin{array}{lllllll}
    \textstyle \vphantom{\big|^{|^|}}
    \frac{\var S_{\delta}}{\var{\gB^{\gix{k}}_{ \ZZm{r}}}}
     &\!\!=\!\!& 0
     \,:
    \\[1.3ex]
    \textstyle
    \frac{\var S_{\delta}}{\var{\dcgB^{\gix{k}}_{ \ZZm{r}}}}
     &\!\!=\!\!& 0
     \,:
  \end{array}
   \qquad
  \left\{
  \begin{array}{lllllll}
    \displaystyle
        \dcgC^{\gix{k-1}}_{ \ZZn{r+1}} P_{\sublead}^{\ZZn{r+1}\ZZm{r+1}} \gamma_{\Zm}
        +
        \dcgC^{\gix{k+1}}_{ \ZZn{r-1}} P_{\lead}^{\ZZn{r-1}\ZZm{r-1}} \gamma^{\Zm}   
     &=& 0
    \,,
    \\[1.5ex]
    \displaystyle
        \gamma_{\Zn} P_{\sublead}^{\ZZn{r+1}\ZZm{r+1}} \gC^{\gix{k-1}}_{ \ZZm{r+1}}
        +
        \gamma^{\Zn} P_{\lead}^{\ZZn{r-1}\ZZm{r-1}} \gC^{\gix{k+1}}_{ \ZZm{r-1}}
     &=& 0
     \,.
     \phantom{\big|_{|}}
  \end{array}
  \right.
 \label{gf_constraints_min-rank_p=p}
 \end{equation}
We mixed the two representations (\ref{min-rank_gf_p=p}) to enable projectors to act on fields directly. The upper equations for the Dirac-conjugated ghost fields decouple from the lower subsystem for non-conjugated fields. As discussed in Section~\ref{Sect:BVst}, these equations are solved chainwise, as only ghosts from each chain $r{\,+\,}k = const$ from Fig.~\ref{BVp_Triangular_diagram} interact in (\ref{gf_constraints_min-rank_p=p}).


\def\gfChainSF
{
\hspace{-30pt}
\begin{picture}(220,250)
 \put(194.5,243){\line(+3,-4){8}}
 \put(194.8,243){\line(+3,-4){8}}
\put(201,222){$\gC^{\gix{0}}_{ \ZZn{\ro}}$}
\put(180,190){\vector(3,4){15}} 
\put(201,218){\line(-3,-4){15}}
 \put(214,217.3){\line(+3,-4){9}} 
 \put(214.3,217.3){\line(+3,-4){9}} 
%
\put(142,181){$\dcgB^{\gix{1}}_{ \ZZn{\ro-1}}$,\,$\gC^{\gix{1}}_{ \ZZn{\ro-1}}$}
%
%
%
\put(150,150){\vector(3,4){17}}
\put(171,178){\vector(-3,-4){17}}
\put(110,140){$\dcgB^{\gix{2}}_{ \ZZn{\ro-2}}$,\,$\gC^{\gix{2}}_{ \ZZn{\ro-2}}$}
%
\put(106,92){\vector(3,4){30.5}}
\put(140.5,138){\vector(-3,-4){30.5}}
   {\color{white}
    \linethickness{14pt}
    \put(100,116){\line(1,0){100}}
    \linethickness{1pt}
   }
\put(117,115){$\ldots$}
\if{ 
\put(110,101){$\gC^{\gix{\ro-2}}_{ \ZZn{2}}$} 
\put(93,74){\vector(3,4){15}}
\put(111,98){\vector(-3,-4){15}}
%
\put(81,62){$\dcgB^{\gix{\ro-1}}_{\Zn}$\!\!,\,$\gC^{\gix{\ro-1}}_{\Zn}$}
\put(63,34){\vector(3,4){15}}
\put(81,58){\vector(-3,-4){15}}
%
\put(54,23){$\dcgB^{\gix{\ro}}_{\Zo}$\!,\,$\gC^{\gix{\ro}}_{\Zo}$}
}\fi
\put(75,81){$\dcgB^{\gix{\ro-2}}_{\ZZn{2}}$\!\!,\,$\gC^{\gix{\ro-2}}_{\ZZn{2}}$}
\put(78,54){\vector(3,4){15}}
\put(96,78){\vector(-3,-4){15}}
%
\put(46,42){$\dcgB^{\gix{\ro-1}}_{\Zn}$\!\!,\,$\gC^{\gix{\ro-1}}_{\Zn}$}
\put(48,14){\vector(3,4){15}}
\put(66,38){\vector(-3,-4){15}}
%
\put(27,3){$\dcgB^{\gix{\ro}}_{\Zo}$\!,\,$\gC^{\gix{\ro}}_{\Zo}$}
%
\end{picture}
}


\def\gfSubchainsSF
{
\hspace{-30pt}
\begin{picture}(220,250)
%
%
 \put(239.5,243){\line(+3,-4){8}}
 \put(239.8,243){\line(+3,-4){8}}
\put(246,222){$\gC^{\gix{0}}_{ \ZZn{\ro}}$}
\put(224,190){\vector(3,4){15}} 
\put(224,190){\line(3,4){22}} 
 \put(259,217.3){\line(+3,-4){9}} 
 \put(259.3,217.3){\line(+3,-4){9}} 
  \put(219,206.5){\MgfO{ \hspace{-10mm} X^{\ZZn{\ro-1}\ZZn{\ro}} }}
%
\put(212,181){$\dcgB^{\gix{1}}_{ \ZZn{\ro-1}}$}
%
%
  \put(204,162){\MgfS{ \hspace{-20mm} \dc{X}^{\ZZn{\ro-2}\ZZn{\ro-1}} }}
%
\put(195,150){\line(3,4){17}}
\put(216,178){\vector(-3,-4){17}}
\put(183,141){$\gC^{\gix{2}}_{ \ZZn{\ro-2}}$}
\put(151,92){\vector(3,4){30.7}}
\put(185.5,138){\line(-3,-4){30.5}}
   {\color{white}
    \linethickness{18pt}
    \put(155,117){\line(1,0){25}}
    \put(145,99){\line(1,0){25}}
    \linethickness{1pt}
   }
%
%
  \if{
     \put(194.5,243){\line(+3,-4){8}}
     \put(194.8,243){\line(+3,-4){8}}
    \put(201,222){$\gC^{\gix{0}}_{ \ZZn{\ro}}$}
    \put(180,190){\vector(3,4){15}} 
    \put(201,218){\line(-3,-4){15}}
     \put(214,217.3){\line(+3,-4){9}} 
     \put(214.3,217.3){\line(+3,-4){9}} 
      \put(175,204){\MgfO{ \hspace{-10mm} X^{\ZZn{\ro-1}\ZZn{\ro}} }}
    %
    \put(167,181){$\dcgB^{\gix{1}}_{ \ZZn{\ro-1}}$}
    %
    %
      \put(156,160){\MgfS{ \hspace{-20mm} \dc{X}^{\ZZn{\ro-2}\ZZn{\ro-1}} }}
    %
    \put(150,150){\line(3,4){17}}
    \put(171,178){\vector(-3,-4){17}}
    \put(138,141){$\gC^{\gix{2}}_{ \ZZn{\ro-2}}$}
  }\fi
\put(106,92){\vector(3,4){30.5}}
\put(140.5,138){\line(-3,-4){30.5}}
   {\color{white}
    \linethickness{18pt}
    \put(100,114){\line(1,0){40}}
    \put(115,132){\line(1,0){40}}
    \linethickness{1pt}
   }
%
%
\put(94,81){$\dcgB^{\gix{\ro-2}}_{ \ZZn{2}}$} 
\put(78,54){\line(3,4){15}}
\put(96,78){\vector(-3,-4){15}}
  \put(74,66){\MgfO{\hspace{-5mm} \dc{X}^{\Zn\ZZn{2}} }}
\put(66,42){$\gC^{\gix{\ro-1}}_{ \Zn}$}
\put(48,14){\vector(3,4){15}}
\put(66,38){\line(-3,-4){15}}
  \put(44,26){\MgfS{\hspace{-1mm} X^{\Zn} }}
\put(39,3){$\dcgB^{\gix{\ro}}_{\Zo}$}
%
%
\put(196,92){\vector(3,4){30.5}}
\put(230.5,138){\vector(-3,-4){30.5}}
   {\color{white}
    \linethickness{18pt}
    \put(190,114){\line(1,0){30}}
    \linethickness{18pt}
    \put(216,129){\line(1,0){30}}
    \linethickness{1pt}
   }
%
%
\put(184,81){$\gC^{\gix{\ro-2}}_{ \ZZn{2}}$} 
\put(168,54){\vector(3,4){15}}
\put(186,78){\line(-3,-4){15}}
  \put(164,66){\MgfO{\hspace{-5mm} X^{\Zn\ZZn{2}} }}
\put(155,42){$\dcgB^{\gix{\ro-1}}_{ \Zn}$}
\put(138,14){\line(3,4){15}}
\put(156,38){\vector(-3,-4){15}}
  \put(134,26){\MgfS{\hspace{-1mm} \dc{X}^{\Zn} }}
\put(129,3){$\gC^{\gix{\ro}}_{\Zo}$}
%
%
%
\put(243,181){$\gC^{\gix{1}}_{ \ZZn{\ro-1}}$}
  \put(289,157){\MgfS{ \hspace{-20mm} X^{\ZZn{\ro-2}\ZZn{\ro-1}} }}
\put(221,150){\vector(3,4){17}}
\put(242,178){\line(-3,-4){17}}
%
%
\put(177,92){\line(3,4){30.5}}
\put(211.5,138){\vector(-3,-4){30.5}}
   {\color{white}
    \linethickness{18pt}
    \put(180,117){\line(1,0){25}}
    \put(168,100){\line(1,0){25}}
    \linethickness{1pt}
   }
%
%
%
%
\if{
    %
    %
%
\put(198,181){$\gC^{\gix{1}}_{ \ZZn{\ro-1}}$}
%
%
  \put(246,158){\MgfS{ \hspace{-20mm} X^{\ZZn{\ro-2}\ZZn{\ro-1}} }}
%
\put(176,150){\vector(3,4){17}}
\put(197,178){\line(-3,-4){17}}
\put(167,140){$\dcgB^{\gix{2}}_{ \ZZn{\ro-2}}$}
%
}\fi
%
%
\put(132,92){\line(3,4){30.5}}
\put(166.5,138){\vector(-3,-4){30.5}}
   {\color{white}
    \linethickness{18pt}
    \put(132,114){\line(1,0){25}}
    \put(145,132){\line(1,0){25}}
    \linethickness{1pt}
   }
%
%
\put(121,81){$\gC^{\gix{\ro-2}}_{ \ZZn{2}}$} 
\put(104,54){\vector(3,4){15}}
\put(122,78){\line(-3,-4){15}}
  \put(132,64){\MgfO{\hspace{-5mm} X^{\Zn\ZZn{2}} }}
\put(92,42){$\dcgB^{\gix{\ro-1}}_{ \Zn}$}
\put(74,14){\line(3,4){15}}
\put(92,38){\vector(-3,-4){15}}
  \put(91,24){\MgfS{\hspace{-1mm} \dc{X}^{\Zn} }}
\put(65,3){$\gC^{\gix{\ro}}_{\Zo}$}
%
%
\put(222,92){\vector(3,4){30.5}}
\put(256.5,138){\line(-3,-4){30.5}}
   {\color{white}
    \linethickness{18pt}
    \put(230,114){\line(1,0){60}}
    \put(230,132){\line(1,0){60}}
    \linethickness{1pt}
   }
%
%
\put(211,81){$\dcgB^{\gix{\ro-2}}_{ \ZZn{2}}$} 
\put(194,54){\line(3,4){15}}
\put(212,78){\vector(-3,-4){15}}
  \put(222,64){\MgfO{\hspace{-5mm} \dc{X}^{\Zn\ZZn{2}} }}
\put(182,42){$\gC^{\gix{\ro-1}}_{ \Zn}$}
\put(164,14){\vector(3,4){15}}
\put(182,38){\line(-3,-4){15}}
  \put(181,24){\MgfS{\hspace{-1mm} X^{\Zn} }}
\put(155,3){$\dcgB^{\gix{\ro}}_{\Zo}$}
%
%
 \put(212,140){$\dcgB^{\gix{2}}_{ \ZZn{\ro-2}}$}
\if{
 \put(117,115){$\ldots$}
 \put(142.5,115){$\ldots$}
 \put(162.5,115){$\ldots$}
 \put(189,115){$\ldots$}
 \put(207,115){$\ldots$}
 \put(233,115){$\ldots$}
}\fi
\if{
 \put(130,115){$\ldots$}
 \put(176,115){$\ldots$}
 \put(220,115){$\ldots$}
}\fi
 \put(125,110){$\ldots$}
 \put(181,120){$\ldots$}
 \put(215,110){$\ldots$}
 \put(114,114){\text{\footnotesize(\,$\ro$\, odd\,)}}
 \put(204,114){\text{\footnotesize(\,$\ro$\, even\,)}}
%
%
    \put(44,107.5){\line(-1,0){15}}
    \put(44,113){\line(-1,0){15}}
    \put(49,110.5){\line(-3,2){10}}
    \put(49,110.5){\line(-3,-2){10}}
\end{picture}
}

\begin{figure}[htbp]
  \centering
  \small
  \gfChainSF
  \gfSubchainsSF
  \caption{\normalsize \hspace{24ex} Splitting of a gauge-fixing chain
  \\ \phantom{.} \\
  \small \it
   A gauge-fixing chain, which originates from the minimal ghost $\gC^{\gix{0}}_{\ZZn{\ro}}$ and ends at $\gC^{\gix{\ro}}$. The chain splits into two subchains of equations relating ghosts within  $\gC^{\gix{\ro-2n}}_{\ZZn{2n}}$ and $\gC^{\gix{\ro-1-2n}}_{\ZZn{2n+1}}$ families. The configuration of the subchains' endings depends on the parity of the chain's length.
  }
 \label{fig:gf_chain_p=p} 
\end{figure}

\newpar


Consider a chain from the lower subsystem for non-conjugated fields, Fig.~\ref{fig:gf_chain_p=p}, with the $\gC^{\gix{0}}_{\ZZm{\ro}}$ as the root and $\gC^{\gix{\ro}}_{\Zo}$ at the other end.
The equations couple every other fields, such that the chain splits into two subchains, which couple fields of the same ghost number (with degrees of the same parity)
 \begin{equation}
  \begin{array}{l}
    \gC^{\gix{\ro}}_{ \Zo},\,\quad \gC^{\gix{\ro-2}}_{ \ZZm{2}},\,\quad \gC^{\gix{\ro-4}}_{ \ZZm{4}}\,, \,\quad \ldots
    \,;
    \\[0.5ex]
    \gC^{\gix{\ro-1}}_{ \Zm},\,\quad \gC^{\gix{\ro-3}}_{ \ZZm{3}},\,\quad \gC^{\gix{\ro-5}}_{ \ZZm{5}}\,,\,\quad \ldots
    \,.
    \\
  \end{array}
 \label{subchains_p=p} 
 \end{equation}
In each chain, the fields have unique generations $k$ depending on their degrees $r$: $k=\ro-r$, allowing us to omit the generation index temporarily.

The iterative solution starts from the end (bottom) of each subchain, solving an equation involving $\gC^{\gix{\ro}}_{\Zo}$ or $\gC^{\gix{\ro-1}}_{\Zm}$, and finishes with solving equation involving the antighost  $\gC^{\gix{1}}_{\ZZm{\ro-1}}$ or the minimal ghost $\gC^{\gix{0}}_{\ZZm{\ro}}$, depending on the parity of the chain's length. Thus, in the upper subchain (\ref{subchains_p=p}), the first equation to solve is
 $
   \gamma_{\Zn} P_{\sublead}^{\ZZn{2} \ZZm{2}} \gC^{\gix{{\HIDE{\ro-2}}}}_{ \ZZm{2}}
        +
   \gamma^{\Zn} \gC^{\gix{{\HIDE{\ro}}}}_{\Zo}
    = 0
 $,\,
which is equivalent to
 $
  \HIDE{+} 
  \nfrac{d-2}{2} \gCC_{\rix{2}}^{\gix{{\HIDE{\ro-2}}}\Zn}
        +
   \gamma^{\Zn} \gCC^{\gix{{\HIDE{\ro}}}}_{\rix{0} \Zo}
    = 0
 $.
Since irreducible spinorial fields are coefficients of different matrix basis elements, they both vanish.
The first equation in the second subchain is
 $
   \gamma_{\Zn} P_{\sublead}^{\ZZn{1} \ZZm{1}} \gC^{\gix{{\HIDE{\ro-1}}}}_{ \ZZm{1}}
    \equiv
    - 
    d \,\gCC_{\rix{1}}^{\gix{{\HIDE{\ro-1}}}\Zo}
    = 0
 $,\,
which force the lower-degree field to vanish.

Each field  $\gC^{\hgix{{\ro-r}}}_{ \ZZm{r}}$ in either subchain, except those at the extremes, enters the subchain's subsystem of equations twice: as a lower-degree field\HIDE{ with leading-parity projector} and as a higher-degree field\HIDE{ with subleading parity projector}
 \begin{equation}
  \begin{array}{|@{\hspace{12pt}}ll}
    \hspace{9em}\ldots \hspace{14em}\ldots
   \\[-1.2ex]
   \displaystyle
    \gamma_{\Zn} P_{\sublead}^{\ZZn{r+2}\ZZm{r+2}} \gC^{\hgix{\ro-r-2}}_{\ZZm{r+2}}
        +
    \gamma^{\Zn} P_{\lead}^{\ZZn{r}\ZZm{r}} \uwave{ \gC^{\hgix{\ro-r}}_{\ZZm{r}} }
    =
    \sum_{n=0}^{[(r+1)/2]} \!\!\!
    d_{\rix{r+2}}^{\rix{r+1-2n}}
    \gamma^{\ZZn{2n}}
    \gCC^{\hgix{\ro-r-2} \ZZn{r+1-2n}}_{\rix{r+2}}
        +
    \sum_{n=0}^{[r/2]} \;
    \gamma^{\ZZn{2n+1}}
    \uwave{ \gCC^{\hgix{\ro-r} \ZZn{r-2n}}_{\rix{r}}{}^{} }
    = 0
    \,,
   \\[0.3em]
   \displaystyle
    \gamma_{\Zn} P_{\sublead}^{\ZZn{r}\ZZm{r}} \uwave{ \gC^{\hgix{\ro-r}}_{ \ZZm{r}}}
        +
    \gamma^{\Zn} P_{\lead}^{\ZZn{r-2}\ZZm{r-2}} \gC^{\hgix{\ro-r+2}}_{ \ZZm{r-2}}
    =
   \sum_{n=0}^{[(r-1)/2]} \!\!\!
    d_{\rix{r}}^{\rix{r-1-2n}}
    \gamma^{\ZZn{2n}}
    \uwave{ \gCC^{\hgix{\ro-r} \ZZn{r-1-2n}}_{\rix{r}}}
        +
   \sum_{n=0}^{[(r-2)/2]} \!\!\!
    \gamma^{\ZZn{2n+1}}
     \gCC^{\hgix{\ro-r+2} \ZZn{r-2-2n}}_{\rix{r-2}}
    = 0
    \,,
    \\[-0.9ex]
    \hspace{9em}\ldots \hspace{14em}\ldots
  \end{array}
 \end{equation}
where the numerical constants\,
  $d_{\rix{r}}^{\rix{s}} =
  \msp^{r} 
  \frac{r-s}{r} (d{-}r{+}s{+}1)$.
The lower equation forces all subleading-parity irreducible components of $\gC^{\hgix{{\ro-r}}}_{ \ZZm{r}}$ to vanish, whereas the upper equation forces to vanish the leading-parity components of this field. Solving the subchain equations iteratively from the lowest degrees till the equation for fields of degrees $r$ and $r{\,+\,}2$ (before solving it), one gets that the $\gC^{\hgix{\ro-r+2n}}_{\ZZm{r-2n}}$ fields for $n{\,>\,}0$ are forced to vanish, whereas for  $ \gC^{\hgix{\ro-r}}_{\ZZm{r}}$ only the subleading-parity components are constrained to zero: $P_{\sublead}^{\ZZn{r}\ZZm{r}} \gC^{\hgix{{\ro-r}}}_{\ZZm{r}} = 0$. By resolving other equations from both subchains, one gets the analogous intermediate results.

One subchain originates from the antighost  $\gC^{\gix{1}}_{ \ZZm{\ro-1}}$, while the other subchain --- from the minimal ghost $\gC^{\gix{0}}_{ \ZZm{\ro}}$, and these fields enter only the final equation within each subchain:
 \begin{equation}
  \begin{array}{l}
   \displaystyle
    \gamma_{\Zn} P_{\sublead}^{\ZZn{\ro-1}\ZZm{\ro-1}} \dashuline{ \gC^{\gix{1}}_{ \ZZm{\ro-1}}}
        +
    \gamma^{\Zn} P_{\lead}^{\ZZn{\ro-3}\ZZm{\ro-3}} \gC^{\gix{3}}_{ \ZZm{\ro-3}}
    = 0
    \,,
   \\[0.3em]
    \gamma_{\Zn} P_{\sublead}^{\ZZn{\ro}\ZZm{\ro}} \dotuline{ \gC^{\gix{0}}_{ \ZZm{\ro}}}
        +
    \gamma^{\Zn} P_{\lead}^{\ZZn{\ro-2}\ZZm{\ro-2}} \gC^{\gix{2}}_{ \ZZm{\ro-2}}
    = 0 \,.
    \\[-0.3em]
  \end{array}
 \label{final_Eqs_in_chain} 
 \end{equation}
Solutions of (\ref{final_Eqs_in_chain}) complete the reduction within a chain, resulting in all extraghosts from both subchains being constrained to zero, while the roots of the subchains --- the minimal ghost or original field and the antighost --- are constrained to zero only in their subleading-parity irreducible components. Their leading-parity components --- $P_{\lead}^{\ZZn{\ro}\ZZm{\ro}} \gC^{\gix{0}}_{\ZZm{\ro}}$,\; $P_{\lead}^{\ZZn{\ro-1}\ZZm{\ro-1}} \gC^{\gix{1}}_{\ZZm{\ro-1}}$ --- are not fixed by the reduction.

By aggregating the results from all gauge-fixing chains of fields, the solution of the reduction constraints (\ref{gf_constraints_min-rank_p=p}) for the minimal-rank gauge fixing (\ref{min-rank_gf_p=p}),
 \begin{equation}
  \begin{array}{ll}
   \hspace{2.5em} \text{Fixed and reduced}
    \\[0.2em]
   \hline \vphantom{\Big|}
     P_{\sublead} {}_{\ZZn{p}}^{\:\ZZm{p}} \psi_{\ZZm{p}}
     \,,\: 
     \dcpsi_{\ZZm{p}} P_{\sublead} {}_{\,\ZZn{p}}^{\ZZm{p}}
     \:=\: 0\,,
    \\[0.6em]
     P_{\sublead} {}_{\ZZn{r}}^{\:\ZZm{r}} \gC^\gix{0}_{\ZZm{r}}
     \,,\: 
     \dcgC^\gix{0}_{\ZZm{r}} P_{\sublead} {}_{\,\ZZn{r}}^{\ZZm{r}}
     \:=\:0\,,
    \\[0.6em]
     P_{\sublead} {}_{\ZZn{r}}^{\:\ZZm{r}} \gC^\gix{1}_{\ZZm{r}}
     \,,\: 
     \dcgC^\gix{1}_{\ZZm{r}} P_{\sublead} {}_{\,\ZZn{r}}^{\ZZm{r}}
     \:=\:0\,,
    \\[0.5em]
     \gC^\gix{k}_{\ZZn{r}}
     \,,\: 
     \dcgC^\gix{k}_{\ZZn{r}}
     \:=\: 0\,,
     \\[0.3em]
     \hline
   \end{array}
   \hspace{-2ex}
  \begin{array}{cc}
   & \vphantom{\text{F}}
    \\[0.2em]
   \vphantom{\Big|}
     & \text{\small $\begin{array}{l}  (\,k=0 ;\; r=p\,)\,,\end{array}$}
  \if{
    \\[0.6em]
     &  \text{\small $\begin{array}{l} (\,k = 0,\,1 ;\; 0\leq r \leq p{-}1\,)\end{array}$}
  }\fi
    \\[0.6em]
     &  \text{\small $\begin{array}{l} (\,k = 0 ;\; 0\leq r \leq p{-}1\,)\end{array}$}
    \\[0.6em]
     &  \text{\small $\begin{array}{l} (\,k = 1 ;\; 0\leq r \leq p{-}1\,)\end{array}$}
    \\[0.5em]
     & \text{\small $\begin{array}{l} (\,2\leq k < p{-}r\,;\; 0\leq r \leq p{-}2\,)\end{array}$}
     \\[0.3em]
 \if{
     \\[-0.95em]
     & \text{\small $\begin{array}{l} (\,1 \leq k \leq 2\,;\; 0 \leq r \leq p{-}1\,)\end{array}$}   \!\!\!\!
       \\[0.5em]
     & \text{\small $\begin{array}{l} (\,1 \leq k \leq 2\,;\; 0 \leq r \leq p{-}1\,)\end{array}$}   \!\!\!\!
       \\[0.5em]
     & \text{\small $\begin{array}{l} (\,3 \leq k \leq p{-}r\,;\; 0 \leq r \leq p{-}3\,)\end{array}$}    \!\!\!\!
     \\
 }\fi
   \end{array}
    \hspace{0ex}
   \begin{array}{ll}
   \hspace{3.5em} \text{Free} \; ( \breve{\CPhi} )
    \\[0.2em]
   \hline \vphantom{\Big|}
     P_{\lead} {}_{\ZZn{p}}^{\:\ZZm{p}} \psi_{\ZZm{p}}
     \,,\,\;
     \dcpsi_{\ZZm{p}} P_{\lead} {}_{\,\ZZn{p}}^{\ZZm{p}}
     \,,
    \\[0.6em]
     P_{\lead} {}_{\ZZn{r}}^{\:\ZZm{r}} \gC^\gix{0}_{\ZZm{r}}
     \,,\,\;
     \dcgC^\gix{0}_{\ZZm{r}} P_{\lead} {}_{\,\ZZn{r}}^{\ZZm{r}}
     \,,
    \\[0.6em]
     P_{\lead} {}_{\ZZn{r}}^{\:\ZZm{r}} \gC^\gix{1}_{\ZZm{r}}
     \,,\,\;
     \dcgC^\gix{1}_{\ZZm{r}} P_{\lead} {}_{\,\ZZn{r}}^{\ZZm{r}}
     \,,
    \\[0.5em]
    \\[0.3em]
     \hline
 \if{
    \\[-0.95em]
    \\[0.5em]
    \\[0.5em]
     \\
 }\fi
    \end{array}
 \label{gf_min-rank_reduction_surface} 
 \end{equation}
confirmes the result observed from the action structure: all irreducible components of ghosts and original fields from the column vanish.
Ultimately, the reduction surface is parameterized by the irreducible components of the original fields, minimal ghosts, and antighosts, with leading-parity degrees.


  \subsection{Full-rank gauge fixing and physical space}
   \label{SSect:Gauge_fixing_full_p=p}
    \hspace{\parindent}
A natural covariant choice for the algebraic full-rank gauge fixing is
 \begin{equation} 
  \begin{array}{lllllll}
    & 
      X^{\ZZm{r}\ZZn{r+1}}
    \;=\;  
    \gamma_{\Zm} \GG^{\ZZm{r+1}\ZZn{r+1}}
    \,,
    \quad
    & \dc{X}^{\ZZn{r+1}\ZZm{r}}
    \;=\; \GG^{\ZZn{r+1}\ZZm{r+1}} \gamma_{\Zm}
    \,,
  \end{array}
 \label{full-rank_gf_p=p} 
 \end{equation}
where the matrix operator $ \GG^{\ZZm{r+1}\ZZn{r+1}} $ (\ref{def_index_operators}), covariantly raises indices of antisymmetrized groups, and the index $\Zm$ of the gamma matrices contracts according to the convention (\ref{def_contraction_order_convention}). These are the full-rank matrices with $\rank X^{\ZZm{r}\ZZn{r+1}} = \rank \dc{X}^{\ZZn{r+1}\ZZm{r}} = \binom{d}{r}$.\footnote{As before, rank counts the tensorial rank of the objects, each component of which is an endomorphism in the space of spinors.
}

The gauge fermion (\ref{gauge_fermion_gen_p=p}) is then expressed as
 \begin{equation}
  \begin{array}{lllrrl}
   \varPsi[\varPhi]
   &\!=\!&
    \displaystyle
   \dcgC^{\gix{1}}{}^{\ZZm{p-1}}
    \gamma^{\Zm} \psi_{\ZZm{p}}
    +
    \dcpsi^{\ZZm{p}}
    \gamma_{\Zm} \gC^{\gix{1}}_{ \ZZm{p-1}}
    +
     \sum_{r=0}^{p-2}
     \sum_{k=1}^{p-r}
    \big(
    \dcgC^{\gix{k}}{}^{\ZZm{r}} \gamma^{\Zm} \gC^{\gix{k-1}}_{ \ZZm{r+1}}
    +
    \dcgC^{\gix{k-1}}{}^{\ZZm{r+1}} \gamma_{\Zm} \gC^{\gix{k}}_{ \ZZm{r}}
    \big)
   \,,
  \end{array}
 \label{gaugeFermion_full-rank_p=p} 
 \end{equation}
where the indices of Dirac-conjugated fields are raised by the spacetime metric.

\newpar

The ghost part (\ref{gf_gh action_p=p}) of the gauge-fixed minimal action for the full-rank gauge (\ref{full-rank_gf_p=p}),
 \begin{eqnarray}
  \begin{array}{lllllll}
   & S^{}_{\delta\,\text{gh}} [\varPhi]
    & \!\!=\!\! &   \displaystyle
    \!\!\int \!d^dx\,
       \,i \sum_{r=0}^{p-1}  
      \big( \dcgC^{\gix{1} \ZZm{r}} \gamma^\Zm
           D_{\Zm} \gC^{\gix{0}}_{ \ZZm{r}}
         {\,-\,}
          \dcgC^{\gix{0} \ZZm{r}} \dc{D}^{\Zm}
            \gamma_{\Zm} \gC^{\gix{1}}_{ \ZZm{r}}
      \big)
     \,,
  \end{array}
 \label{gf_gh action_full-rank_p=p} 
 \end{eqnarray}
when expressed in the basis of irreducible fields (\ref{decomp_p=p}), is given by
 \begin{eqnarray}
 \hspace{-1.0ex}
  \begin{array}{@{\hspace{2pt}}l@{\hspace{2pt}}llllll}
   & S^{}_{\delta\,\text{gh}} [\CPhi]
  \if{
    &\!\!=\!\!&
    \displaystyle
    \!\!\int \!d^dx\,
        i \sum_{r=0}^{p-1}  
      \big( \dcgC^{\gix{1}}_{ \ZZn{r}}\gamma_{\Zn} P_{\sublead}^{\ZZn{r+1}\ZZm{r+1}} D_{\Zm} \gC^{\gix{0}}_{ \ZZm{r}}
      -
       \dcgC^{\gix{0}}_{ \ZZn{r}}\dc{D}_{\Zn} P_{\sublead}^{\ZZn{r+1}\ZZm{r+1}} \gamma_{\Zm} \gC^{\gix{1}}_{ \ZZm{r}} \big)
     \,,
    \\[.3em]
  }\fi
    &\!\!\!=\!\!\!&
    \displaystyle
    \!\!\int \!d^dx\,
        i \sum_{r=0}^{p-1} 
          \hspace{-1pt} \sum_{s=0}^{r}
         \cnc[r]{s}
     \big(
       \dcgCC^{\gix{1} \ZZm{s}}_{\rix{r}}
         \big( \slashed{\nabla}
              {\,-\,} i  
              \mAdS  
              \msp^{r-s} (d{-}2s)
         \big)
       \gCC^{\gix{0}}_{\rix{r} \ZZm{s}}
      \!-
       \dcgCC^{\gix{0} \ZZm{s}}_{\rix{r}}
         \big( \dc{\slashed{\nabla}}
              {\,+\,} i 
              \mAdS  
              \msp^{r-s} (d{-}2s)
         \big)
       \gCC^{\gix{1}}_{\rix{r} \ZZm{s}}
     \big)
   \hspace{-5ex}
    \\[.3em] 
     &&\!\!\!\!&
   \displaystyle
    - 
    \!\!\int \!d^dx\,
        i  \sum_{r=0}^{p-1}  
          \sum_{s={\color{black}1}}^{{\color{black}r}} 
    \textstyle
       \cnc[r]{s}
       \nfrac{(r{-}s{+}2)(d{-}2s)}{(d{-}r{-}s)} 
      \big(
       \dcgCC^{\gix{1} \ZZm{s}}_{\rix{r}}
        {\nabla}^\Zm
       \gCC^{\gix{0}}_{\rix{r} \ZZm{s+1}}
       \,-
       \dcsc[s-1]{s}
       \dcgCC^{\gix{0} \Zr\ZZm{s+1}}_{\rix{r}} \dc{\nabla}_{\!\Zm} \gCC^{\gix{1}}_{\rix{r} \ZZm{s}}
      \big)
    \\[.3em]
     &&\!\!\!\!&
    \displaystyle
    + 
     \!\!\int \!d^dx\,
        i \sum_{r=0}^{p-1}  
          \sum_{s={\color{black}1}}^{{\color{black}r}} 
         \cnc[r]{s} (d{-}2s) 
     \big(
       \dcgCC^{\gix{1} \ZZm{s}}_{\rix{r}}
        {\nabla}_{\!\Zm}
       \gCC^{\gix{0}}_{\rix{r} \ZZm{s-1}}
      -
       \dcsc[s]{s-1}
       \dcgCC^{\gix{0} \ZZm{s-1}}_{\rix{r}}
        \,\dc{\nabla}^\Zm
       \gCC^{\gix{1}}_{\rix{r} \ZZm{s}}
       \big)
     \,,
    \\[.3em]
  \end{array}
 \label{Sgh_gf_full-rank_p=p} 
 \end{eqnarray}
where $\mAdS \equiv \pm \frac{1}{2}\, \sqrr$.
The overall normalizing coefficients are
 \begin{equation}
  \cnc[r]{s}
  \:=\:
   \dcsc[r]{s}
   \msp^{ \frac{r(r{\color{black}-}1)}{2} - \frac{s(s{\color{black}-}1)}{2}}
   \msp^{r}\, 
   \nfrac{s!\,(r{-}s{+}1)!}{(r{+}1)!}  \nfrac{(d{-}2s{-}1)!}{(d{-}r{-}s{-}1)!}
   \,,
 \label{Sgh_gf_full-rank_p=p_coef}
 \end{equation}
where $\dcsc[r]{s}$ are the sign factors from the definitions of the Dirac-conjugated irreducible fields (\ref{decomp_p=p}). For the choice (\ref{def_dcsc}), the relative coefficients read
 $
  \dcsc[s]{s-1}
  \defeq {\dcsc[r]{s-1}}/{\dcsc[r]{s}}
  = \dcsc[s-1]{s} = \msp^{s-1}
 $.\,

Similar to the irreducible representation of the original action (\ref{So_in_components_p=p}), the ghost part of the minimal action comprises a sum of diagonal terms with a Dirac-like massive operator and sub- and super-diagonal kinetic contributions. Moreover, the degree-dependent spectrum of masses coincides with that of the original fields --- the absolute mass value has the same degree-dependent coefficient $d{\,-\,}2s$ for degree-$s$ irreducible fields.

\newpar

The gauge-fixed auxiliary part (\ref{gf_aux_action_p=p}) of the action for the full-rank gauge (\ref{full-rank_gf_p=p}),
 \begin{eqnarray}
  \begin{array}{llll}
   & S^{}_{\delta\,\text{aux}} [\varPhi]
    &\!\!=\!\!&
    \displaystyle
   \!\!\int \!d^dx\,
    \sum_{r=0}^{p-1} \sum_{k=1}^{p-r} \!
      \Big(
        \dcgC^{\gix{k-1}}{}^{\ZZm{r+1}} \gamma_{\Zm} \gB^{\gix{k}}_{ \ZZm{r}}
        +
        \dcgC^{\gix{k+1}}{}^{\ZZm{r-1}} \gamma^{\Zm} \gB^{\gix{k}}_{ \ZZm{r}}
    \\
     &&& \displaystyle
      \hspace{32mm}
      \!+
        \dcgB^{\gix{k}}{}^{\ZZm{r}} \gamma^{\Zm} \gC^{\gix{k-1}}_{ \ZZm{r+1}}
        +
        \dcgB^{\gix{k}}{}^{\ZZm{r}}  \gamma_{\Zm} \gC^{\gix{k+1}}_{ \ZZm{r-1}}
     \Big)
    \,,
  \end{array}
 \label{gfAction_aux_full-rank_p=p}
 \end{eqnarray}
when expressed in terms of irreducible fields, is given by
 \begin{eqnarray}
  \begin{array}{llll}
   & S^{}_{\delta\,\text{aux}} [\CPhi]
    &\!\!=\!\!&
    \displaystyle
   \!\!\int \!d^dx\,
    \sum_{r=0}^{p-1} \sum_{k=1}^{p-r} \!
      \Big(
        \sum_{s=0}^{r}
        \anc[r+1]{s}
        \dcgCC^{\gix{k-1} \ZZm{s}}_{\rix{r+1}} \gCB^{\gix{k}}_{\rix{r} \ZZm{s}}
        +
        \sum_{s=0}^{r-1}
        \dcsc[r-1]{r} 
        \anc[r]{s}
        \dcgCC^{\gix{k+1} \ZZm{s}}_{\rix{r-1}}  \gCB^{\gix{k}}_{\rix{r} \ZZm{s}}
    \\
     &&& \displaystyle
      \hspace{32mm}
      \!+
        \sum_{s=0}^{r}
        \dcsc[r]{r+1}  
        \anc[r+1]{s}
        \dcgCB^{\gix{k} \ZZm{s}}_{\rix{r}} \gCC^{\gix{k-1}}_{\rix{r+1} \ZZm{s}}
        +
        \sum_{s=0}^{r-1}
        \anc[r]{s}
        \dcgCB^{\gix{k} \ZZm{s}}_{\rix{r}} \gCC^{\gix{k+1}}_{\rix{r-1} \ZZm{s}}
     \Big)
    \,,
  \end{array}
 \label{gfAction_aux_in_components_full-rank_p=p}
 \end{eqnarray}
where the normalizing coefficients are
 \begin{eqnarray}
  \anc[r]{s}
   &=& 
       \dcsc[r]{s}
       \msp^{\frac{r(r-1)}{2}-\frac{s(s-1)}{2} }
       \msp^{r-s} \, 
       \nfrac{s! \,(r{-}s)!}{r!} \nfrac{(d{-}2s)!}{(d{-}r{-}s)!}
       \,,
 \label{base_potential_coef}
 \end{eqnarray}
and, for the choice (\ref{def_dcsc}), the relative sign coefficients read
 $
   \dcsc[r]{r+1}
    \defeq {\dcsc[r]{s}}/{\dcsc[r+1]{s}}
    = \msp^{r}
 $.\,

The structure of the decomposed auxiliary part\HIDE{ of the gauge-fixed action},
for the full-rank gauge is more involved than that for the minimal-rank gauge. Fields do not decouple into noninteracting pairs. Instead, the fields are systematically coupled within gauge-fixing subchains. Determining the gauge-fixed set of fields and the fields parameterizing the reduction surface, is more efficiently achieved by directly solving the reduction equations.
\if{
 $
  \frac{\var S_{\delta}}{\var{\gB^{\gix{k}}_{ \ZZm{r}}}}
   {\,=\,} 0
   ,\;
  \frac{\var S_{\delta}}{\var{\dcgB^{\gix{k}}_{ \ZZm{r}}}}
   {\,=\,} 0
 $.
}\fi


    \subsubsection*{Physical space and reduced gauge-fixed fields}
     \hspace{\parindent}
The algebraic delta-function gauge-fixing allows the subsequent reduction of gauge-fixed fields\HIDE{ in the gauge-fixed action} by solving
 \begin{equation}
  \begin{array}{lllllll}
    \textstyle
    \frac{\var S_{\delta}}{\var{\gB^{\gix{k}}_{ \ZZm{r}}}}
     &\!\!=\!\!& 0
     \,:
    \\[0.7em]
    \textstyle
    \frac{\var S_{\delta}}{\var{\dcgB^{\gix{k}}_{ \ZZm{r}}}}
     &\!\!=\!\!& 0
     \,:
  \end{array}
    \qquad
  \left\{
  \begin{array}{lllllll}
    \displaystyle
        \dcgC^{\gix{k-1}}_{ \ZZm{r+1}} \gamma^{\Zm}
        +
        \dcgC^{\gix{k+1}}_{ \ZZm{r-1}} \gamma_{\Zm}
     &=& 0
    \,,
    \\[0.7em]
    \displaystyle
        \gamma^{\Zm} \gC^{\gix{k-1}}_{ \ZZm{r+1}}
        +
        \gamma_{\Zm} \gC^{\gix{k+1}}_{ \ZZm{r-1}}
     &=& 0
     \,.
     \phantom{\big|_{|}}
  \end{array}
  \right.
  \qquad
  \begin{array}{l}
    \text{\small$(\,1 \tleq k \tleq r\,,\;\;
    \,0 \tleq r \tleq p{\,-}1\,)$}\,,
  \end{array}
 \label{gf_constraints_full-rank_p=p} 
 \end{equation}
where, in accordance with the convention (\ref{def_contraction_order_convention}), the contracted gamma matrices carry the first (left) index of $\ZZn{r+1}$ family in the upper equation and the last (right) index of the $\ZZm{r+1}$ family in the lower equation. The explicit solution of (\ref{gf_constraints_full-rank_p=p}) constructively determines the physical (non-fixed) subset of fields, which parameterizes the reduction surface.

As before, the gauge-fixing couples ghosts with Dirac-conjugated Lagrange multipliers and vice versa, resulting in the decoupling of subsets of the upper and lower equations (\ref{gf_constraints_full-rank_p=p}). Based on the general properties of the BV gauge-fixing discussed in Section~\ref{Sect:BVst}, it follows that within each subfamily, the set of coupled fields splits into chains, depicted in Fig.~\ref{BVp_Triangular_diagram}. Moreover, each chain decouples into two subchains, which connect fields of the same ghost number (as well as the same Grassmann parity, and the same parity of tensor rank), as shown in Fig.~\ref{fig:gf_chain_p=p}.

Consider an arbitrary chain of (non-conjugated) fields with a root vertex associated with a minimal ghost of degree $\ro$ (i.e. level $p{\,-\,}\ro\,$).
Within a chain, the fields' generation indices $k$ are uniquely determined by their degrees $r$ as $k=\ro{\,-\,}r$, allowing us to omit the generation index while analyzing the chain. The set of equations (\ref{gf_constraints_full-rank_p=p}) splits into subchains
 \begin{equation}
  \begin{array}{l}
    \gC^{\gix{\ro-1}}_{ \Zm},\, \gC^{\gix{\ro-3}}_{ \ZZm{3}},\, \gC^{\gix{\ro-5}}_{ \ZZm{5}}\,,\, \ldots
    \quad:\quad
    \\[0.5em]
    \gC^{\gix{\ro}}_{ \Zo},\, \gC^{\gix{\ro-2}}_{ \ZZm{2}},\, \gC^{\gix{\ro-4}}_{ \ZZm{4}}\,, \,\ldots
    \quad:\quad
    \\
  \end{array}
  \begin{array}{lllllll}
    \displaystyle
    \big| \quad
        \gamma^{\Zm} \gC^{\hgix{{\ro-2n-1}}}_{ \ZZm{2n+1}}
        +
        \gamma_{\Zm} \gC^{\hgix{{\ro-2n+1}}}_{ \ZZm{2n-1}}
     \!&\!=\!& 0
    \,,
    \\[0.5em]
    \displaystyle
    \big| \quad
        \gamma^{\Zm} \gC^{\hgix{{\ro-2n-2}}}_{ \ZZm{2n+2}}
        +
        \gamma_{\Zm} \gC^{\hgix{{\ro-2n}}}_{ \ZZm{2n}}
     \!&\!=\!& 0
     \,.
     \phantom{\big|_{|}}
     \\
  \end{array}
 \end{equation}
The iterative solution starts from the bottom of each subchain with a field $\gC^{\gix{\ro}}_{ \Zo}$ or $\gC^{\gix{\ro-1}}_{ \Zm}$, and finishes at its root vertex: either associated with the antighost  $\gC^{\gix{1}}_{ \ZZm{\ro-1}}$ or with the minimal ghost $\gC^{\gix{0}}_{ \ZZm{\ro}}$, depending on the parity of the chain's length.

Each ghost field  $\gC^{\hgix{{\ro-r}}}_{ \ZZm{r}}$ in either subchain, except those at the extremes $\gC^{\gix{\ro}}_{ \Zo}$, $\gC^{\gix{1}}_{ \ZZm{\ro-1}}$, $\gC^{\gix{0}}_{ \ZZm{\ro}}$, enters the subchain twice --- as a lower-degree and as a higher-degree field:
 \begin{equation}
  \begin{array}{|@{\hspace{12pt}}lll}
   \hspace{8ex}\ldots &\hspace{20ex}\ldots
   \\[-1.3ex]
    \gamma^{\Zm} \gC^{\hgix{{\ro-r-2}}}_{ \ZZm{r+2}}
        +
    \gamma_{\Zm} \uwave{ \gC^{\hgix{{\ro-r}}}_{ \ZZm{r}} }
   \!\!&\displaystyle
    =\:
    \sum_{s=0}^{r+1}
    d_{\rix{r+2}}^{\rix{s}}
    \gamma_{\ZZm{r+1-s}}
    \gCC^{\hgix{{\ro-r-2}}}_{\rix{r+2}}{}_{\ZZm{s}}
        +
    \sum_{s=0}^{r} \;
    \gamma_{\ZZm{r+1-s}}
    \uwave{ \gCC^{\hgix{{\ro-r}}}_{\rix{r}}{}_{\ZZm{s}} }
    = 0
    \,,
   \\[0.3em]
    \gamma^{\Zm} \uwave{ \gC^{\hgix{{\ro-r}}}_{ \ZZm{r}}}
        +
    \gamma_{\Zm} \gC^{\hgix{{\ro-r+2}}}_{ \ZZm{r-2}}
   \!\!&\displaystyle
   =\:
   \sum_{s=0}^{r-1}
    d_{\rix{r}}^{\rix{s}}
    \gamma_{\ZZm{r-1-s}}
    \uwave{ \gCC^{\hgix{{\ro-r}}}_{\rix{r}}{}_{\ZZm{s}}}
        +
   \sum_{s=0}^{r-2}
    \gamma_{\ZZm{r-1-s}}
     \gCC^{\hgix{{\ro-r+2}}}_{\rix{r-2}}{}_{\ZZm{s}}
    = 0
    \,,
    \\[-0.7em]
   \hspace{8ex}\ldots &\hspace{20ex}\ldots
  \end{array}
 \label{gf_full_subchain} 
 \end{equation}
where the numerical constants read
  $d_{\rix{r}}^{\rix{s}} =
  \msp^{r}  
  \frac{r-s}{r} (d{-}r{+}s{+}1)$.

We iteratively solve equations one by one by comparing the irreducible fields at identical matrix basis elements. If an irreducible component of the lower-degree field is yet unfixed, the equation fixes it in terms of either the corresponding component of the higher-degree field, or, if the latter is absent, it is set to zero. If a component of a certain basis element of the lower-degree field is already fixed to zero at the previous step, then the corresponding component of the higher-degree field is fixed (to zero) at the current step.

The crucial properties of the constraints for any subchain at any level are:
\begin{itemize}
\setlength\itemsep{0ex}
  \item The left sum for the decomposed higher-degree field does \emph{not} contain the leading-degree irreducible component, which is eliminated by a contracting gamma matrix.
  \item Only the subleading-degree irreducible component of the higher-degree field enters the equation at the unit-matrix basis element due to a higher upper limit in the left sum, which forces this\HIDE{ subleading-degree irreducible} component to vanish.
  \item When resolving subchain equations for fields from the lowest-degree to highest-degree ones, after solving the equation with fields of degrees $r{\,-\,}2$ and $r$, the subleading-parity irreducible components of $\gC^{\hgix{{\ro-r}}}_{ \ZZm{r}}$ are fixed to zero: $P_{\sublead} {}_{\ZZn{r}}^\ZZm{r} \gC^{\hgix{{\ro-r}}}_{ \ZZm{r}}=0$, whereas components of the leading-parity degrees are free. All irreducible components of all lower-degree fields from the subchain --- $\gC^{\hgix{{\ro-r+2n}}}_{ \ZZm{r-2n}}$,\, $(n{\,>\,}0)$ --- are fixed to either zero or proportional to the yet unfixed leading-parity components of  $\gC^{\hgix{{\ro-r}}}_{\ZZm{r}}$.
\end{itemize}
The first two properties are evident from the irreducible-field representation (\ref{gf_full_subchain}). The third property relies on the first two and can be proved by induction.

The iterative solution of the subchain with $\gC^{\gix{\ro}}_{ \Zo}$ (see Fig.~\ref{fig:gf_chain_p=p}) starts with
 \begin{equation}
  \HIDE{+} 
  \nfrac12 (d{-}2) \gCC^{\gix{\ro-2}}_{\rix{2} \Zm}
  + 
  (d{-}1)\gamma_\Zm \gCC^{\gix{\ro-2}}_{\rix{2} \Zo}
  + \gamma_\Zm \gCC^{\gix{\ro}}_{\rix{0} \Zo}
  = 0
  \,,
 \end{equation}
which expresses $\gC^{\gix{\ro}}_{ \Zo} \equiv \gCC^{\gix{\ro}}_{\rix{0} \Zo}$ in terms of the irreducible component $\gCC^{\gix{\ro-2}}_{\rix{2} \Zo}$ (of $\gC^{\gix{\ro-2}}_{ \ZZm{2}}$) and simultaneously forces the subleading component $\gCC^{\gix{\ro-2}}_{\rix{2} \Zm}$ (of $\gC^{\gix{\ro-2}}_{ \ZZm{2}}$) to vanish.
The iterative solution of the subchain with the field $\gC^{\gix{\ro-1}}_{ \Zm}$ starts with
 \begin{equation}
  - 
  d\, \gCC^{\gix{\ro-1}}_{\rix{1} \Zo}
  = 0
  \,,
 \end{equation}
which just forces the subleading component $ \gCC^{\gix{\ro-1}}_{\rix{1} \Zo}$ (of $\gC^{\gix{\ro-1}}_{ \Zm}$) to vanish.
These boundary cases confirm the validity of the properties for the starting equations of each subchain and provide base cases for the induction, enabling proof of the third property for any level.

At the iteration step when solving the constraint relating fields of degrees $r$ and $r{\,-\,}2$, all unfixed irreducible components (of leading-parity degrees) of the lower-degree field $ \gC^{\hgix{{\ro-r+2}}}_{ \ZZm{r-2}}$ are expressed in terms of the corresponding leading-parity irreducible components of $ \gC^{\hgix{{\ro-r}}}_{ \ZZm{r}}$, which remain unfixed within this equation. The first property ensures that the leading-degree component of $\gC^{\hgix{{\ro-r}}}_{ \ZZm{r}}$ does not enter the equation and also remains unfixed. At the same time, the basis gamma-matrix elements of the equation with the subleading-parity irreducible components of $\gC^{\hgix{{\ro-r+2}}}_{ \ZZm{r-2}}$, which were fixed to zero at previous steps, now fix the corresponding subleading-parity components of $\gC^{\hgix{{\ro-r}}}_{ \ZZm{r}}$ to zero. This is true for all components, except the subleading-degree component itself, which, however, is also fixed to zero due to the second property.

After the next iteration step, solving the equation relating the components of $\gC^{\hgix{{\ro-r}}}_{ \ZZm{r}}$ and $\gC^{\hgix{{\ro-r-2}}}_{ \ZZm{r+2}}$ --- the upper equation in (\ref{gf_full_subchain})) --- one similarly obtains:
 \begin{equation}
  \begin{array}{|@{\hspace{8pt}}ll}
   \gCC^{\hgix{{\ro-r}}}_{\rix{r} \ZZm{r-2n}}
   \:\sim\:
   \gCC^{\hgix{{\ro-r-2}}}_{\rix{r+2} \ZZm{r-2n}}
   \,,
   & \text{ these components of $\gC^{\hgix{{\ro-r-2}}}_{ \ZZm{r+2}}$ remain free}
   \,,
   \\
   \gCC^{\hgix{{\ro-r-2}}}_{\rix{r+2} \ZZm{r+2}}
   \,,
   & \text{ highest-rank component is yet absent and so remains free}
   \,,
   \\
   \gCC^{\hgix{{\ro-r-2}}}_{\rix{r+2} \ZZm{r-2n-1}}
   \:\sim\:
   \gCC^{\hgix{{\ro-r}}}_{\rix{r} \ZZm{r-2n-1}}
   = 0
   \,,
   & \text{ these components of $\gC^{\hgix{{\ro-r-2}}}_{ \ZZm{r+2}}$ are fixed to zero}
   \,,
   \\
   \gCC^{\hgix{{\ro-r-2}}}_{\rix{r+2} \ZZm{r+1}}
   = 0
   \,,
   & \text{ these component is fixed to zero by the equation}
   \,.
  \end{array}
  \label{iteration_r_r+2} 
 \end{equation}
Before this step, all components of lower-degree fields were either fixed to zero or expressed via proportion to the components of $\gC^{\hgix{{\ro-r}}}_{ \ZZm{r}}$ with leading-parity degrees. After this step, the letter components are expressed via proportion to the leading-parity degree components of  $\gC^{\hgix{{\ro-r-2}}}_{ \ZZm{r+2}}$.

The iterative solution is completed when a leading-degree field in the equation is either the root antighost  $\gC^{\gix{1}}_{ \ZZm{\ro-1}}$ or the root minimal ghost $\gC^{\gix{0}}_{ \ZZm{\ro}}$. By aggregating the outcomes from both subchains, one gets that subleading-parity components of all ghosts ($r\tleq\ro$) are fixed to zero, whereas the leading-parity components of the extraghosts ($r\tleq\ro-2$) are fixed, being expressed in terms of free leading-parity components of the antighost and the minimal ghost.
The procedure is analogous for all chains of the fields and Dirac-conjugated fields. This also applies to the longest gauge-fixing chain (the leftmost on the triangular diagram in Fig.~\ref{BVp_Triangular_diagram}), which starts from the original fields, playing the role of the minimal ghosts of the $0$-th level. Similar to genuine minimal ghosts of levels $l\tgeq1$, the reduction eliminates gauge-fixed subleading-parity components os the field, leaving irreducible components of degrees $p,\,p{-}2,\,p{-}4,\,\ldots$ as physical degrees of freedom.

To summarize, the reduction in the case of full-rank gauge (\ref{min-rank_gf_p=p}) implies
 \begin{equation}
  \begin{array}{ll}
   \hspace{2.5em} \text{Fixed and reduced}
    \\[0.2em]
   \hline \vphantom{\Big|}
     \quad P_{\sublead} {}_{\ZZn{p}}^{\:\ZZm{p}} \psi_{\ZZm{p}}
     \,,\: 
     \dcpsi_{\ZZm{p}} P_{\sublead} {}_{\,\ZZn{p}}^{\ZZm{p}}
     \:=\: 0\,,
    \\[0.6em]
     \quad P_{\sublead} {}_{\ZZn{r}}^{\:\ZZm{r}} \gC^\gix{0}_{\ZZm{r}}
     \,,\: 
     \dcgC^\gix{0}_{\ZZm{r}} P_{\sublead} {}_{\,\ZZn{r}}^{\ZZm{r}}
     \:=\: 0\,,
    \\[0.6em]
     \quad P_{\sublead} {}_{\ZZn{r}}^{\:\ZZm{r}} \gC^\gix{1}_{\ZZm{r}}
     \,,\: 
     \dcgC^\gix{1}_{\ZZm{r}} P_{\sublead} {}_{\,\ZZn{r}}^{\ZZm{r}}
     \:=\: 0\,,
    \\[0.5em]
     P_{\lead} {}_{\ZZn{r}}^{\:\ZZm{r}} \gC^\gix{k}_{\ZZn{r}}
     \,,\: 
     \dcgC^\gix{k}_{\ZZn{r}} P_{\lead} {}_{\,\ZZn{r}}^{\ZZm{r}}
     \,=\,  f (\breve{\varPhi}) \!\!
     \\[0.5em]
     \quad P_{\sublead} {}_{\ZZn{r}}^{\:\ZZm{r}} \gC^\gix{k}_{\ZZn{r}}
     \,,\: 
     \dcgC^\gix{k}_{\ZZn{r}}  P_{\sublead} {}_{\,\ZZn{r}}^{\ZZm{r}}
     \:=\:  0
     \\[0.3em]
     \hline
   \end{array}
   \hspace{-2ex}
  \begin{array}{cc}
   & \vphantom{\text{F}}
    \\[0.2em]
   \vphantom{\Big|}
     & \text{\small $\begin{array}{l}  (\,k=0 ;\; r=p\,)\,,\end{array}$}
    \\[0.6em]
     &  \text{\small $\begin{array}{l} (\,k = 0 ;\; 0\leq r \leq p{-}1\,)\end{array}$}
    \\[0.6em]
     &  \text{\small $\begin{array}{l} (\,k = 1 ;\; 0\leq r \leq p{-}1\,)\end{array}$}
    \\[0.5em]
     & \text{\small $\begin{array}{l} (\,2\leq k < p{-}r\,;\; 0\leq r \leq p{-}2\,)\end{array}$}
    \\[0.5em]
     & \text{\small $\begin{array}{l} (\,2\leq k < p{-}r\,;\; 0\leq r \leq p{-}2\,)\end{array}$}
     \\[0.3em]
   \end{array}
   \hspace{0ex}
   \begin{array}{ll}
   \hspace{3.5em} \text{Free} \; (\breve{\varPhi})
    \\[0.2em]
   \hline \vphantom{\Big|}
     P_{\lead} {}_{\ZZn{p}}^{\:\ZZm{p}} \psi_{\ZZm{p}}
     \,,\,\;
     \dcpsi_{\ZZm{p}} P_{\lead} {}_{\,\ZZn{p}}^{\ZZm{p}}
     \,,
    \\[0.6em]
     P_{\lead} {}_{\ZZn{r}}^{\:\ZZm{r}} \gC^\gix{0}_{\ZZm{r}}
     \,,\,\;
     \dcgC^\gix{0}_{\ZZm{r}} P_{\lead} {}_{\,\ZZn{r}}^{\ZZm{r}}
     \,,
    \\[0.6em]
     P_{\lead} {}_{\ZZn{r}}^{\:\ZZm{r}} \gC^\gix{1}_{\ZZm{r}}
     \,,\,\;
     \dcgC^\gix{1}_{\ZZm{r}} P_{\lead} {}_{\,\ZZn{r}}^{\ZZm{r}}
     \,,
    \\[0.5em]
    \\[0.5em]
     \\[0.3em]
     \hline
    \end{array}
 \label{gf_full-rank_reduction_surface_p=p} 
 \end{equation}
The subleading-parity components of all ghosts and original fields are fixed to zero, the leading-parity components of extraghosts ($k \tgeq 2$) are fixed being expressed in terms of free fields --- the leading-parity components of the antighosts, minimal ghosts, and original fields.
Similar to the minimal-rank gauge fixing, the reduction surface is parameterized by the irreducible components of the original fields, minimal ghosts, and antighosts, with leading-parity degrees.


  \subsection{Reduced gauge-fixed action and effective action for arbitrary $p$} 
   \label{}
    \hspace{\parindent}
The explicit reduction of the gauge-fixed fields shows that for both the minimal-rank and full-rank gauges the physical subspace is the same.
After the reduction in the gauge-fixed action only the leading-parity components of the original fields, minimal ghosts and antighosts survive:
 \begin{eqnarray}
   \breve{\varPhi}\,:\quad\!
   P_{\lead} {}_{\ZZn{p}}^{\:\ZZm{p}} \psi_{\ZZm{p}},\,
     P_{\lead} {}_{\ZZn{r}}^{\:\ZZm{r}} \gC^\gix{0}_{\ZZm{r}},\,
     P_{\lead} {}_{\ZZn{r}}^{\:\ZZm{r}} \gC^\gix{1}_{\ZZm{r}}\,;\,
     P_{\lead} {}_{\ZZn{p}}^{\:\ZZm{p}} \dcpsi_{\ZZm{p}},\,
     P_{\lead} {}_{\ZZn{r}}^{\:\ZZm{r}} \dcgC^\gix{0}_{\ZZm{r}},\,
     P_{\lead} {}_{\ZZn{r}}^{\:\ZZm{r}} \dcgC^\gix{1}_{\ZZm{r}}
    \quad \leftrightarrow \quad
  \nonumber
  \\[1ex]
    \quad \leftrightarrow \quad
       \breve{\CPhi}\,:\quad\!
     \CPsi_{\ZZm{p-2n}},\,
     \gCC^\gix{0}_{\rix{r} \ZZm{r-2n}},\,
     \gCC^\gix{1}_{\rix{r} \ZZm{r-2n}}\,;\,
     \dcCPsi_{\ZZm{p-2n}},\,
     \dcgCC^\gix{0}_{\rix{r} \ZZm{r-2n}},\,
     \dcgCC^\gix{1}_{\rix{r} \ZZm{r-2n}}
     \,.\qquad
 \label{gf_phys_space_p=p} 
 \end{eqnarray}
where degrees $r$ of the ghost fields run $\,0 \tleq r \tleq p{\hspace{1pt}-\hspace{1pt}}1\,$.
The gauge-fixed action being reduced on the reduction surface considerably simplify, and the results of the reduction for both cases coincide.
This unique reduced gauge-fixed action in terms of irreducible fields reads
 \begin{eqnarray}
   \begin{array}{lllllll}
    &\breve{S}^{}_{\delta} [\breve{\CPhi}]
    \:\defeq\: S^{}_{\delta} [\varPhi(\breve{\CPhi})]
    &\!\!=\!\!&   \displaystyle
     \!\int \!d^dx\,
     \,\Bigl\{\,
     i\sqrt{|\GG|}
      \sum_{n=0}^{[p/2]}
     \bnc[p]{p-2n}
    \, \dcCPsi^{\ZZm{p-2n}}
        \slashed{\nabla}_{\!\oix{p-2n}}
    \, \CPsi_{\ZZm{p-2n}}
    \\
    &&&
    \qquad\quad
    +
    \displaystyle
     \,i
      \sum_{r=0}^{p-1} 
      \sum_{n=0}^{[r/2]}
        \cnc[r]{r-2n}
   \textstyle
       \dcgCC^{\gix{1} \ZZm{r-2n}}_{\rix{r}}
        \slashed{\nabla}_{\!\oix{r-2n}}
       \gCC^{\gix{0}}_{\rix{r} \ZZm{r-2n}}
    \\
    &&&
    \qquad\quad
    -
    \displaystyle
     \,i
      \sum_{r=0}^{p-1} 
      \sum_{n=0}^{[r/2]}
        \cnc[r]{r-2n}
   \textstyle
       \dcgCC^{\gix{0} \ZZm{r-2n}}_{\rix{r}}
         \dc{\slashed{\nabla}}_{\!\oix{r-2n}}
       \gCC^{\gix{1}}_{\rix{r} \ZZm{r-2n}}
   \Bigr\}
   \,.
  \end{array}
  \label{S_red_p=p} 
 \end{eqnarray}
where we use the short-hand notation for the Dirac-like massive operators introduced in (\ref{nabla-s_p=3})
 \begin{equation}
  \slashed{\nabla}_{\!\oix{s}}
  \,\defeq\,
  \slashed{\nabla}
  \,\mp 
  \nfrac{i}{2}\, \sqrr\,(d{-}2s)
  \,,
  \qquad
  \dc{\slashed{\nabla}}_{\!\oix{s}}
  \,\defeq\,
  \dc{\slashed{\nabla}}
  \,\pm 
  \nfrac{i}{2}\, \sqrr\,(d{-}2s)
  \,,
 \label{nabla-s_p=p} 
 \end{equation}
assuming $\slashed{\nabla}$ acts to the right and $\dc{\slashed{\nabla}}$ acts to the left.
 The normalizing coefficients $\bnc[p]{p-2n}$, (\ref{So_in_components_p=p_coef}), and $\cnc[r]{r-2n}$, (\ref{Sgh_gf_min-rank_p=p_coef}), (\ref{Sgh_gf_full-rank_p=p_coef}), are
 \begin{equation}
   \bnc{p-2n}
   \:=\:
    \, \dcsc[p]{p-2n}
    \msp^{\frac{p(p-1)}{2}} \msp^{p-n} 
    \, (p{-}2n)!\, \nfrac{(d{-}2p{+}4n{-}1)!}{(d{-}2p{-}1)!}
    \,,
 \label{Sred_p=p_coef}
 \end{equation}
 \begin{equation}
   \cnc[r]{r-2n}
   \:=\:
     \dcsc[r]{r-2n}
     \msp^{r-n} 
     \nfrac{(2n{+}1)!(r{-}2n)!}{(r{+}1)!} \nfrac{(d{-}2r{+}4n{-}1)!}{(d{-}2r{+}2n{-}1)!}
    \,,
 \label{Sgh_gf_min-rank_p=p_coef}
 \end{equation}
in which for sign conventions (\ref{def_dcsc}), $\dcsc[p]{p-2n}\, = \dcsc[r]{r-2n} = \msp^{n}$.


The original action (\ref{def_S0_p=p})\HIDE{ as a part of the minimal gauge-fixed BV action}, in components --- (\ref{So_in_components_p=p}),  being reduced on the same physical space remains the same for both gauges: subleading parity diagonal components vanish, as well as vanish all nondiagonal contributions. The latter couple a leading-parity component with a subleading-parity component and thus inevitably vanish on the reduction surface.
The ghost part of the minimal action (\ref{gf_gh action_p=p}), in components --- (\ref{Sgh_gf_min-rank_p=p}) and (\ref{Sgh_gf_full-rank_p=p}), being the functional of only the original fields, minimal ghosts and antighosts, when reduced to the physical space (\ref{gf_phys_space_p=p}) become identical for both gauges. Reduction wipes out subleading-parity diagonal terms in the full-rank action (\ref{Sgh_gf_full-rank_p=p}), and all nondiagonal contributions in (\ref{Sgh_gf_min-rank_p=p}) and (\ref{Sgh_gf_full-rank_p=p}).
The auxiliary part of the (\ref{gf_aux_action_p=p}) for both gauges vanish since it is proportional to the reductions constraints on ghost and original fields generated by variations with respect to Lagrange multipliers.

Gauge fixings with both gauges are proper because the resultant reduced gauge-fixed action is nondegenerate on the reduced physical space.

\newpar

For the algebraic gauge-fixing the quantization on the reduced physical space of fields (\ref{gf_phys_space_p=p}) is defined via the reduced gauge-fixed action (\ref{S_red_p=p})
 \begin{eqnarray}
   Z^{\scriptscriptstyle(p)}
   = \int D \breve{\varPhi} \,\exp \Bigl\{ i\breve{S}_{\delta}[\breve{\varPhi}] \Bigr\}
   \,.
 \end{eqnarray}
The action is quadratic in fields and block-diagonal. Further integration over the reduced space generate determinants
 \begin{equation}
   \Delta_{\oix{s}}
   \:\defeq\:
   {\rm Det}_{\oix{s}} \slashed{\nabla}_{\!\oix{s}}
   \:=\:
   {\rm Det}_{\oix{s}} \dc{\slashed{\nabla}}_{\!\oix{s}}
   \,,
 \label{def_Delta_p=p}
 \end{equation}
where subscript at ${\rm Det}$ denote the degree of the irreducible field's space over which the determinant is defined. In this model, the parameter of the mass spectrum, subscripted at $\slashed{\nabla}$, always coincides with the parameter of the functional space. We do not distinguish determinants of operator $ \slashed{\nabla}_{\!\oix{s}}$ and its Dirac conjugate.

The leading-parity components $s=p{\,-\,}2n$ of the original odd fields $\psi_{\ZZn{p}}$, $\dcpsi_{\ZZn{p}}$ via functional integration generate determinants $\Delta_{\oix{p-2n}}$ for \,$0 \tleq 2n \tleq p$. Each determinant appears once.
Integrating the leading-parity components of the ghost fields $\gC^{\gix{0}}_{ \ZZn{r}},\gC^{\gix{1}}_{ \ZZn{r}},\dcgC^{\gix{0}}_{ \ZZn{r}},\dcgC^{\gix{1}}_{ \ZZn{r}}$ where \,$0\tleq r \tleq p {\hspace{1pt}-\hspace{1pt}} 1$,\;  at each level $l= p{\,-\,}r$:\, $1 \tleq l \tleq p$\, generate determinants $\Delta_{\oix{p-l-2n}}$ for each integer $n$:\, $0 \tleq 2n \tleq r = p{\,-\,}l$. For given $l$ and $n$ each determinant appears twice being generated by two (Dirac conjugated) pairs of ghosts with these quantum numbers.
 %
In total, there are $s{\,+\,}1$ different pairs of degree-$s$ irreducible fields, thus generating $s{\,+\,}1$ determinants $\Delta_{\oix{s}}$.
\\

The following table illustrates the number of terms of degree $r$ irreducible components in the reduced action, which equals the multiplicity of determinants $\Delta_{\oix{r}}$ after functional integration.
{
\newcommand{\sminus}{{-}}
\newcommand{\fGC}[2][]{{C_{\scriptscriptstyle\gix{#1}}^{\scriptscriptstyle\gix{#2}}}}
\newcommand{\dcfGC}[2][]{{\dc{C}_{\scriptscriptstyle\gix{#1}}^{\scriptscriptstyle\gix{#2}}}}
\newcommand{\xxx}[1][]{{} {{{#1}}}}
\def\DetTable
{
 \begin{equation}
  \begin{array}{|l@{\hspace{1pt}}l@{\hspace{2pt}}|ccc@{\hspace{6pt}}c@{\hspace{6pt}}c|@{\hspace{2pt}}c@{\hspace{2pt}}|}
   \hspace{4.6 em}\textbf{Odd fields and components} \hspace{-20 em}
   \\
   \hline
   & \quad\text{{\footnotesize \text{level} $\to$}}
   & \scriptstyle{l\,=\,0}
   & \scriptstyle{l\,=\,2}
   & \scriptstyle{l\,=\,4}
   & \ldots
   & \scriptstyle{l\,=\,2[\frac{p}2]}
   &
  \\[0.2em]
   & \genfrac{}{}{0pt}{}{ \text{component} }{ \text{$\downarrow$ \;degree $s$ } }
   & \dcpsi, \psi
   & \genfrac{}{}{0pt}{}{ \dcgC^{\gix{0}}, \gC^{\gix{1}} }{ \dcgC^{\gix{1}}, \gC^{\gix{0}} }
   & \genfrac{}{}{0pt}{}{ \dcgC^{\gix{0}}, \gC^{\gix{1}} }{ \dcgC^{\gix{1}}, \gC^{\gix{0}} }
   & \ldots
   & \genfrac{}{}{0pt}{}{ \dcgC^{\gix{0}}, \gC^{\gix{1}} }{ \dcgC^{\gix{1}}, \gC^{\gix{0}} }
   & \genfrac{}{}{0pt}{}{ \text{total \#} }{\text{of terms} }
   \\[0.5em]
  \hline
   & \;\; p
   & \xxx[1]
   &
   &
   & \ldots
   &
   & 1
  \\
   & \;\; p\sminus2
   & \xxx[1]
   & \xxx[2]
   &
   & \ldots
   &
   & 3
  \\
   & \;\; p\sminus4
   & \xxx[1]
   & \xxx[2]
   & \xxx[2]
   & \ldots
   &
   & 5
  \\
   & \;\; \ldots
   & \ldots
   & \ldots
   & \ldots
   & \ldots
   & \ldots
   & \ldots
  \\
   & p \!\!\!\!\mod 2
   & \xxx[1]
   & \xxx[2]
   & \xxx[2]
   & \ldots
   & \xxx[2]
   & 2\big[\tfrac{p}{2}\big]{+}1
   \\[0.3em]
   \hline
  \end{array}
  \quad
  \begin{array}{|l@{\hspace{1pt}}l@{\hspace{2pt}}|cc@{\hspace{10pt}}c@{\hspace{6pt}}c|@{\hspace{2pt}}c@{\hspace{2pt}}|}
   \hspace{4.7 em}\textbf{Even fields and components} \hspace{-20 em}
   \\
   \hline
   & \quad\text{{\footnotesize level$\to$}}
   & \scriptstyle{l\,=\,1}
   & \scriptstyle{l\,=\,3}
   & \ldots
   & \scriptstyle{l\,=\,2[\frac{p+1}2]-1}
   &
  \\[0.2em]
   & \genfrac{}{}{0pt}{}{ \text{component} }{\text{$\downarrow$ \;degree $s$ } }
   & \genfrac{}{}{0pt}{}{ \dcgC^{\gix{0}}, \gC^{\gix{1}} }{ \dcgC^{\gix{1}}, \gC^{\gix{0}} }
   & \genfrac{}{}{0pt}{}{ \dcgC^{\gix{0}}, \gC^{\gix{1}} }{ \dcgC^{\gix{1}}, \gC^{\gix{0}} }
   & \ldots
   & \genfrac{}{}{0pt}{}{ \dcgC^{\gix{0}}, \gC^{\gix{1}} }{ \dcgC^{\gix{1}}, \gC^{\gix{0}} }
   & \genfrac{}{}{0pt}{}{ \text{total \#} }{\text{of terms} }
   \\[0.5em]
\hline
  &&&&&&
  \\[-0.7em]
   & \;\; p\sminus1 \vphantom{I^I{^I}}
   & \xxx[2]
   &
   & \ldots
   &
   & 2
  \\
   & \;\; p\sminus3
   & \xxx[2]
   & \xxx[2]
   & \ldots
   &
   & 4
 \if{
  \\
   & \;\; p\sminus5
   & \xxx[2]
   & \xxx[2]
   & \xxx[2]
   & \ldots
   &
   & 6
 }\fi
  \\[0.3em]
   & \;\; \ldots
   & \ldots
   & \ldots
   & \ldots
   & \ldots
   & \ldots
  \\[0.3em]
   &  p\sminus1 \!\!\!\!\!\mod\! 2
   & \xxx[2]
   & \xxx[2]
   & \ldots
   & \xxx[2]
   & 2[\frac{p+1}2]
   \\[0.3em]
   \hline
  \end{array}
  \nonumber
 \end{equation}
}
\begin{figure}[htbp]
  \centering
  \small
  \DetTable
  \caption{\normalsize \hspace{20ex} Table of determinants in effective action
  \\ \phantom{.} \\
  \small \it
   The table visualize determinant contributions from different irreducible fields from the physical space (\ref{gf_phys_space_p=p}). We separate fields by their Grassmann parity into two tables. In each table rows list degrees $s$ of nonfixed components, the columns organize fields by levels $l$. The numbers in cells of particular row and column denotes the number of different pairs of fields in the gauge-fixed action (\ref{S_red_p=p}), empty cells imply $0$. The numbers in right cells is the total number of different bilinear terms producing $\Delta_{\oix{s}}$.
  }
 \label{fig:DetTable_p=p}
\end{figure}

Thus finally
 \begin{equation}
 \label{final}
  Z^{\scriptscriptstyle(p)}
   \;=\;
  {\displaystyle \prod_{n=0}^{[p/2]} (\Delta_{\oix{p-2n}})^{2n+1} }{\displaystyle \prod_{m=0}^{[(p-1)/2]} \!\!\!\! (\Delta_{\oix{p-1-2m}})^{-2m-2} },
 \end{equation}
where $[N/2]$ in the upper limits mean the integer part of a number $N/2$.
The relation (\ref{final}) can be transformed to the form
\begin{eqnarray}
  Z^{\scriptscriptstyle(p)}
  \;=\;
  \frac{\Delta_{\oix{p}}\;\Delta_{\oix{p-2}}^3\;\Delta_{\oix{p-4}}^5 \,\cdots\, \Delta_{\oix{p-2[p/2]}}^{2[p/2]+1} \vphantom{\big|}}
  {\Delta_{\oix{p-1}}^2\;\Delta_{\oix{p-3}}^4 \cdots \Delta_{\oix{p-2[}\oix{(p-1)/2}\oix{]-1}}^{2[{(p-1)}/2]{+2}} \vphantom{\big|^b}}
  \,,
   \label{arb_p_result_EA_COPY}
 \end{eqnarray}
This is our final result for the generating functional, from which one can express the effective action for the fermionic $p$-form theory with arbitrary stage of reducibility. The functional determinants $\Delta_{\oix{p}}$ are given by
(\ref{def_Delta_p=p}). Relation (\ref{arb_p_result_EA_COPY}) confirms a conjecture about general structure of the effective action put forward in \cite{BBKN4:2025}.



\section{Conclusion \label{Sect:concl}}
  \hspace{\parindent}
Let us summarize our results. We have studied the quantization of the totally antisymmetric rank-$p$ tensor-spinor field theory (fermionic $p$-form model). This theory, recently proposed in \cite{Buchbinder:2009pa}, provides a novel example of a $(p{-}1)$-stage-reducible gauge theory, which is consistently defined in $AdS_d$  for $2p {\,<\,} d$ and is dual to fermionic $(d{-}p{-}2)$-form model. We performed the quantization for arbitrary $p$ in the framework of the field-antifield Batalin-Vilkovisky method, suitably adapted to this model. This enabled us to derive the functional integral for the partition function with the correct ghost structure. Since the theory\HIDE{ under consideration} is free in quantum fields, the quantum effective action follows directly and is expressed as the sequence of functional determinants of specific Dirac-type differential operators acting on $p$-forms in AdS space. The final result is given by relation (\ref{arb_p_result_EA_COPY}).


As a byproduct, the fermionic $p$-form models serve as examples demonstrating the validity of a broader class of gauge fermions. We show that a unique, nondegenerate gauge-fixed action can be achieved under relaxed rank conditions on the gauge-fixing coefficients of the gauge fermion --- extending beyond those in the seminal works \cite{Batalin:1983ggl, Gomis:1994he}. Two distinct covariant gauges are considered: one satisfying the minimal rank conditions, and another using full-rank gauge-fixing coefficients. The reduction procedures are analyzed in detail, proving equivalence of the gauge-fixed actions on the reduction surface.\footnote{It can also be shown that contributions to the quantum measure upon passing to the reduced representation, being ultralocal in our examples, coincide.}

\newpar



The obtained results raise several interesting questions for future study related to the fermionic $p$-form models. Below we list some of them.

\begin{itemize}
\setlength\itemsep{0ex}
\item{
    The development of the canonical formulation of the fermionic $p$-form model, the derivation of its first-class constraints, and the implementation of BFV-BRST quantization.}
\item{
    The calculation of the functional determinants defining the effective action (\ref{arb_p_result_EA_COPY}). We expect this to be achievable using the general technique of \cite{Cam} for analyzing spectra of operators in AdS space. This would allow the effective action to be expressed in explicit form. Notably, (\ref{arb_p_result_EA_COPY}) takes the form of a ratio of products of functional determinants, suggesting possible cancellations that could simplify a final expression.}
\item{The investigation of quantum equivalence between dual fermionic $p$-form theories.}
\item{The construction of a supersymmetric extension of the fermionic $p$-form theory and its quantization.}
\item{The study of the massive fermionic $p$-form theory proposed in \cite{Buchbinder:2009pa}. Although this theory is not gauge-invariant, we expect reformulated as a gauge theory by introducing appropriate Stueckelberg fields, which\HIDE{ apparently} would themselves be spinor-valued differential forms. The resulting reducible gauge system could then be quantized along lines similar to the present case.}

\end{itemize}

We plan to address these issues in forthcoming works.

\section*{Acknowledgements}
  \hspace{\parindent}
The research of A.O.\,Barvinsky and D.V.\,Nesterov was supported by the Russian Science Foundation grant No. \href{https://rscf.ru/en/project/23-12-00051/}{23-12-00051}.


\appendix


\section{Batalin-Vilkovisky formalism for theories with reducible generators}
 \label{Sect:BVst}
 %
    \hspace{\parindent}
Consider a gauge theory of fields $\psi^i$ with action $S_0[\psi]$, invariant under the gauge transformations $\var \psi^i = R^j_{\zza{1}}\lambda^{\zza{1}}$:
 \begin{eqnarray}
   S_0[\psi]\;:
   &\qquad&
   S_0[\psi]_{,j} R^j_{\zza{1}}
   \,=\, 0
   \,,
 \label{BV_So} 
 \end{eqnarray}
where a field index after comma in subscripts denote the variational derivative with respect to this field. In (\ref{BV_So}) and throughout we use the condensed DeWitt notation, where field indices such as $i$ and $\zza{1}$ combine discrete labels with spacetime coordinates. Repeated (contracted) indices imply both summation over discrete values and integration over spacetime.

In a general gauge theory the generators  $R^i_{\zza{0}}$ may depend on the fields and must satisfy the consistency condition
 \begin{eqnarray}
  R^i_{\zza{1},j} R^j_{\zzb{1}}
  - (-1)^{\varepsilon_{\zza{1}}\varepsilon_{\zzb{1}}}
  R^i_{\zzb{1},j} R^j_{\zza{1}}
  =
  R^i_{\zzc{1}}T^{\zzc{1}}_{\zza{1}\zzb{1}}
  -
  S_{0,j} E^{ij}_{\zza{1}\zzb{1}}
  \,.
 \end{eqnarray}
In general, the structure functions $F^{\zzc{1}}_{\zza{1}\zzb{1}}$ and $E^{ij}_{\zza{1}\zzb{1}}$ may depend on the fields $\psi^i$.\footnote{
 For the spinorial $p$-form field theory in AdS space studied in Sections~\ref{Sect:p=3} and \ref{Sect:p=p}, the generators $R^i{\zza{1}}$ are field-independent. The algebra is therefore abelian and closed, with
 $ 
  F^{\zzc{1}}_{\zza{1}\zzb{1}} = 0
  \,, \;
  E^{ij}_{\zza{1}\zzb{1}} = 0
  \,.
 \nonumber
 $ 
}

The gauge generators $R^i_{\zza{1}}$ may be linear dependent, admitting nontrivial zero-eigenvalue eigenvectors
 \begin{eqnarray}
   Z^{\zza{1}}_{\,\zza{2}}\,:
   \qquad
    R^i_{\zza{1}} Z^{\zza{1}}_{\,\zza{2}}
   =\,
    S_{0,j} V^{ji}_{{\scriptscriptstyle(1)} \zza{2}}
   \,.
  \label{BV_first_reducibility_generator} 
 \end{eqnarray}
Here the \emph{reducibility generators} $Z^{\zza{1}}_{\,\zza{2}}$ and the structure functions $V^{ji}_{\zza{2}}$ may again depend on the fields.
If $Z^{\zza{1}}_{\,\zza{2}}$ are full rank and have no nontrivial zero modes, the theory is said to be \emph{first-stage reducible}. Otherwise, there are nontrivial eigenvectors of  $Z^{\zza{1}}_{\,\zza{2}}$ with zero eigenvalue (at least on shell) which are denoted by $Z^{\zza{2}}_{\,\zza{3}}$. If these have no further zero modes, then the theory is said to be a \emph{second-stage reducible} gauge theory. Otherwise this process continues up to some
$(p{-}1)$-th \emph{stage of reducibility} if the chain terminates at $Z^{\zza{p-1}}_{\,\zza{p}}$, which has no nontrivial zero modes:
 \begin{eqnarray}
  R^i_{\zza{1}} Z^{\zza{1}}_{\,\zza{2}}
  = S_{0,j} V^{ji}_{{\scriptscriptstyle(1)} \zza{2}}
   \,, \quad\;\;
  Z^{\zza{1}}_{\,\zza{2}} Z^{\zza{2}}_{\,\zza{3}}
  = S_{0,j} V^{ji}_{{\scriptscriptstyle(2)} \zza{3}}
   \,, \quad\;\;
  \ldots
   \,, \quad\;\;
  Z^{\zza{p-2}}_{\,\zza{p-1}} Z^{\zza{p-1}}_{\,\zza{p}}
  = S_{0,j} V^{ji}_{{\scriptscriptstyle(p-1)} \zza{p}}
   \,.
 \label{BV_reducibility_generators_p} 
 \end{eqnarray}
On-shell ranks of such gauge and reducibility generators imply:
 \begin{eqnarray}
  \rank R^i_{\zza{1}}
  = \mmm_{1} 
  \,,\quad
  \rank Z^{\zza{l-1}}_{\,\zza{l}}
  = \mmm_{l} 
  \,,\qquad
  \mmm_{l} \defeq \nnn_{l}-\nnn_{l+1}+ \nnn_{l+2} - ... +\msp^{p-l}\nnn_{p}
  \,,
 \label{BV_generator_ranks} 
 \end{eqnarray}
where $\nnn_{0}$ is the range of index $i$, and $\nnn_{l}$ for $l\tgeq1$ are ranges of corresponding indices $\zza{l}$.

The theory of fermionic $p$-form field in AdS space provides a particular example of the\HIDE{ abelian} gauge theory with $p{\,-\,}1$ stages of reducibility. This theory is abelian and free (quadratic)\HIDE{ which grants a simplification in general}, though such theories experience all features of BV formalism manifesting in one-loop approximation.\footnote{
 For field-independent gauge generators $R^i_{\zza{1}}$, one may always choose off-shell, field independent null-vectors $Z^{\zza{1}}_{\,\zza{2}}$. The same holds for higher-stage reducibility generators leading to
 $ 
  R^i_{\zza{1}} Z^{\zza{1}}_{\,\zza{2}}
  \!= 0
   \,, \;\,
  Z^{\zza{1}}_{\,\zza{2}} Z^{\zza{2}}_{\,\zza{3}}
  \!= 0
   \,, \;
  \ldots
   \,, \;\,
  Z^{\zza{p-2}}_{\,\zza{p-1}} Z^{\zza{p-1}}_{\,\zza{p}}
  \!= 0
  .
 $ 
}

 \subsubsection*{BV extended configuration space}
  \hspace{\parindent}
To quantize a gauge theory with $p{\,-\,}1$ stages of reducibility using the BV method, the field space is extended from $\psi^i$ to $\varPhi^I$. In addition to the original fields $\psi^i$, the \emph{minimal} extension implies inclusion of ghost fields $\gC_{\gix{0}}^{\zza{l}}$:
 \begin{equation}
  \gC_{\gix{0}}^{\zza{l}},
  \qquad
  \alpha_{l} = 1,2,\ldots, \nnn_{l},
  \qquad
  (\,l=1,2, \ldots, p\,),
 \end{equation}
where by $\nnn_{l}$ we denote the range of indices of these level $l$ ghosts, which characterize the number of components of the latter.
For proper gauge fixing, the configuration space is also extended with the \emph{auxiliary} fields of two types: ghosts $\gC_{\gix{k}}^{\zza{l}}$ and fields $\gB_{\gix{k}}^{\zza{l}}$, often referred as Lagrange-multiplier fields. At each level $l$, these auxiliary fields have $l$ generations, labeled by $k$:
 \begin{equation}
  \begin{array}{lll|ll}
   \gC_{\gix{k}}^{\zza{l}},\; \gB_{\gix{k}}^{\zza{l}},
    \qquad
   (\,l=1,2, \ldots, p,  \quad   k=1,2, \ldots, l\,).
  \end{array}
 \end{equation}
Fields belonging to the same level $l$ share the same index range $\alpha_{l} = 1,2,\ldots, \nnn_{l}$. The original fields, minimal ghosts, and auxiliary fields together are called the \emph{nonminimal} BV fields, or \emph{BV fields} for short.

The BV configuration space carries two gradings. \emph{Grassmann} parity is the $\mathbb{Z}_2$-grading,\, $\epsilon(FG) = \epsilon(F)+\epsilon(G)\,\mod\,2$:
  \begin{align}
    &\epsilon(\psi^i)\equiv\epsilon_i
    \,,
    && 
    \epsilon(R^i_{\zza{1}}) = \epsilon_i + \epsilon_{\zza{1}}
    \,,
    && \epsilon(Z^{\zza{l-1}}_{\,\zza{l}}) = \epsilon_{l-1} + \epsilon_{l}
    \,,
    \nonumber
    \\
    &\epsilon(\gC_{\gix{k}}^{\zza{l}})=\epsilon_{\zza{l}}\! + l
    \,,
    &&\epsilon(\gB_{\gix{k}}^{\zza{l}})=\epsilon_{\zza{l}}\! + l + 1
    \,,
  \end{align}
where $\epsilon_{l}$ are the parities of the gauge parameters $\glambda^{\zza{l}}$ in the original theory (\ref{BV_So}).
The second grading is the \emph{ghost number}, an additive $\mathbb{Z}$-grading,\, $\gh{FG} = \gh{F}+\gh{G}$, defined by:
 \begin{align}
  & \gh{\psi^i}=0
   \,,
  && \gh{\gC_{\gix{k}}^{\zza{l}}}
   =
   \left[
     \begin{array}{lr}
       l-k\,, & (k \text{ even}), \\
       k-l-1\,, & (k \text{ odd}), \\
     \end{array}
   \right.
  && \gh{\gB_{\gix{k}}^{\zza{l}}}
    = \gh{\gC_{\gix{k}}^{\zza{l}}} +1
    \,.
 \label{BV_ghost_numbers} 
 \end{align}

Finally, for each field $\varPhi^I$ an associated \emph{antifield} $\varPhi^\af_I$\, is introduced, with gradings
  \begin{align}
   &\epsilon(\varPhi^\af_I)=\epsilon(\varPhi^I)+1
   \,,
   &\gh{\varPhi^\af_I}=-\gh{\varPhi^I}-1
   \,.
  \end{align}

The resultanting\HIDE{ even-dimensional} nonminimal set of BV fields and antifields is summarized in  Table \ref{Table:BVst-Improved}.
\def \TblBVstGhNumImproved
{
\vspace{-0.3em}
  \begin{center}
\hspace{-50pt}
    \begin{tabular}{rc|c|c@{\hspace{2pt}} c@{\hspace{2pt}}||c|c@{\hspace{2pt}} c@{\hspace{2pt}}|}
    \cline{3-8}
    &
    & \quad$\varPhi^I$\quad  & $\gh{\varPhi^I}$ && \quad$\varPhi^\af_I$\!\quad  & $\gh{\varPhi^\af_I}$ &
    \vphantom{\Big|}
    \\[0.1em]
    \cline{3-8}
    \cline{3-8}
    \multirow{10}{0cm}{} 
    &
    & \multicolumn{6}{c|}{\textit{minimal sector}} \\
    \cline{3-8}
    \vphantom{\Big|}
    {\small level $0$}
    &
    &  $\psi^i $
       & $0$  &
    &  $\psi^{\af}_i$
       & $-1$ &
    \\
    \vphantom{\Big|} 
    {\small levels $l=1, ... \,,p$}
    &
    &  $ \gC_{\gix{0}}^{\zza{l}}$
      & $l$  &
    & $ \gC_{\gix{0}}^{\af}{}_{\zza{l}}$
      & $-l{\,-\,}1$  &
    \\[0.5ex]
    \cline{3-8}
    &
    & \multicolumn{6}{c|}{\textit{auxiliary sector }}
    \\
    &
    & \multicolumn{6}{c|}{\textrm{\small generations ${k}$:\, $1\tleq k \tleq l$}}
    \\
    \cline{3-8}
    \cline{3-8}
    \vphantom{\Big|}
    {\small levels $l=1, ... \,,p$}
    &
    &  $ \gC_{\gix k}^{\zza{l}}$
      & $ (-)^k(l{-}k{+}\tfrac12) -\tfrac12 $  &
    & $ \gC_{\gix k}^{\af}{}_{\zza{l}}$
      & $ - (-)^k(l{-}k{+}\tfrac12) -\tfrac12 $   &
    \\[0.5em]
    \vphantom{\Big|}
    {\small levels $l=1, ... \,,p$}
    &
    & $\gB_{\gix k}^{\zza{l}}$
      & $ (-)^k(l{-}k{+}\tfrac12) + \tfrac12 $   &
    &    &   &
    \\[0.5ex]
    \cline{3-8}
  \end{tabular}
 \end{center}
 \vspace{-2ex}
}
\begin{table}[h!]
 \centering
  \TblBVstGhNumImproved
 \caption{BV-extended configuration space}
 \label{Table:BVst-Improved} 
\end{table}
\vspace{-1em}

%
\def \TriangleDN
{
 \hspace{-2.5cm}
 \begin{picture}(300,250)
  \put(172,222){\text{\small $\psi^i$}}
  \put(151,191.3){\vector(3,4){15}}
  \put(171,218){\line(-3,-4){15}}
   \put(200,190){\line(-3,4){21}}
   \put(200.3,190){\line(-3,4){21}}
    \put(191,204){\text{\small \MGen{R^i_{\zza{1}}} }}
  \put(128,181.5)
    {\text{\small $\gC_{\gix{1}}^{\zza{1}}$\!,\hspace{1pt}$\gB_{\gix{1}}^{\zzb{1}}$}}
  \put(199,181)
    {\text{\small $\gC_{\gix{0}}^{\zza{1}}$}}
  \put(123,154){\vector(3,4){15}}
  \put(141,178){\vector(-3,-4){15}}
  \put(183,154){\vector(3,4){15}}
  \put(201,178){\line(-3,-4){15}}
   \put(229,151){\line(-3,4){20}}
   \put(229.3,151){\line(-3,4){20}}
    \put(221,165){\text{\small \MGen{Z^{\zza{1}}_{\:\zza{2}}} }}
  \put(98,142.5)
    {\text{\small $\gC_{\gix{2}}^{\zza{2}}$\!,\hspace{1pt}$\gB_{\gix{2}}^{\zzb{2}}$}}
  \put(158,142.5)
    {\text{\small $\gC_{\gix{1}}^{\zza{2}}$\!,\hspace{1pt}$\gB_{\gix{1}}^{\zzb{2}}$}}
  \put(228,142)
    {\text{\small $\gC_{\gix{0}}^{\zza{2}}$}}
  \if{
  \put(94,149.5)
    {\text{\small $\gC^{\gix{2}}_{ \ZZm{p-2}}$ \phantom{,$\gB^{\gix{2}}_{ \ZZm{p-2}}$}}}
  \put(82,135.5)
    {\text{\small \phantom{$\gC^{\gix{2}}_{ \ZZm{p-2}}$,}$\gB^{\gix{2}}_{ \ZZm{p-2}}$}}
  \put(154,149.5)
    {\text{\small $\gC^{\gix{1}}_{ \ZZm{p-2}}$ \phantom{,$\gB^{\gix{1}}_{ \ZZm{p-2}}$}}}
  \put(142,135.5)
    {\text{\small \phantom{$\gC^{\gix{1}}_{ \ZZm{p-2}}$,}$\gB^{\gix{1}}_{ \ZZm{p-2}}$}}
  \put(228,142)
    {\text{\small $\gC^{\gix{0}}_{ \ZZm{p-2}}$}}
  }\fi
  \put(93,114){\vector(3,4){15}}
  \put(111,138){\vector(-3,-4){15}}
  \put(153,114){\vector(3,4){15}}
  \put(171,138){\vector(-3,-4){15}}
  \put(213,114){\vector(3,4){15}}
  \put(231,138){\line(-3,-4){15}}
   \put(259,111){\line(-3,4){20}}
   \put(259.3,111){\line(-3,4){20}}
    \put(251,125){\text{\small \MGen{Z^{\zza{2}}_{\:\zza{3}}} }}
  \put(69,103)
    {\text{\small $\gC_{\gix{3}}^{\zza{3}}$\!,\hspace{1pt}$\gB_{\gix{3}}^{\zzb{3}}$}}
  \put(130,103)
    {\text{\small $\gC_{\gix{2}}^{\zza{3}}$\!,\hspace{1pt}$\gB_{\gix{2}}^{\zzb{3}}$}}
  \put(189,103)
    {\text{\small $\gC_{\gix{1}}^{\zza{3}}$\!,\hspace{1pt}$\gB_{\gix{1}}^{\zzb{3}}$}}
  \put(258,102)
    {\text{\small $\gC_{\gix{0}}^{\zza{3}}$}}
  %
  \put(63,74){\vector(3,4){15}}
  \put(81,98){\vector(-3,-4){28}}
  \put(81,98){\line(-3,-4){31.5}}
  \put(123,74){\vector(3,4){15}}
  \put(141,98){\vector(-3,-4){28}}
  \put(141,98){\line(-3,-4){28}}
  \put(141,98){\line(-3,-4){31.5}}
  \put(188.5,81.5){\vector(3,4){9}}
  \put(201,98){\line(-3,-4){9}}
  \put(243,74){\vector(3,4){15}}
  \put(261,98){\line(-3,-4){22}}
  \put(268,80){\line(-3,-4){18}}
  \put(268,80){\vector(-3,-4){14}}
   \put(268.5,97){\line(3,-4){32}}
   \put(268.8,97){\line(3,-4){32}}

  \put(24.5,44) 
    {\text{\small $\gC_\gix{p}^{\zza{p}}$,\,$\gB_\gix{p}^{\zza{p}}$}}
  \put(80,44)
    {\text{\small $\gC_\gix{p-1}^{\zza{p}}$,\,$\gB_\gix{p-1}^{\zza{p}}$}}
  \put(152,46)
    {\text{\small $\ldots$}}
  \put(197,46)
    {\text{\small $\ldots$}}
  \put(228,44)
    {\text{\small $\gC_\gix{1}^{\zza{p}}$,\,$\gB_\gix{1}^{\zza{p}}$}}
  \put(300,44)
    {\text{\small $\gC_\gix{0}^{\zza{p}}$}}
  %
   {\color{white}
    \linethickness{14pt}
    \put(40,75){\line(1,0){300}}
    \linethickness{1pt}
   }
  \put(296,65){\text{\small\MGen{Z^\zza{p-1}_{\,\zza{p}}}}}
  \put(60,74) {\text{\small $...$}}
  \put(120,74) {\text{\small $...$}}
  \put(180,74) {\text{\small $...$}}
  \put(240,74) {\text{\small $...$}}
  \put(258,74) {\text{\small $...$}}
  \put(281,74) {\text{\small $...$}}
  \put(358,74) {\text{\small $...$}}
  \put(396,74) {\text{\small $...$}}
  \put(346,242) {\text{level (l)\;\;\; \#}}
  \put(387,250){\vector(0,-1){10}}
  \put(335,35){\vector(-1,0){20}}
  \put(290,21){\text{generation} $(k)$}
    \multiput(350,220)(-5,0){22}{\line(1,0){3}}
    \put(360,220) {\text{\small $0$ \qquad\: $\nnn_0$}}
    \multiput(350,181)(-5,0){17}{\line(1,0){3}}
    \put(360,181) {\text{\small $1$ \qquad\: $\nnn_1$}}
    \multiput(350,142)(-5,0){12}{\line(1,0){3}}
    \put(360,142) {\text{\small $2$ \qquad\: $\nnn_2$}}
    \multiput(350,102)(-5,0){7}{\line(1,0){3}}
    \put(360,102) {\text{\small $3$ \qquad\: $\nnn_3$}}
    \multiput(350,44)(-5,0){2}{\line(1,0){3}}
    \put(361,44) {\text{\small $p$ \qquad\: $\nnn_p$}}
  %
 \end{picture}
\vspace{-28pt}
}

  \subsubsection*{BV master action}
   \hspace{\parindent}
On the BV space of fields  $\varPhi^I$ and antifields $\varPhi^\af_I$, the odd symplectic structure is defined via the antibracket
\vspace{-2mm}
 \begin{eqnarray}
  (F,G)
  \:\equiv\:
  \frac{\partial_{r} F}{\partial\varPhi^I}\frac{\partial_l G}
   {\partial\varPhi^\af_I} - \frac{\partial_r F}{\partial\varPhi^\af_I}
  \frac{\partial_l G}{\partial\varPhi^I}
  \,,
 \label{BV_antibracket} 
 \end{eqnarray}
where $F$ and $G$ are functionals of $\varPhi^I$ and $\varPhi^\af_I$.
The classical action $S_0[\psi]$ is then extended to the BV action $S_{\iBV}[\varPhi,\varPhi^\af]$, which by definition satisfies the following master equation
 \begin{eqnarray}
  (S_{\iBV},S_{\iBV})
  \:=\:
  0
 \label{master_equation} 
 \end{eqnarray}
along with the correspondence condition $S_{\iBV}[\varPhi,\varPhi^\af]\big|_{\varPhi^\af{=\,}0} {\:=\:}S_0[\psi]$, and additional boundary conditions encoding the reducible gauge structure of the original theory imposed through terms bilinear in the minimal ghost fields and the antifields of the minimal sector. We do not discuss the general solution for the BV\HIDE{ master} action here and refer to \cite{Batalin:1983ggl,Henneaux:1992ig}. However, the structure sufficient for the one-loop approximation and complete for \emph{abelian} gauge theories is already present in the ansatz:
 \begin{eqnarray}
  S_{\iBV}
  [\varPhi,\varPhi^\af]
  &=&
  S_0[\psi]
  + S_{\text{gh}}[\varPhi_{min},\varPhi_{min}^\af]
  + S_{\text{aux}}[\varPhi_{aux},\varPhi_{aux}^\af]
  \,,
 \label{BV_master_action} 
 \end{eqnarray}
where $S_0[\psi]$ is the original action (\ref{BV_So}), and
\vspace{-1ex}
 \begin{eqnarray}
  &&
  S_{\text{gh}}[\varPhi_{min},\varPhi_{min}^\af]
  \:=\:
  \psi^\af_i R^i_{\,\zza{1}}\gC_{\gix{0}}^{\zza{1}}
  + \sum_{l=1}^{p-1} \gC_{\gix{0}\,\zza{l}}^\af Z^{\zza{l}}_{\,\zza{l+1}}\gC_{\gix{0}}^{\zza{l+1}}
  + \ldots
  \,,
 \label{BV_gh_action} 
 \\[-0.8em]
  &&
  S_{\text{aux}}
  [\varPhi_{aux},\varPhi_{aux}^\af]
  \:=\:
  \sum_{l=1}^p \sum_{k=1}^l
  \gC^\af_{\gix{k}\,\zza{l}} \gB_{\gix{k}}^{\zza{l}}
  \,.
 \label{BV_aux_action} 
 \end{eqnarray}
The minimal action, $ S_{\text{min}}[\varPhi_{min},\varPhi_{min}^\af] \defeq S_0[\psi] + S_{\text{gh}}[\varPhi_{min},\varPhi_{min}^\af]$,\, is important as a consistent solution of the master equation (\ref{master_equation}) on the minimal set of BV fields and antifields.
Only the ghost part of the minimal action extends to higher powers of antifields and compensating ghosts. This extension is defined by solving (\ref{master_equation}) iteratively on the minimal BV space starting from the boundary conditions given by $S_0[\psi] + \psi^\af_i R^i_{\,\zza{1}}\gC_{\gix{0}}^{\zza{1}} + \sum_{l=1}^{p-1} \gC_{\gix{0}\,\zza{l}}^\af Z^{\zza{l}}_{\,\zza{l+1}}\gC_{\gix{0}}^{\zza{l+1}}$,\, \cite{Batalin:1983ggl,Henneaux:1992ig}.\footnote{
 The $\psi$-dependent coefficients at higher powers of antifields and ghost fields are proportional to higher-order structure functions of the original gauge algebra. The solution is unique up to (anti)canonical transformations. Provided a basis of generators is chosen, the boundary conditions (\ref{BV_gh_action}) encode only these gauge and reducibility generators, $R^j_{\zza{1}}$, $Z^{\zza{l}}_{\zza{l+1}}$, and not the structure functions of the gauge algebra.}
As mentioned above, for the purposes of this article the standard ``abelian'' ansatz (\ref{BV_master_action}-\ref{BV_aux_action}) is sufficient.
The auxiliary action $S_{\text{aux}}[\varPhi_{aux},\varPhi_{aux}^\af]$ is a functional of the fields and antifields from the separate auxiliary sector. It has the simple universal form (\ref{BV_aux_action}) and it also provides a consistent solution of (\ref{master_equation}).

\newpar

{ 
\newcommand{\PPhi}{\mathsf{\Phi}} 
\newcommand{\gRR}{\mathcal{R}} 

The minimal and auxiliary actions, as well as their sum, are solutions of the master equation. Each of them possesses a global BRST symmetry and an extended BV local symmetry \cite{Batalin:1981jr,Batalin:1983ggl,Henneaux:1992ig,Gomis:1994he}. In terms of the total set of BV fields and antifields $\PPhi^A = (\varPhi^I,\varPhi^\af_I)$,\, $A=1,\ldots,2N$,\, $I=1,\ldots,N$,\, the master equation (\ref{master_equation}) for the BV action reads
 \begin{eqnarray}
  \frac{{\partial_{r}} S_{\iBV} } {\partial \PPhi^A}
  \zeta^{AB}
  \frac{{\partial_{l}} S_{\iBV}} {\partial \PPhi^B}
  = 0,\;\quad\;
  \zeta^{AB}
  =\left[
     \begin{array}{@{\hspace{2pt}}c@{\hspace{6pt}}c@{\hspace{6pt}}}
       0 & \; 1 \\
       -1 & \; 0 \\
     \end{array}
   \right].
 \label{master_equation2}
 \end{eqnarray}
By differentiating (\ref{master_equation2}) one obtains the Noether identities
 \begin{eqnarray}
  \frac{{\partial_{r}} S_{\iBV}} {\partial \PPhi^A}
   \gRR^A_{\,B}
  = 0,
  \;\quad\;
  \gRR^A_{\,B}
   =
   \zeta^{AC}
  \frac{\partial_{l} \partial_{r} S_{\iBV}} {\partial \PPhi^C  \partial \PPhi^B}
  \,,
 \label{BV_NoetherIdmaS} 
 \end{eqnarray}
which involve $2N$ gauge generators $\gRR^A_{\,C}$ of the extended local BV gauge symmetry $\gvar[\iBV] \PPhi^A = \gRR^A_{\,B} \Sigma^B$.
These generators, however, form a reducible set. Indeed, the second variational derivative of (\ref{master_equation2}) shows their $N$-reducibility on shell,
 $
  \zeta^{-1}_{DB}
  \gRR^B_{\,A} \gRR^A_{\,C}
  \propto \partial S_{\iBV}/\partial \PPhi^A = 0
 $.\,
Thus, the master action $S_{\iBV}[\varPhi,\varPhi^\af]$
admits $N$ on-shell independent symmetries --- half the dimension of the BV configuration space $(\varPhi^I,\varPhi^\af_I)$.\footnote{
 Nilpotence of $2N \times 2N$ matrix implies that its rank is at most $N$. A \emph{proper} solution requires the Hessian $ \partial_{l} \partial_{r} S_{\iBV}/\partial \PPhi^A \partial \PPhi^B$ to have rank $N$. Otherwise, the set of Noether identities (\ref{BV_NoetherIdmaS}) is incomplete, and there are more than $N$ gauge generators, if the Hessian admits more than $N$ zero modes.
}
} 

  \subsubsection*{Gauge fixing}
   \hspace{\parindent}
To fix the\HIDE{ extended} BV local gauge symmetry (\ref{BV_NoetherIdmaS}) of the master action $S_{\iBV}[\varPhi\HIDE{^I},\varPhi^\af\HIDE{_I}]$, the universal gauge-fixing prescription is traditionally applied. It implies the elimination of the antifields according to\HIDE{ the rule}
 \begin{eqnarray}
  \varPhi^\af_I
  \:=\:
  \frac{\var \varPsi[\varPhi]}{\var \varPhi^I}
  \,,
 \label{BV_gauge_fixing} 
 \end{eqnarray}
where $\varPsi[\varPhi^I]$ is a fermionic functional depending only on the BV fields $\varPhi^I$, but not on the antifields.

A gauge-fixing fermion that implements a delta-function type gauge fixing is independent of the Lagrange multipliers $\gB_{\gix{k}}^{\zza{s}}$ and has the form
 \begin{eqnarray}
  \varPsi_{\delta}[\varPhi\HIDE{^I}]
   \:=\:
  \gC^{\zza{1}}_{\gix{1}} \chi_{\zza{1}}\!(\psi) 
    +\sum_{l=2}^{p} \sum_{k=1}^{l}
    \gC^{\zza{l}}_{\gix{k}} X^{\gix{k}}_{\zza{l}\zzb{l-1}} \gC^{\zzb{l-1}}_{\gix{k-1}}
    \,.
 \label{BV_gauge_fermion} 
 \end{eqnarray}
In general, the gauge-fixing operators $X^{\gix{k}}_{\zza{s}\zzb{s-1}}$ may depend on original fields $\psi^i$. It is convenient to choose a \emph{linear gauge}, parameterized by field-independent operators $X^{\gix{k}}_{\zza{s}\zzb{s-1}}$, with
 \begin{equation}
  \chi_{\zza{1}}\!(\psi)
   \:=\: X^{\gix{1}}_{\zza{1} i} \psi^i
   \: (\,\equiv\, \psi^i X^{\gix{1}}_{i\, \zza{1}}\, )
   \,.
 \label{BV_linear_gauge} 
 \end{equation}
In this case, gauge fixing does not introduce higher vertices into the theory, and the gauge fermion makes explicit that fields $\psi^i$ act as the minimal ghosts of level $0$, identifying $\psi^i {\:\leftrightarrow\:} \gC^{\zza{0}}$,\, $i {\:\leftrightarrow\:} \zza{0}$.
 %

\newpar

It is convenient to illustrate various structures related to gauge fixing by the triangular diagram in Fig.~\ref{fig:BVst_Triangular_diagram}, which also serves as an alternative visualization of the BV configuration space.

The structure of the gauge fermion (\ref{BV_gauge_fermion}) can be read off the triangular diagram in Fig.~\ref{fig:BVst_Triangular_diagram} as follows.
Each left-downward line\footnote{We reserve arrows on these lines to differ operator and its transposed version, which are distinguished later in the gauge-fixed action.
}
between vertices of levels $l$ and $l\,{-}\,1$ corresponds to a gauge-fixing operator $X^{\gix{k}}_{\zza{l}\zzb{l-1}}$, including the upper-left line with $ X^{\gix{1}}_{\zza{1}i} \equiv \chi_{\zza{1} ,i}$. The gauge fermion (\ref{BV_gauge_fermion}) is the sum of terms assigned to all such left-downward lines, each involving an operator $X^{\gix{k}}_{\zza{l}\zzb{l-1}}$ that couples the ghosts $\gC_{\gix{k-1}}^{\zza{l-1}}$ and $\gC_{\gix{k}}^{\zza{l}}$ from the two connected vertices. For nonlinear $\chi_{\zza{1}}(\psi)$ the only modification arises in the first term in (\ref{BV_gauge_fermion}) associated with the upper line of the left chain, which takes the nonlinear form $\gC_{\gix{1}}^{\zza{1}} \chi_{\zza{1}}\!(\psi)$.

In general, gauge-fixing operators that connect the vertices between neighbouring levels in different chains can be chosen independently. The combination of the index types corresponding to the connected levels and the generation index\footnote{We assign to each operator a generation index equal to that of the lower vertex with the higher generation number.} uniquely identifies the line, and hence the corresponding operator $X^{\gix{k}}_{\zza{l}\zzb{l-1}}$, in the triangular diagram (\ref{fig:BVst_Triangular_diagram}).
Nevertheless, it is often possible to choose gauge-fixing operators identical at each level and omit the generation index in the operator notation.

 \begin{figure}[htbp]
  \centering
  \TriangleDN  
  \caption{ \qquad\qquad\quad Triangular diagram: BV fields and gauge fixing
  \\ \phantom{.} \\
  \small \it
  BV fields are associated with the vertices of this triangular tree diagram. The slanted right-downward lines connect vertices of the minimal sector fields with the original fields at the top. The remaining vertices extending left-downward correspond to fields of the auxiliary sector, where each vertex denotes a pair consisting of a ghost and a Lagrange multiplier. The fields on each horizontal line correspond to the same {level}, which in our notation is encoded by a unique index type and given by the index subscript.
  Within each level, as well as along each chain proceeding left-downward, the vertices are distinguished by a generation number $k$.
  Chains of fields with $l{-}k=\mathrm{const}$, originating from a minimal sector field and extending left-downward, correspond to subsets of fields linked by the standard BV gauge-fixing procedure.
  }
 \label{fig:BVst_Triangular_diagram} 
\end{figure}

\newpar

Applying the gauge-fixing prescription (\ref{BV_gauge_fixing}) to the BV action yields the gauge-fixed action in the form
 \begin{eqnarray}
   S_{\textrm{$\delta$}}[\varPhi\HIDE{^I}]
   &\!\defeq\!&
   S_{\iBV}[\varPhi\HIDE{^I},\var \varPsi_{\delta}/\var \varPhi\HIDE{^I}]
   \;=\;
   S_0[\psi]
   + S_{\delta\,\text{gh}}[\varPhi]
   + S_{\delta\,\text{aux}}[\varPhi]
   \,,
 \label{BV_gf_action} 
 \end{eqnarray}
where the original action $S_0[\psi]$ remains unchanged, while the\HIDE{ gauge-fixed} ghost and auxiliary parts take the form
 \begin{eqnarray}
    S_{\delta\,\text{gh}}[\varPhi]
   &\!=\!&
     \gC_{\gix{1}}^{\zza{1}}
      \chi_{\zza{1},i} R^i_{\zzb{1}}
     \gC_{\gix{0}}^{\zzb{1}}
    +
    \sum_{l=2}^{p}
     \gC_{\gix{1}}^{\zza{l}}
     X^{\gix{1}}_{\zza{l}\zzc{l-1}} Z^{\zzc{l-1}}_{\zzb{l}}
     \gC_{\gix{0}}^{\zzb{l}}
    + \ldots
   \,, \quad\quad
 \label{BV_gh_gf} 
  \\[-0.8ex]
   S_{\delta\,\text{aux}}[\varPhi]
   &\!=\!&
    \big(
    \chi_{\zzb{1}}
    \! + \gC^{\zza{2}}_{\gix{2}} X^{\gix{2}}_{\zza{2}\zzb{1}}
    \big)
    \gB_{\gix{1}}^{\zzb{1}}
    +
    \sum_{l=2}^{p}
    \sum_{k=1}^{l}
    \big(\gC^{\zza{l-1}}_{\gix{k-1}} X^{\gix{k}}_{\zza{l-1}\zzb{l}}
    \! + \gC^{\zza{l+1}}_{\gix{k+1}} X^{\gix{k+1}}_{\zza{l+1}\zzb{l}} \big)
    \gB_{\gix{k}}^{\zzb{l}}
    \,.
 \label{BV_aux_gf} 
 \end{eqnarray}
For the linear gauge $\chi_{\zza{1}} = \psi^i X^{\gix{1}}_{i \zza{1}}$,\, the first terms in the ghost and auxiliary actions can be absorbed into the sums as the $l{\,=\,}k{\,=\,}1$ contributions, by identifying\, $\zza{0} \leftrightarrow i$,\,  $\gC^{\zza{0}}_{\gix{0}} \leftrightarrow \psi^i$, $X^{\gix{1}}_{\zza{1}\zzb{0}} \leftrightarrow  \chi_{\zza{1} ,i}$.
 %
 %
Gauge fixing mixes the minimal and auxiliary sectors, clarifying the role of the latter: introducing auxiliary ghosts with negative ghost numbers enables the gauge fermion, which has ghost number $-1$, to encode proper gauge conditions.

\newpar


\def \auxgfDiagramDNshort
{
\hspace{-20pt}
\newcommand{\tmpskip}{\hspace{20pt}}
\begin{picture}(280,180)
\put(93,154){\line(3,4){9}} 
 %
\put(202,154){\line(3,4){9}} 
\put(231,153){\line(3,4){9}} 
%
\put(64,142.5) 
  {\text{\small $\gC_{\gix{k-1}}^{\zza{\lll-1}}$,\hspace{1pt}$\gB_{\gix{k-1}}^{\zzb{\lll-1}}$}}
\put(183,142.5) 
  {\text{\small $\gB_{\gix{k-1}}^{\zzb{\lll-1}}$}}
\put(213,142) 
  {\text{\small $\gC_{\gix{k-1}}^{\zza{\lll-1}}$}}
\put(63,114){\vector(3,4){15}} 
\put(81,138){\vector(-3,-4){15}} 
 %
\put(173,114){\line(3,4){15}}  
\put(191,138){\vector(-3,-4){15}} 
\put(198,114){\vector(3,4){15}} 
\put(216,138){\line(-3,-4){15}} 
%
 \put(122,123){\MgfX{\tmpskip X^{\gix{k}}_{\zza{\lll} \zzb{\lll-1}}}} 
 \put(187,120){\MgfX{\tmpskip X^{\gix{k}}_{\zza{\lll-1} \zzb{\lll}}}} 
%
\put(37,103) 
  {\text{\small $\gC_{\gix{k}}^{\zza{\lll}}$,\hspace{1pt}$\gB_{\gix{k}}^{\zzb{\lll}}$}}
 %
\put(161,103) 
  {\text{\small $\gC_{\gix{k}}^{\zza{\lll}}$}}
\put(187,102) 
  {\text{\small $\gB_{\gix{k}}^{\zzb{\lll}}$}}
%
    \put(104,97.5){\line(-1,0){15}}
    \put(104,103){\line(-1,0){15}}
    \put(109,100.5){\line(-3,2){10}}
    \put(109,100.5){\line(-3,-2){10}}
\put(33,74){\vector(3,4){15}} 
\put(51,98){\vector(-3,-4){15}} 
 %
\put(143,74){\vector(3,4){15}} 
\put(161,98){\line(-3,-4){15}} 
\put(168,74){\line(3,4){15}} 
\put(186,98){\vector(-3,-4){15}} 
 \put(91,82){\MgfX{\tmpskip X^{\gix{k+1}}_{\zza{\lll} \zzb{\lll+1}}}} 
 \put(157,81){\MgfX{\tmpskip X^{\gix{k+1}}_{\zza{\lll+1} \zzb{\lll}}}} 
%
\put(10,62) 
  {\text{\small $\gC_{\gix{k+1}}^{\zza{\lll+1}}$\!,\hspace{1pt}$\gB_{\gix{k+1}}^{\zzb{\lll+1}}$}}
  %
\put(125,62) 
  {\text{\small $\gB_{\gix{k+1}}^{\zzb{\lll+1}}$}}
\put(153,62) 
  {\text{\small $\gC_{\gix{k+1}}^{\zza{\lll+1}}$}}
%
%
\put(14,46){\line(3,4){9}} 
 %
\put(123,46){\line(3,4){9}} 
\put(149,46){\line(3,4){9}} 
%
%
\put(115,40){.}
\put(120,40){.}
\put(125,40){.}
\put(165,170){.}
\put(170,170){.}
\put(175,170){.}
%
%
\put(60,15){\text{Structure of $S_{\delta\,\text{aux}}$}} 
\end{picture}
\hspace{-60pt}
 \label{Pic:auxgfDiagramDNshort}
}


\def \ghgfDiagramDNshort
{
\hspace{-100pt}
\newcommand{\tempskip}{\hspace{-3pt}}
\begin{picture}(280,200)
\if{
\put(180,221.0){\text{\small $\psi^i$}}
\put(159.5,192.3){\vector(3,4){13}}
\put(178.5,217.6){\line(-3,-4){14}}
\put(208,190){\line(-3,4){20}}
  \put(154,207){\MgfX{\displaystyle \chi}}
  \put(154,205){\MgfX{\phantom{\chi}\displaystyle \HIDE{\chi}\vphantom{\big|}_{\zzb{1} ,i}}}
  \put(198,205){\text{\small \MGen{R^{i}_{\zza{1}}}}}
{\color{blue}
 \put(182.5,193.05){\oval(30,20)[t]}  
 \put(182.5,193){\oval(29.2,19.6)[t]}  
 \put(182.4,202.9){\vector(1,0){5}}
  {\color{white}
  \put(196.8,193){\line(1,0){1}}
  \put(168.2,193){\line(-1,0){1}}
 }
}
\put(145,180)
  {\text{\small $\gC_{\gix{1}}^{\zzb{1}}$}}
\put(205,181)
  {\text{\small $\gC_{\gix{0}}^{\zza{1}}$}}
%
\put(144,177){\line(-3,-4){9}}
\put(187,154){\vector(3,4){15}}
\put(205,178){\line(-3,-4){15}}
\put(234,151){\line(-3,4){20}}
  \put(181,167){\MgfX{\tempskip \gfX{1}_{\zzb{2} \zza{1}}}}
  \put(224,166){\text{\small \MGen{Z^{\zza{1}}_{\,\zza{2}}}}}
{\color{blue}
 \put(210,153.05){\oval(30,20)[t]}  
 \put(210,153){\oval(29.2,19.6)[t]}  
 \put(208,162.9){\vector(1,0){5}}
  {\color{white}
  \put(224.3,153){\line(1,0){1}}
  \put(195.7,153){\line(-1,0){1}}
 }
}
}\fi
\put(230,151){\line(-3,4){9}}
%
\put(230,142)
  {\text{\small $\gC_{\gix{0}}^{\zza{\lll-1}}$}}
%
\put(213,114){\vector(3,4){15}}
\put(231,138){\line(-3,-4){15}}
\put(259,111){\line(-3,4){20}}
  \put(205,127){\MgfX{\tempskip X^{\gix{1}}_{\zzb{\lll} \zza{\lll-1}}}}
  \put(248.5,126){\text{\small \MGen{Z^{\zza{\lll-1}}_{\,\zza{\lll}}}}}
{\color{blue}
 \put(235.5,110.05){\oval(52,20)[t]}  
 \put(235.5,110){\oval(51.2,19.6)[t]}  
 \put(233.4,119.9){\vector(1,0){5}}
  {\color{white}
  \linethickness{3pt}
  \put(210.2,111){\line(-1,0){1}}
  \put(260.8,111){\line(1,0){1}}
  \linethickness{1pt}
 }
}
%
%
\put(200,102)
  {\text{\small $\gC_{\gix{1}}^{\zzb{\lll}}$}}
\put(255,102)
  {\text{\small $\gC_{\gix{0}}^{\zza{\lll}}$}}
%
%
\put(200,99){\line(-3,-4){9}}
\put(237,74){\vector(3,4){15}}
\put(255,98){\line(-3,-4){15}}
\put(285,71){\line(-3,4){20}}
 \put(231,87){\MgfX{\tempskip X^{\gix{1}}_{\zzb{\lll+1} \zza{\lll}}}}
  \put(277,86){\text{\small\MGen{Z^{\zza{\lll}}_{\,\zza{\lll+1}}}}}
{\color{blue}
 \put(260.5,70.05){\oval(54,20)[t]}  
 \put(260.5,70){\oval(53.2,19.6)[t]}  
 \put(258.4,79.9){\vector(1,0){5}}
  {\color{white}
  \linethickness{3pt}
  \put(234.2,71){\line(-1,0){1}}
  \put(286.8,71){\line(1,0){1}}
  \linethickness{1pt}
 }
}
%
%
\put(224,61)
  {\text{\small $\gC_{\gix{1}}^{\zzb{\lll+1}}$}}
\put(281,61)
  {\text{\small $\gC_{\gix{0}}^{\zza{\lll+1}}$}}
\put(214,46){\line(3,4){9}}
\put(272,46){\line(3,4){9}}
\put(300,46){\line(-3,4){9}}
\put(165,150){.}
\put(170,150){.}
\put(175,150){.}
\put(165,80){.}
\put(170,80){.}
\put(175,80){.}
\put(185,40){.}
\put(190,40){.}
\put(195,40){.}
\put(245,40){.}
\put(250,40){.}
\put(255,40){.}
\put(190,15){\text{Structure of $S_{\delta\,\text{gh}}$}} 
\end{picture}
\hspace{30pt}
 \label{Pic:ghgfDiagramDNshort}
}

\begin{figure}[htbp]
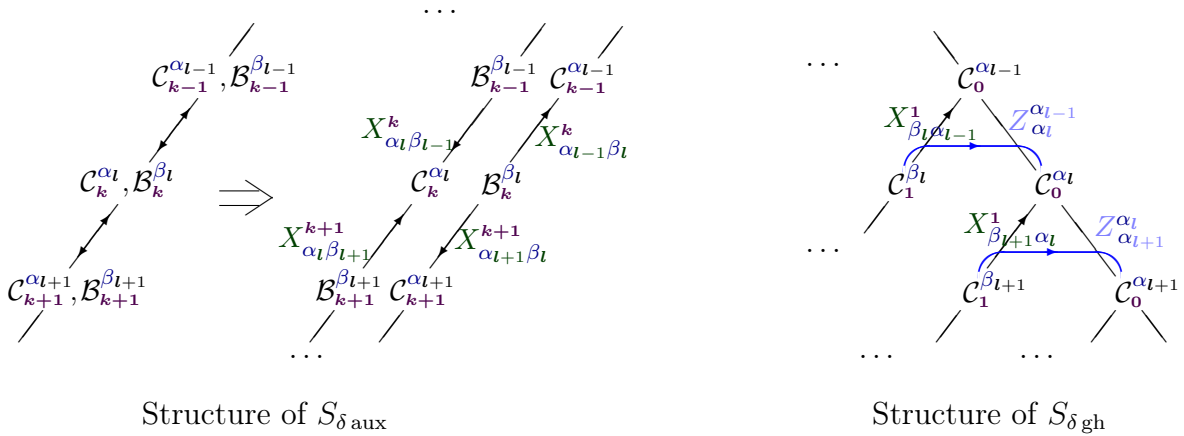

\vspace{-20pt}
  \centering
  \auxgfDiagramDNshort
  \ghgfDiagramDNshort
  \vspace{-1em}
  \caption{Structure of the gauge-fixed action
  } 
 \label{fig:DN_aux_gf_diagram} 
 \label{fig:DN_gh_gf_diagram} 
\end{figure}

\if{ 

\def \ghgfDiagramDNlong
{
\newcommand{\tempskip}{\hspace{-3pt}}

\begin{picture}(300,250)
\put(180,221.0){\text{\small $\psi^i$}}
\put(159.5,192.3){\vector(3,4){13}}
\put(178.5,217.6){\line(-3,-4){14}}
\put(208,190){\line(-3,4){20}}
  \put(154,207){\MgfX{\displaystyle \chi}}
  \put(154,205){\MgfX{\phantom{\chi}\displaystyle \HIDE{\chi}\vphantom{\big|}_{\zzb{1} ,i}}}
  \put(198,205){\text{\small \MGen{R^{i}_{\zza{1}}}}}
{\color{blue}
 \put(182.5,193.05){\oval(30,20)[t]}  
 \put(182.5,193){\oval(29.2,19.6)[t]}  
 \put(182.4,202.9){\vector(1,0){5}}
  {\color{white}
  \put(196.8,193){\line(1,0){1}}
  \put(168.2,193){\line(-1,0){1}}
 }
}
\put(145,180)
  {\text{\small $\gC_{\gix{1}}^{\zzb{1}}$}}
\put(205,181)
  {\text{\small $\gC_{\gix{0}}^{\zza{1}}$}}
%
\put(144,177){\line(-3,-4){9}}
\put(187,154){\vector(3,4){15}}
\put(205,178){\line(-3,-4){15}}
\put(234,151){\line(-3,4){20}}
  \put(181,167){\MgfX{\tempskip \gfX{1}_{\zzb{2} \zza{1}}}}
  \put(224,166){\text{\small \MGen{Z^{\zza{1}}_{\,\zza{2}}}}}
{\color{blue}
 \put(210,153.05){\oval(30,20)[t]}  
 \put(210,153){\oval(29.2,19.6)[t]}  
 \put(208,162.9){\vector(1,0){5}}
  {\color{white}
  \put(224.3,153){\line(1,0){1}}
  \put(195.7,153){\line(-1,0){1}}
 }
}
%
%
\put(171,142)
  {\text{\small $\gC_{\gix{1}}^{\zzb{2}}$}}
\put(230,142)
  {\text{\small $\gC_{\gix{0}}^{\zza{2}}$}}
%
\put(171,138){\line(-3,-4){9}}
\put(213,114){\vector(3,4){15}}
\put(231,138){\line(-3,-4){15}}
\put(259,111){\line(-3,4){20}}
  \put(205,127){\MgfX{\tempskip X^{\gix{1}}_{\zzb{3} \zza{2}}}}
  \put(248.5,126){\text{\small \MGen{Z^{\zza{2}}_{\,\zza{3}}}}}
{\color{blue}
 \put(235.5,113.05){\oval(30,20)[t]}  
 \put(235.5,113){\oval(29.2,19.6)[t]}  
 \put(233.4,122.9){\vector(1,0){5}}
  {\color{white}
  \put(249.8,113){\line(1,0){1}}
  \put(221.2,113){\line(-1,0){1}}
 }
}
%
%
\put(200,102)
  {\text{\small $\gC_{\gix{1}}^{\zzb{3}}$}}
\put(255,102)
  {\text{\small $\gC_{\gix{0}}^{\zza{3}}$}}
%
%
\put(200,99){\line(-3,-4){9}}
\put(238,74){\vector(3,4){15}}
\put(256,98){\line(-3,-4){15}}
\put(286,71){\line(-3,4){20}}
 \put(231,87){\MgfX{\tempskip X^{\gix{1}}_{\zzb{4} \zza{3}}}}
  \put(277,86){\text{\small\MGen{Z^{\zza{3}}_{\,\zza{4}}}}}
{\color{blue}
 \put(261.5,73.05){\oval(30,20)[t]}  
 \put(261.5,73){\oval(29.2,19.6)[t]}  
 \put(259.4,82.9){\vector(1,0){5}}
  {\color{white}
  \put(275.8,73){\line(1,0){1}}
  \put(247.2,73){\line(-1,0){1}}
 }
}
%
%
\put(223,61)
  {\text{\small $\gC_{\gix{1}}^{\zzb{4}}$}}
\put(281,61)
  {\text{\small $\gC_{\gix{0}}^{\zza{4}}$}}
\put(214,46){\line(3,4){9}}
\put(272,46){\line(3,4){9}}
\put(300,46){\line(-3,4){9}}
%
%
%
\put(170,45){.}
\put(170,40){.}
\put(170,35){.}
\put(35,15){\text{Structure of $S_{\delta\,\text{gh}}$}} 
\end{picture}
}

}\fi 

The triangular diagram from Fig.~\ref{fig:BVst_Triangular_diagram} again helps to trace the structure of the ghost and auxiliary parts of the gauge-fixed action --- (\ref{BV_gh_gf}) and (\ref{BV_aux_gf}). The part of $S_{\delta\,\text{gh}}$ that is bilinear in ghosts is the sum of bilinear combination of the antighost $\gC_{\gix{1}}^{\zzb{l}}$ and the minimal ghost $\gC_{\gix{0}}^{\zzb{l}}$, coupled by the operator $X^{\gix{1}}_{\zza{l}\zzc{l-1}} Z^{\zzc{l-1}}_{\zzb{l}}$ at each level $1 \tleq l \tleq p$. Graphically, this corresponds to composing the upward lines from the vertices of the antighost and the minimal ghost, as shown on the right in Fig.~\ref{fig:DN_gh_gf_diagram}.


The gauge-fixed auxiliary action is the sum of terms associated with the edges of the left-downward chains in Fig.~\ref{fig:BVst_Triangular_diagram}, each of which splits into two subchains, as illustrated in Fig.~\ref{fig:DN_aux_gf_diagram}. The $S_{\delta,\text{aux}}$ collects contributions from every edge of these subchains. In each case, the operator $X^{\gix{k}}_{\zza{l-1}\zzb{l}}$ from one subchain with upward arrow couples the fields $\gC^{\zza{l-1}}_{\gix{k-1}}$ and $\gB_{\gix{k}}^{\zzb{l}}$ from neighboring vertices connected by the line. In the other subchain with downward arrow, the coupling is between $\gC^{\zza{l}}_{\gix{k}}$ and $\gB_{\gix{k-1}}^{\zzb{l-1}}$ by the transposed operator $X^{\gix{k}}_{\zza{l}\zzb{l-1}}$.
From the bottom of the triangle, both subchains terminate at vertices of level $p$. On the top edge of each chain in Fig.~\ref{fig:BVst_Triangular_diagram} there is only one upward arrow: this edge belongs to only one subchain ending at vertex with the minimal ghost $\gC^{\zza{\lo}}_{\gix{0}}$, and corresponds in the gauge-fixed action to the term $\gC^{\zza{\lo}}_{\gix{0}} X^{\gix{1}}_{\zza{\lo}\zzb{\lo+1}}  \gB_{\gix{1}}^{\zzb{\lo+1}}$. The other subchain is shorter and originates from the vertex of level $(\lo{\,+\,}1)$ associated with the antighost $\gC^{\zza{\lo+1}}_{\gix{1}}$.

The splitting --- first into chains, then within the gauge-fixed action into subchains --- follows from grading properties with respect to the ghost number. The chains in  Fig.~\ref{fig:BVst_Triangular_diagram} contain ghosts with non-overlapping ghost numbers, and each chain is uniquely labeled by the value of $l{\,-\,}k$. Such chain aggregates ghosts $\gC_{\gix{k}}^\zza{l}$ with ghost numbers $l{\,-\,}k$ (for even $k$) and $-(l{\,-\,}k){\,-\,}1$ (for odd $k$), as can be inferred from Table~\ref{Table:BVst-Improved}. At the same time, the chain aggregates $\gB_{\gix{k}}^{\zza{l}}$ with ghost numbers $l{\,-\,}k{\,+\,}1$ (for even $k$) and $-(l{\,-\,}k)$ (for odd $k$). Thus, bilinear combinations in the gauge fermion (with ghost number $-1$) are possible for ghosts from adjacent vertices within a chain, and in the\HIDE{ auxiliary part of the} gauge-fixed action (with ghost number $0$) bilinear combinations are possible for ghosts and Lagrange multipliers from adjacent vertices. The alternation of the two different ghost numbers within each chain naturally leads to a split into two subchains.


  \subsubsection*{Quantization}
   \hspace{\parindent}
Provided the gauge-fixed action (\ref{BV_gf_action}) is \emph{nondegenerate}, the theory can be quantizes through the generating functional
 \begin{eqnarray}
  Z \:=\:
   \!\int \! D \varPhi\,
   \exp \Bigl\{ i S_{\delta}[\varPhi] \Bigr\}
  \,.
 \label{BV_Z} 
 \end{eqnarray}
Nondegeneracy of $S_{\delta}[\varPhi]$ is related to complete proper gauge fixing: the operators $X^{\gix{k}}_{\zza{l}\zzb{l-1}}$ must be chosen so as to provide complete and admissible fixing of the gauge freedom of the BV action (\ref{BV_master_action}). This imposes certain rank restrictions on the operators $X^{\gix{1}}_{\zza{l}\zzc{l-1}} Z^{\zzc{l-1}}_{\zzb{l}}$ from the ghost part and $X^{\gix{k}}_{\zza{l}\zzb{l-1}}$ from the auxiliary part of the gauge-fixed action.
These conditions will be analysed in the next subsection.

\newpar

Depending on a particular choice of $X^{\gix{k}}_{\zza{l}\zzb{l-1}}$, some fields in the gauge-fixed action may remain nondynamical and can be integrated with an ultralocal contribution to the measure.\footnote{
 This occurs when some or all of the gauge-fixing operators are algebraic matrices. For dynamical (differential) gauge-fixing operators, the integration introduces determinants of differential operators into the measure of (\ref{BV_Z_red}).
} This implies reducing the space of fields $\varPhi \to \breve{\varPhi}$ and constraining the gauge-fixed action in the exponent to the corresponding reduction surface ${S}[{\varPhi}] \to \breve{S}[\breve{\varPhi}]$, so that the generating functional takes the form
 \begin{eqnarray}
  Z \:=\:
   \int\! D \breve{\varPhi}\,
   \exp \Bigl\{ i \breve{S}_{\delta}[\breve{\varPhi}] \Bigr\}
   \,,
   \qquad\qquad
   \breve{S}_{\delta}[\breve{\varPhi}]
   = S_{\delta}[\varPhi(\breve{\varPhi})]
   \,.
 \label{BV_Z_red} 
 \end{eqnarray}

Alternatively, algebraically gauge-fixed fields may be eliminated already at the classical level, prior to quantization. A subset of nondynamical variables which can be expressed in terms of the remaining fields from their variational equations, can be reduced directly by substituting solutions of these equations into the gauge-fixed action (\ref{BV_gf_action}) \cite{Henneaux:1992ig}. This yields the reduced gauge-fixed action $\breve{S}[\breve{\varPhi}]$, which appear in (\ref{BV_Z_red}). For nondegenerate $S_{\delta}[\varPhi]$, classically equivalent reduced action $S_{\delta}[\varPhi(\breve{\varPhi})]$ is also nondegenerate and is suitable to quantizing the theory via (\ref{BV_Z_red}).

\newpar

The reduction is typically performed for the nondynamical subset consisting of certain Lagrange multipliers together with the corresponding ghosts and original fields. Together, these form the classically reducible set of fields. For purely algebraic gauge-fixing operators, the reduction in the action amounts to solving the variational equations with respect to the Lagrange multiplier fields:
 \begin{eqnarray}
  \frac{\var S_{\delta}}{\var \gB_{\gix{k}}^{\zzb{l}}} \equiv \frac{\var S_{\delta\, \text{aux}}}{\var \gB_{\gix{k}}^{\zzb{l}}}
  = 0
  \,:\qquad
  \Big| \quad
   \gC^{\zza{l-1}}_{\gix{k-1}} X^{\gix{k}}_{\zza{l-1}\zzb{l}}
   \! + \gC^{\zza{l+1}}_{\gix{k+1}} X^{\gix{k+1}}_{\zza{l+1}\zzb{l}}
    = 0
   \,,
   \qquad \text{\small $(1 \tleq l \tleq p ,\; 1 \tleq k \tleq l)$\,. }
 \label{BV_gf_constraints} 
 \end{eqnarray}
and substituting the general solution of these equations into the gauge-fixed minimal action, $S_0[\psi] + S_{\delta\,\text{gh}}[\psi,\gC_{\gix{0}},\gC_{\gix{1}}]$. The auxiliary part $S_{\delta\,\text{aux}}[\varPhi]$ vanishes on the reduction surface, since for delta-function gauge fixing it is proportional to the right-hand sides of (\ref{BV_gf_constraints}). Under proper gauge fixing, all extraghosts $k\tgeq2$ are fixed and eliminated by the reduction, along with a portion of the antighosts, minimal ghosts, and original fields. The part of the latter fields which remains unfixed, constitute the residual space $\breve{\varPhi}$, often referred as \emph{physical} space.

\newpar

The above brief review summarizes the folklore of the Batalin-Vilkovisky approach to quantizing gauge theories with reducible sets of generators. However, the precise criteria and scope of proper gauge fixing leading to a nondegenerate gauge-fixed action are much less transparent\HIDE{ and fully understandable}.
In their seminal paper \cite{Batalin:1983ggl},  Batalin and Vilkovisky proposed the so-called \emph{redundant} gauge-fixing scheme, which requires
 $\rank  X^{\gix{k}}_{\zza{l}\,\zza{l-1}}
  \!= \nnn_{l}-\nnn_{l+1}+ \nnn_{l+2} - ... +\msp^{p-l}\nnn_{p}
  = \mmm_{l} $,\,
following from the conditions that the kernels of $X^{\gix{k}}_{\zza{l}\,\zza{l-1}}$ and $X^{\gix{k}}_{\zza{l}\,\zza{l+1}}$ do not intersect and form the complementary linear subspaces in $\nnn_{l}$-dimensional linear space. However, the latter conditions is not necessary and the choice corresponds to the minimal possible ranks of admissible gauge-fixing operators.
Alternative options, such as full-rank $X^{\gix{k}}_{\zza{l}\,\zza{l-1}}$, are also possible, enabling simpler covariant formulations and offering specific applications \cite{Barvinsky:2022guw}.\footnote{For example, within the framework of restricted gauge theories, the use of full-rank gauge fixing allows one to establish a nontrivial relation between the one-loop effective actions of the restricted and parent gauge theories \cite{Barvinsky:2022guw}.}
In the main part of the article we apply the two extremal cases of gauges --- defined by the full-rank and minimal-rank gauge-fixing operators --- to the theory of antisymmetric tensor-spinor field.



\end{document}